%% file: kep.tex
\documentclass[apjl]{emulateapj}
\usepackage{comment}
\usepackage{ifthen}
\usepackage{amsmath}
\usepackage{amsfonts}
\usepackage{amssymb}

\input{newcommand_gen.tex}

\input{newcommand_spec_4.tex}
\input{newcommand_spec_5.tex}
\input{newcommand_spec_6.tex}
\input{newcommand_spec_7.tex}
\input{newcommand_spec_8.tex}

\newboolean{emulateapj}
\setboolean{emulateapj}{true}

\newboolean{astroph}
\setboolean{astroph}{true}

\shortauthors{Kipping \& Bakos}
\shorttitle{Kepler-4b through Kepler-8b}
\ifthenelse{\boolean{emulateapj}}{
    \newcommand{\titledag}{$\dagger$}
}{
    \newcommand{\titledag}{\dagger}
}

\begin{document}

\title{An Independent Analysis of Kepler-4b through
	Kepler-8b\altaffilmark{\titledag}}

\author{
	{\bf David Kipping \& G\'asp\'ar Bakos}
}
\altaffiltext{1}{Harvard-Smithsonian Center for Astrophysics,
	Cambridge, MA, dkipping@cfa.harvard.edu}

\altaffiltext{$\dagger$}{
Based on archival data of the \kept\ telescope. 
}


\begin{abstract}

We present two independent, homogeneous, global analyses of the transit \lcs,
radial velocities and spectroscopy of \kepivb, \kepvb, \kepvib,
\kepviib\ and \kepviiib, with numerous differences over the previous
methods.  These include: i) improved decorrelated parameter fitting set
used, ii) new limb darkening coefficients, iii) time stamps modified to
BJD for consistency with RV data, iv) two different methods for
compensating for the long integration time of Kepler LC data,
v) best-fit secondary eclipse depths and excluded upper limits, vi)
fitted mid-transit times, durations, depths and baseline fluxes for
individual transits.  We make several determinations not found in the
discovery papers: 
i) We detect a secondary eclipse for \kepviib\ of
depth $(47 \pm 14)$\,ppm and statistical significance 3.5-$\sigma$. 
We conclude reflected light is a much more plausible origin than
thermal emission and determine a geometric albedo of 
$A_g = (0.38 \pm 0.12)$. 
ii) We find that an eccentric orbit model for the
Neptune-mass planet \kepivb\ is detected at the 2-$\sigma$ level with
$e = (0.25 \pm 0.12)$. If confirmed, this would place \kepivb\ in a 
similar category as GJ 436b and HAT-P-11b as an eccentric, Neptune-mass 
planet. 
iii) We find weak evidence for a secondary eclipse in Kepler-5b of 2-$\sigma$
significance and depth $(26 \pm 17)$\,ppm. The most plausible explanation
is reflected light caused by a planet of geometric albedo 
$A_g = (0.15 \pm 0.10)$. 
iv) A 2.6-$\sigma$ peak in \kepvib\ TTV periodogram is detected and is not
easily explained as an aliased frequency. We find that mean-motion
resonant perturbers, non-resonant perturbers and a companion extrasolar
moon all provide inadequate explanations for this signal and the most
likely source is stellar rotation. 
v) We find different impact parameters relative to the discovery papers in
most cases, but generally self-consistent when compared to the two methods 
employed here. 
vi) We constrain the presence of mean motion resonant planets for all five
planets through an analysis of the mid-transit times. 
vii) We constrain the presence of extrasolar moons for all five planets.
viii) We constrain the presence of Trojans for all five planets.
\end{abstract}



\section{Introduction}
\label{sec:introduction}

The \emph{Kepler Mission} was successfully launched on March 7th 2009
and began science operations on May 12th of the same year. Designed to
detect Earth-like transits around Sun-like stars, the required
photometric precision is at the level of $20$\,ppm over 6.5 hours
integration on 12$^{\mathrm{th}}$ magnitude stars and early results
indicate this impressive precision is being reached \citep{bor09}.
In 2010, the first five transiting exoplanets (TEPs) to be 
discovered by the
\emph{Kepler Mission} were announced by the Kepler Science Team
\citep{bor10a}, known as \kepivb, \kepvb, \kepvib, \kepviib\ and
\kepviiib\ (\citet{bor10b}; \citet{koc10}; \citet{dun10};
\citet{lat10}; \citet{jen10a}). These expanded the sample of known
transiting exoplanets to about 75 at the time of announcement. 

The main objective of the \emph{Kepler Mission} is to discover
Earth-like planets, but the instrument naturally offers a vast array of
other science opportunities including detection of gas giants, searches
for thermal emission \citep{deming:2005} and/or reflection from 
exoplanets, detection of orbital phase curves \citep{knutson:2007} and 
ellipsoidal variations \citep{wel10}, asteroseismology \citep{christ:2010}
and transit timing \citep{agol:2005}, to name a few. Confirmation and 
follow-up of exoplanet transits is known to be a resource intensive 
activity and since the detection of new planets is Kepler's primary 
objective, it is logical for many of these other scientific tasks to be 
conducted by the astronomical community as a whole.

Independent and detailed investigations of the Kepler photometry
provides an ``acid-test'' of the methods employed by the Kepler Science
Team. Indeed, the distinct analysis of any scientific measurement
has always been a fundamental corner stone of the scientific method. In
this paper, we present two independent analyses of the discovery
photometry for the first five Kepler planets. We aim to not only test
the accuracy of the methods used in the discovery papers, but also test
our own methods by performing two separate studies. Both
methods will be using the same original data, as published in the
discovery papers.

Some additional data-processing tasks are run through the Kepler
reduced data, which we were not used in the original analyses presented
in the discovery papers, and are discussed in \S3. In this section, we
also discuss the generation of new limb darkening coefficients and
methods for compensating for the long integration time of the Kepler
long-cadence photometry. We also perform individual transit fits for
all available Kepler transits in order to search for transit timing
variations (TTV), transit duration variations (TDV) and other possible
changes. These will be used to provide a search for perturbing planets
and companion exomoons.

\section{Fitting Methodology}
\label{sec:meth}

\subsection{Method A}

In this work, the two of us adopt different algorithms for fitting the
observational data.  The first code has been written by D. Kipping and
is a Markov Chain Monte Carlo (MCMC) code (for a brief introduction on
this method, consult appendix A of \citet{teg04}) with a
Metropolis-Hastings algorithm, written in Fortran 77/90.  
The \lc\ model is generated using the analytic quadratic limb darkening
\citet{mandel:2002} routine, treating the specific stellar intensity with:

\begin{equation}
\frac{I_{\mu}}{I_1} = 1 - u_1 (1-\mu) - u_2 (1-\mu)^2
\end{equation}

The quadratic limb darkening coefficients
$u_1$ and $u_2$ are known to be highly correlated \citep{pal:2008b}, 
but a principal component analysis provides a more efficient fitting set:

\begin{align}
u_1' &= u_1 \cos 40^{\circ} - u_2 \sin 40^{\circ} \\
u_2' &= u_1 \sin 40^{\circ} + u_2 \cos 40^{\circ}
\end{align}

We also implement conditions to ensure the brightness profile is everywhere
positive and monotonically decreasing from limb to center: $u_1>0$ and
$0<u_1+u_2<1$ \citep{car08}.

Quantities relating to time differences are computed by solving
the bi-quartic equation for the various true anomalies of interest
following the methods detailed in \citet{kip08} and \citet{kip10a}.
These quantities include:

\begin{itemize}
\item The transit duration between the first and fourth contact, $T_{1,4}$.
\item The transit duration between the second and third contact, $T_{2,3}$.
\item The ingress and egress transit durations, $T_{12}$ and $T_{34}$ respectively.
\item The secondary eclipse full duration, $S_{1,4}$.
\item The time of the predicted secondary eclipse, $t_{\mathrm{sec}}$.
\item The offset time between when the radial velocity possesses maximum gradient
(for the instance closest to the time of conjunction) and the primary mid-transit,
$\Delta t_{\mathrm{RV}}$.
\end{itemize}

The latter two on this list are particularly important. For example, the RV offset
time provides a strong constraint on $e\cos\omega$. 

The principal fitting parameters for the transit model are
the orbital period, $P$, the ratio-of-radii squared, $p^2$, 
the \citet{kip10a} transit duration equation, $T_1$, 
the impact parameter, $b$, the epoch of mid-transit, $E$ 
and the Lagrange eccentricity parameters $k = e\cos\omega$
and $h = e\sin\omega$. $T_1$ is used as it describes the 
duration between the planet's centre crossing the stellar
limb and exiting under the same condition and this is known
to be a useful decorrelated parameter for light curve fits 
\citep{car08}. We also note that an approximate form is
required as the exact expression requires solving a bi-quartic
equation which is not easily invertible. In contrast, $T_1$
may be inverted to provide $a/R_*$ using \citep{kip10a}:

\begin{align}
\varrho_c &= \frac{1-e^2}{1+e\sin\omega} \\
a/R_* &= \sqrt{\Bigg(\frac{1-b^2}{\varrho_c^2}\Bigg) \csc^2\Bigg[\frac{\pi T_1 \sqrt{1-e^2}}{P \varrho_c^2}\Bigg] + \Bigg(\frac{b^2}{\varrho_c^2}\Bigg)}
\end{align}

Blending is accounted for using the prescription of \citet{kiptin10}
where we modify the formulas for an independent blend rather than a
self-blend.  The nightside pollution effect of the planet is at least
an order of magnitude below the detection sensitivity of \emph{Kepler} due to
the visible wavelength bandpass of the instrument.  Therefore, the only
blending we need account for are stellar blends.  In addition, the
model allows for an out-of-transit flux level ($F_{oot}$) to be fitted for. 
Unlike method B (see later), we fix $F_{oot}$ to be a constant during the
entire orbital phase of the planet. Secondary eclipses are produced
using the \citet{mandel:2002} code with no limb darkening and the
application of a transformation onto the secondary eclipse \lc\ which
effectively squashes the depth by a ratio such that the new depth is
equal to $(F_{pd}/F_\star)$ (planet dayside flux to stellar flux ratio) 
in the selected bandpass. This
technique ensures the secondary eclipse duration and shape are
correctly calculated. The observed flux is therefore modeled as:

\begin{equation}
F_{\mathrm{obs}}(t) = \Big(\frac{F_{\mathrm{model}}(t) + (B - 1)}{B}\Big) F_{oot}
\end{equation}

This model so far accounts for primary and secondary eclipses but an
additional subroutine has been written to model the radial velocity
variations of a single planet. Modeling the RV in conjunction with 
the transit data gives rise to
several potential complications. Firstly, we could try fits using either an
eccentric orbit or a fixed circular orbit model. An eccentric fit will always
find a non-zero eccentricity due to the boundary condition that $e>0$ \citep{lucy71}.
The effects of fitting versus not-fitting for eccentricity are explored 
in this work by presenting both fits for comparison.

A second complication is the possible presence of a linear drift in the
RVs due a distant massive planet. This drifts can give rise to artificial
eccentricity if not accounted for but their inclusion naturally increases
the uncertainty on all parameters coming from the RV fits. Thirdly, an
offset time, $\Delta t_{\mathrm{RV}}$, between when then RV signal has maximum
gradient (for the instance nearest the time of conjunction) and 
the mid-transit time can be included. For eccentric orbits, a non-zero 
offset always exists due to the orbital eccentricity. In our code this 
is calculated and denoted as $\Delta t_{\mathrm{ecc}}$ and is calculated exactly
by solving for the relevant true anomalies and computing the time interval
using the duration function of \citet{kip08}.  However, an additional offset
can also exist, $\Delta t_{\mathrm{troj}}$, which may due to a massive body in a 
Trojan orbit (i.e.~we have $\Delta t_{\mathrm{RV}} = \Delta t_{\mathrm{ecc}} + \Delta t_{\mathrm{RV}}$). This offset was first predicted by \citet{for07} and the
detection of a non-zero value of $\Delta t_{\mathrm{troj}}$ would indicate a Trojan.
For eccentric orbits, the error in $\Delta t_{\mathrm{ecc}}$ is often very large and
dominates the error budget in $\Delta t_{\mathrm{troj}}$, washing out any hints of a 
Trojan.

The question exists as to when one should include these two additional parameters.
Their perenial inclusion would result in very large errors for poorly characterized
orbits and this is clearly not desirable. We therefore choose to run a MCMC + 
simplex fit for all possible models and extract the lowest $\chi^2$ for the RV signal.
This minimum $\chi^2$ is then used to compute the Bayesian Information Criterion, or BIC 
(see \citet{schwarz:1978}; \citet{liddle2007}). BIC compares models with differing
numbers of free parameters ($k$), heavily penalizing those with more and the 
preferred model is given that yielding the lowest BIC, where:

\begin{equation}
\mathrm{BIC} = \chi^2 + k \log N
\end{equation}

Where $N$ is the number of data points. In total, there exists four possible 
models for the circular and four possible models for
the eccentric fit by switching on/off these two parameters. For both the
circular and eccentric fits, we select the model giving the lowest BIC\@. Therefore,
if for example the lowest BIC was that of a zero $\dot{\gamma}$ but non-zero 
$\Delta t_{\mathrm{troj}}$, we would fix the gradient term but let the offset be freely
fitted in the final results. The results of these preliminary investigations can be
found in Table~\ref{tab:BIC} of the Appendix.

The favored solution is generally the eccentric model over the circular model. This is
because the light curve derived stellar density, which we will later use in
our stellar evolution models, has a strong dependence on the eccentricity. In
general, fixing $e=0$ leads to unrealistically small error on $\rho_*$. However,
in some cases sparse RV phase coverage can lead to artificially large $e$ values.

The code is executed as a global routine, fitting all the RV and photometry 
simultaneously with a total of $\geq13$
free parameters: $E$, $p^2$, $T_1$, $b$, $P$, $u_1'$, $u_2'$, $h = e\sin\omega$, 
$k = e\cos\omega$, $\gamma_{\mathrm{rel}}$, $K$, $F_{oot}$
and $(F_{pd}/F_\star)$. The inclusion of $\dot{\gamma}$ and 
$\Delta t_{\mathrm{troj}}$ is reviewed on a case-by-case basis, as discussed.
Additionally, the blending factor,
which we denote as $B$, is allowed to float around its best fit value
in a Gaussian distribution of standard deviation equal to the
uncertainty in $B$. These values are reported in the original discovery
papers.

The final fit is then executed with 125,000 trials with the first 20\%
of trials discarded to allow for burn-in.  This leaves us with 100,000
trials to produce each {\em a posteriori} distribution, which is more
than adequate.  The \citet{gel92} statistic is calculated for all
fitted parameters to ensure good mixing. The final quoted values are
given by the median of the resultant MCMC trials. We define the
\emph{posian} and \emph{negian} as the maximum and minimum confidence
limits of the median respectively. The negian may be found by sorting
the list and extracting the entry which is 34.13\% of the total list
length. The posian is given by the entry which occurs at 68.27\% of the
total list length. These values may be then be used to determine the
confidence bounds on the median.

After the global fitting is complete, we produce the distribution of
the \lc\ derived stellar density and a Gaussian distribution for the
SME-derived effective temperature and metallicity around the values
published in the discovery papers and based on SME analysis of high
resolution spectra. These distributions consist of 100,000 values
and thus may be used to derive 100,000 estimates of the stellar
properties through a YY-isochrone analysis \citep{yi:2001}. For
errors in \teffstar\ and \feh, we frequently adopted double that which
was quoted in the discovery papers, as experience with HAT planets
has revealed that SME often underestimates these uncertainties.
Finally, the planetary parameters and their uncertainties were derived
by the direct combination of the {\em a posteriori} distributions of
the \lc, RV and stellar parameters. The stellar jitter squared is found
by taking the best-fit RV model residuals and calculating the variance
and subtracting the sum of the measurement uncertainties squared.

\subsection{Method B}

The second method builds on the codes developed under the HATNet
project, mostly written in C and shell scripts. These have been used
for the analysis of HATNet planet discoveries, such as
\citet{pal:2008a,bakos:2009,pal:2009b}, and have been heavily modified
for the current case of \kept\ analysis. We used a model for describing
the $F(t)$ flux of the star plus planet system through the full orbit
of the planet, from primary transit to occultation of the planet. In
our initial analysis we assumed a periodic model, and later we
considered the case of variable parameters as a function of transit
number. The $F(t)$ flux was a combination of three terms:
\begin{eqnarray}
\label{eq:methb}
	F(t) & = & \Phi(t,E,P,F_{oot},F_{oos},F_{oom},\zrstar,p,b^2,k,h,w) - \nonumber \\
	     &   & F_{oot}/(1+\epsilon_{pn}+B) \times F_t(t,E,P,\zrstar,p,b^2) - \nonumber \\
         &   & F_{oos}\times\epsilon_{pd}/((1+\epsilon_{pd}+B)p^2) \nonumber \\  
         &   & \times F_s(t,E,P,\zrstar,p,b^2)
\end{eqnarray}

Here $\Phi(.)$ (the first term) is the brightness (phase-curve) of the
star+planet+blend system without the effect of the transit and
occultation. The drop in flux due to the transit is reflected by the
second term, and the equivalent drop due to the occultation of the
planet by the third term. The $\Phi(.)$ phase-curve is a rather simple
step-function of three levels; $F_{oot}$ within $w$ times the duration
of the transit around the transit center, $F_{oos}$ within $w$ times
the duration of the occultation around the center of the occultation,
and $F_{oom}$ in between. For simplicity, we adopted $w=2$. Because the
transit and occultation centers and durations also enter the formula,
$\Phi()$ depends on a number of other parameters, namely the $t$ time,
$E$ mid-transit time, $P$ period, the $\zrstar$ parameter that is
characteristic of the duration of the transit, $p \equiv \rpl/\rstar$
planet to star radius ratio, $b^2$ impact parameter, and $k$ and $h$
Lagrangian orbital parameters. The location and total duration of the
transit and occultation are fully determined by the above parameters.
In general, the combined brightness $F_{oom}$ of the star plus planet
would smoothly vary from $F_{oot}$ outside of the primary transit to
$F_{oos}$ outside of the occultation, typically as a gradual
brightening as the planet shows its star-lit face towards the observer.
If the night-side brightness of the planet is $\epsilon_{pn} \cdot
F_\star$ \citep{kiptin10}, and the day-side brightness of the planet is
$\epsilon_{pd} \cdot F_\star$, and $B F_\star$ is the flux of a blend
contributing light to the system, then $1+B+\epsilon_{pn} = F_{oot}$
and $1+B+\epsilon_{pd} = F_{oos}$ (both equations normalized by
$F_\star$), i.e.~there is a constraint between $F_{oot}$ and $F_{oos}$.
The reason for introducing all three of $F_{oot}$, $F_{oom}$ and
$F_{oos}$ for characterizing the data was because the Kepler \lc, as
provided by the archive, was ``de-trended'', removing the (possible)
brightening of the phase-curve due to the planet. Because of this, we
found that introducing a more complex phase function was not warranted.
We note that while such de-trending may be necessary to eliminate
systematic effects, it also removes real physical variations of small
amplitude.


\begin{figure}[!ht]
\plotone{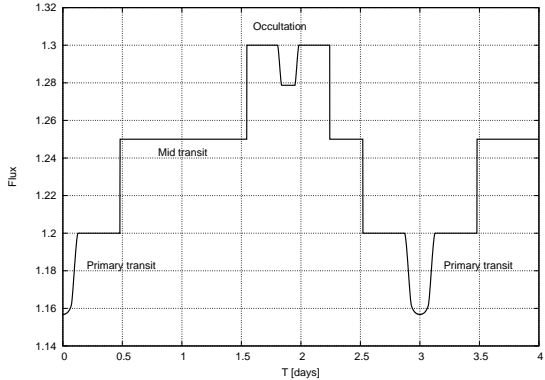}
\caption{
    {\bf Top:} The phase-curve of our simplistic model, showing a
	primary transit, an occultation, and a fix flux level $F_{oom}$ in
	between.  The model was generated with unrealistic planetary
	parameters to aid the visualization: $\rpl/\rstar = 0.2$ (large
	planet-to-star radius ratio), $b^2 = 0.25$, $\zrstar=10$, $E=0$,
	$P=3.0$, $k=0.1$, $h=-0.2$ (eccentric orbit), $\epsilon_{pn}=0.01$,
	$\epsilon_{pd}=0.02$, $B=0.2$, $F_{oot}=1.2$, $F_{oos}=1.3$,
	$F_{oom}=1.25$ and $w=2.0$.  See text for the meaning of these
	parameters. Due to the eccentric orbit, the occultation is clearly
	offset from halfway between the primary transits, and due to the
	lack of limb-darkening, the occultation exhibits a flat bottom.
\label{fig:fitmodeldemo}}
\end{figure}

The second term characterizes the dimming of the combined light of the
star, blend and planet during the primary transit. The loss of stellar
flux $F_t(.)$ was modeled using the analytic formula based on
\citet{mandel:2002} for the eclipse of a star by a planet
\citep{bakos:2009}, where the stellar flux was described by quadratic
limb-darkening. The quadratic limb-darkening coefficients were derived
by evaluating the stellar flux as a function of $\mu\equiv\cos\theta$
(where $\theta$ is the angle between the stellar surface normal vector
and the line-of-sight) using the quadratic $u_1$, $u_2$
coefficients. The formula takes into account
the $B$ blending (constant flux contribution by other sources), the
$\epsilon_{pn}$ night-side illumination of the planet (normalized by
$F_\star$), and the $F_{oot}$ out-of-transit flux level that was
normalized by an arbitrary constant.\footnote{Had the flux been
normalized by the $F_\star$ brightness of the star, without blending
and other sources, $F_{oot}=1$ would hold.}

Finally, the third term characterizes the dimming of the combined light
of the star, blend and planet during the occultation of the planet. The
day-side flux of the planet $F_{pd}$ is gradually lost as it moves
behind the star. This is modeled by a zero limb-darkening model
$F_s(.)$ of the secondary eclipse using the same \citet{mandel:2002}
formalism. This term also takes into account the $B$ blending, and the
$\epsilon_{pd}$ daytime flux of the system.

Our global modeling also included the radial velocity data. Following
the formalism presented by \citet{pal:2009}, the RV curve was
parametrized by an eccentric Keplerian orbit with semi-amplitude $K$,
Lagrangian orbital elements $(k,h)=e\times(\cos\omega,\sin\omega)$, and
systemic velocity $\gamma_{\mathrm{rel}}$.

The Kepler photometry and radial velocity data are connected in a
number of ways. For example, we assumed that there is a strict
periodicity in the individual transit times\footnote{Note that we also
performed a global analysis under method B by allowing the parameters
of the individual transits (such as the transit centers) to vary.}, and
the same $E$ and $P$ ephemeris of the system describes the photometric
and RV variations. Another example: the Lagrangian orbital parameters
$k$ and $h$ were not only determined by the RV data, but also by the
phase of the occultation of the planets in the Kepler data.

Altogether, the 13 parameters describing the physical model were $T_A$
(the $T_c$ of the first transit in the Kepler data), $T_B$ (the same
parameter for the last transit), $K$, $k = e\cos\omega$, $h =
e\sin\omega$, $\gamma_{\mathrm{rel}}$, $\rpl/\rstar$, $b^2$, $\zrstar$,
$\epsilon_{pd}$, $F_{oot}$, $F_{oom}$, and $F_{oos}$. The $B$ blending
factor was kept fixed at the values published in the \kept\ discovery
papers (\citet{bor10b}; \citet{koc10}; \citet{dun10}; \citet{lat10};
\citet{jen10a}). We adopted $\epsilon_{pn}=0$ for the night-side
emission of the planet. Fitting for $B$ and $\epsilon_{pn}$ would
require data with exceptional quality. 

The joint fit on the photometry and radial velocity data was performed
as described in \citet{bakos:2009}.  We minimized \chisq\ in the
parameter space by using a hybrid algorithm, combining the downhill
simplex method \citep[AMOEBA; see][]{press:1992} with the classical
linear least squares algorithm.  Uncertainties for the parameters were
derived using the Markov Chain Monte-Carlo method \citep[MCMC,
see][]{ford:2006}. The {\em a priori} distributions of the parameters
for these chains were chosen from a generic Gaussian distribution, with
eigenvalues and eigenvectors derived from the Fisher covariance matrix
for the best-fit value. The Fisher covariance matrix is calculated
analytically using the partial derivatives given by \citet{pal:2009}.

Stellar parameters were also determined in a Monte-Carlo fashion, using
the tools and methodology described in e.g.~\citet{bakos:2009}.
Basically, we used the MCMC distribution of $\arstar$ and corresponding
$\rhostar$, a Gaussian distribution of \teffstar\ and \feh\ around the
values published in the discovery papers and based on SME analysis of
high resolution spectra. For errors in \teffstar\ and \feh\ we
typically adopted double that of quoted in the discovery papers.
Finally, the planetary parameters and their uncertainties were derived
by the direct combination of the {\em a posteriori} distributions of
the \lc, RV and stellar parameters.

Throughout this text, the final value for any given parameter is the
{\em median} of the distribution derived by the Monte-Carlo Markov
Chain analysis. Error is the standard deviation around the final value.
Asymmetric error-bars are given if the negative and positive standard
deviations differ by 30\%. Parameter tables also list the RV
``jitter,'' which is a component of assumed astrophysical noise
intrinsic to the star that we add in quadrature to the RV measurement
uncertainties in order to have $\chi^{2}/(n-\mathrm{dof}) = 1$ from the
RV data for the global fit.

\subsection{Parameter variation search method}

\subsubsection{Individual fits}

In addition to global fits, we will search for variations in the
\lc\ parameters for each transit. Searching for changes in the
mid-transit time, commonly known as transit timing variations (TTV),
can be used to search for perturbing planets
\citep{agol:2005,holman:2005} or companion moons \citep{sar99,kip09a}.
Similarly, transit duration variations (TDV) provide a
complementary method of searching for exomoons \citep{kip09a,kip09b}.
We also look for depth changes and baseline variations
which may aid in the analysis of any putative signals.

Transit timing and duration measurements are made by splitting the time
series into individual transits and fitting independently.  In these
individual fits, we only fit for $p^2$, $b$, $T_1$, $t_c$
and $F_{oot}$ and float all other parameters around their global best-fit
determined value with their {\em a posteriori} distribution. The limb darkening
coefficients are fixed to the theoretical values for these fits, since epoch
to epoch variation in the LD coefficients is not expected. We choose to 
use method A for the individual fits as the parameter $b$ is more likely
to have a uniform prior than $b^2$ (as used in method B), from
geometric considerations. As for the global fits, we
perform 125,000 trials for each fit with the first 25,000 (20\%) trials being
used as a burn-in.

For the TDV, we define our transit duration as the time for the
sky-projection of the planet's center to move between overlapping the
stellar disc and exiting under the same condition, $T$.  This value was
first proposed by \citet{car08} due to its inherently low correlations
with other parameters. As a result, this definition of the duration may
be determined to a higher precision than the other durations.  $T$ has
the property that its uncertainty is exactly twice that of the
mid-transit time for a trapezoid approximated \lc. In reality, limb
darkening will cause this error to be slightly larger.  Nevertheless,
this property means that a useful definition of TDV is to halve the
actual variations to give TDV = $0.5 (T - \langle T\rangle)$ where
$\langle T\rangle$ is the globally-fitted value of $T$. The factor of
0.5 means that we expect the r.m.s.~scatter of the TDV to be equal to,
or slightly larger than, the TTV scatter. It is also worth noting that
in reality we could use any factor we choose rather than 0.5, because
TDV is really a measurement of the fractional variation in duration and
is not absolute in the sense that TTV is.

\subsubsection{Periodograms}

When analyzing the TTV and TDV signals, we first compute the $\chi^2$
of a model defined by a static system, i.e.~constant duration and
linear ephemeris. This $\chi^2$ value is naturally computed from the
MCMC uncertainties for each transit. In practice, these errors tend to
be overestimates. This is because the MCMC only produces accurate
errors if it is moving around the true global minimum. In reality,
slight errors in the limb darkening laws, the finite resolution of our
integration-time compensation techniques and the error in the assumed
photometric weights themselves mean that our favorite solution is
actually slightly offset from the true solution. Consequently, the MCMC
jumps sample a larger volume of parameter space than if they were
circling the true solution.

Despite the wide-spread use and therefore advocation of the MCMC method
for transit analysis, it is also unclear whether the MCMC method does
produce completely robust errors, particularly in light of the
inevitable hidden systemic effects. Therefore searching for excess
variance in the $\chi^2$ values may not be a completely reliable method
of searching for signals, echoed recently by \citet{gib09}. Concordantly,
in all of our analyses, we compare the $\chi^2$ of a static model
versus that of a signal and compute the F-test. The F-test has the
advantage that it does care not about the absolute $\chi^2$ values,
only the relative changes. Furthermore, it penalizes models which are
overly complex (Occam's razor).

We therefore compute periodograms by moving through a range of periods
in small steps of 1/1000 of the transit period and then fitting for a
sinusoidal signal's amplitude and phase in each case.  We compute the
$\chi^2$ of the new fit and then use an F-test to compute the false
alarm probability (FAP) to give the \emph{F-test periodogram}. Any
peaks below 5\% FAP are investigated further and we define a formal
detection limit of 3-$\sigma$ confidence. Compared to the generalized
Lomb-Scargle periodogram, the F-test is more conservative as it
penalizes models for being overly-complex and therefore offers a more
prudent evaluation of the significance of any signal. Conceptually, we
can see that the difference is that a Lomb-Scargle periodogram computes
the probability of obtaining a periodic signal by chance whereas the
F-test periodogram computes the statistical significance of preferring
the hypothesis of a periodic signal over a null hypothesis.

\subsubsection{Analysis of variance}

In general, there are two types of periodic signals we are interested
in searching for; those of period greater than the sampling rate
(long-period signals) and those of periods shorter than the sampling
rate (short-period signals). An example of a long-period signal would
be an exterior perturbing planet and a short-period signal example
would be a companion exomoon. Periodograms are well suited for
detecting long period signals above the Nyquist frequency, which is
equal to twice the sampling period. Short-period signals cannot be
reliably seeked below the Nyquist frequency. First of all, a period
equal to the sampling rate will always provide a strong peak. Periods
of an integer fraction of this frequency will also present strong
significances (e.g.~1/2, 1/3, 1/4, etc). As the interval between these
aliases becomes smaller, very short-period signals become entwined. If
a genuine short-period signal was present, it would exhibit at least
partial frequency mixing with one of these peaks to create a plethora
of short-period peaks. As a result, peaks in the periodogram below
$P_{\mathrm{Nyquist}}$ cannot be considered as evidence for authentic
signals. Therefore, we only plot our periodograms in the range of
$P_{\mathrm{Nyquist}}$ to twice the total temporal coverage.

To resolve this issue, it is possible to derive an exomoon's period by
taking the ratio of the TTV to TDV r.m.s.~amplitudes
\citep{kip09a,kip09b}. Since both of these signals are very short period, the
data would ostensibly be seen as excess scatter. These excess scatters
can be used to infer the r.m.s.~amplitude of each signal and then
derive the exomoon's period from the ratio. Excess scatter, or
variance, may be searched for by applying a simple $\chi^2$ test of the
data, assuming a null-hypothesis of no signal being present. If a
signal is found, or we wish to place constraints on the presence of
putative signal amplitudes, we use the $\chi^2$-distribution due
to both Gaussian noise plus an embedded high frequency sinusoidal
signal to model the variation.

\subsubsection{Phasing}

An unusual aspect of the individual data sets is that consecutive data
points are 30 minutes apart. Despite accounting for the long
integration times in our modeling (see later in \S3.5), we consider the
possibility that the long cadence data could produce artificial signals
in the parameter variation search. We propose that one method of
monitoring the effects of the long integration time is what we denote
as \emph{phasing}.

Using the globally fitted ephemeris, we know the mid-transit time for
all transits (assuming a non-variable system). We may compare the
difference between this mid-transit time and the closest data point in
a temporal sense. This time difference can take a maximum value of $\pm
15$ minutes (assuming no data gaps) and will vary from transit to
transit. We label this time difference as the \emph{phasing}.  In all
searches for parameter variations, a figure showing the phasing will
also be presented. This allows us to check whether any putative signals
are correlated to the phasing, which would indicate a strong
probability the signal is spurious.

\section{Data processing}

\subsection{Barycentric Julian Dates}

The radial velocity data from Kepler is provided in barycentric Julian
date (BJD$_{\mathrm{UTC}}$) but the photometry is given in heliocentric Julian data
(HJD$_{\mathrm{UTC}}$).  For consistency, we calculate the difference between HJD and
BJD for every time stamp and apply the appropriate correction. We use
the JPL Horizons ephemeris to perform this correction, which is
important given that the difference between HJD and BJD can be a few
seconds. We do not correct times from the UTC system to TDB/TT \citep{eastman:2010}
as the difference is effectively a constant systematic due to leap seconds which 
accumulate over decades timescale. Further, recent Kepler data products (e.g.~Kepler
Data Release 5) use the BJD$_{\mathrm{UTC}}$ system and thus there is a preference
to remain consistent with the probable future data format.

\subsection{Measurement uncertainties}

Unfortunately, the Kepler pipeline output did not include photometric
uncertainties which forces us to evaluate the measurement errors
ourselves. For Kepler-4b, Kepler-5b, Kepler-6b and Kepler-7b, we found
that the r.m.s.~scatter was constant over the observational period. In
order to estimate the standard deviation, we use the median absolute
deviation (MAD) from \citet{gau16} due to its robustness against
outliers. Assuming Gaussian errors, the standard deviation is given by
1.4286 multiplied by the MAD value. This value for the standard
deviation is then assigned to all data points as their measurement
uncertainty.

For Kepler-8b, this approach was no possible since the r.m.s.~scatter
varies with time. To estimate the errors, we calculate the standard
deviation using MAD in 50 point bins, excluding the transits and
secondary eclipses. This binning window is shifted along by one data
point and then repeated until we reach the end of the time series. The
MAD-derived standard deviations are then plotted as a function of
median time stamp from the binning window. Finally, we find that a
quadratic fit through the data provides an excellent estimation of the
trend. This formula is then used to generate all measurement
uncertainties.

\subsection{Outliers}

Although the data has been `cleaned' by the Kepler pipeline
\citep{jen10b}, we run the data through a highly robust outlier
detection algorithm developed by \citet{bea10} using the MAD method.
Essentially, we use MAD to estimate the standard deviation and then
estimate a sigma clipping level. This is found by setting the sigma
clipping level to a value such that if all the errors were Gaussian, we
would only expect to reject one valid data point by accident, which is
a function of the array length.

For each system we use MAD to estimate the outlier-robust standard
deviation to be $100.7$\,ppm, $126.6$\,ppm, 
$121.4$\,ppm and $96.4$\,ppm for Kepler-4b through
7b respectively. For Kepler-8b we find that the standard deviation 
of the photometry is variable as described in the previous subsection.
We present the list of rejected outlier points in Table~2 of the appendix.

\subsection{Limb Darkening}

Accurate limb darkening coefficients were calculated for the Kepler
bandpass for each planet. For the Kepler-bandpass, we used the high
resolution Kepler transmission function found at
http://keplergo.arc.nasa.gov/CalibrationResponse.shtml. We adopted the
SME-derived stellar properties reported in the original discovery
papers. We employed the \citet{kur06} atmosphere model database
providing intensities at 17 emergent angles, which we interpolated
linearly at the adopted $T_{\mathrm{eff}}$ and $\log g$ values. The
passband-convolved intensities at each of the emergent angles were
calculated following the procedure in \citet{cla00}. To compute the
coefficients we considered the following expression:

\begin{equation}
\frac{I(\mu)}{I(1)} = 1-\sum_{k=1}^{2} c_k (1-\mu)^{k}\,,
\end{equation}

where $I$ is the intensity, $\mu$ is the cosine of the emergent angle,
and $c_k$ are the coefficients. The final coefficients resulted from a
least squares singular value decomposition fit to 11 of the 17
available emergent angles. The reason to eliminate 6 of the angles is
avoiding excessive weight on the stellar limb by using a uniform
sampling (10 $\mu$ values from 0.1 to 1, plus $\mu=0.05$), as suggested
by \citet{dia95}. This whole process is performed by a Fortran code
written by I.~Ribas \citep{bea10}. The coefficients are given in the
final parameter tables for each planet.

\subsection{Integration Time}

The Kepler long-cadence (LC) photometry is produced by the onboard
stacking of 270 images of 6.02 seconds exposure. There is a 0.52 second
read-time on-top of the exposure time which leads to a net duty-cycle
of 91.4\%. The duty cycle is sufficiently high that we can consider the
LC photometry to be equivalent to one long exposure of 29.4244 minutes.
As a result of such a long integration, sharp changes in the
photometry, due to say a transit ingress, are smeared out into broader
morphologies. In itself, this does not pose a problem since the nature
of this smearing is well understood and easily predicted.

\citet{kip10b} discusses in detail the consequences of finite
integration times, with particular focus on Kepler's LC photometry. The
effects of the smearing may be accounted for by using a time-integrated
flux model for the \lc, as opposed to an instantaneous flux model. If
we denote the flux predicted at any instant by $F(t)$, the
time-integrated flux by $\tilde{F}(t)$ and the integration time by
$\mathcal{I}$, the observed flux will be given by:

\begin{equation}
\underline{\tilde{F}}(\underline{\tilde{t}}) = \frac{\int_{t=\tilde{t}
-\mathcal{I}/2}^{\tilde{t}+\mathcal{I}/2} F(t) \mathrm{d}t}{\int_{t=
\tilde{t}-\mathcal{I}/2}^{\tilde{t}+\mathcal{I}/2} \mathrm{d}t} \\
\end{equation}

\citet{kip10b} proposes two methods of numerically integrating this
expression; Simpson's composite rule and re-sampling. In this work, both of
us compensate for the effect in different ways. Method A employs
Simpson's composite rule and method B employs re-sampling. In both
cases, we use the expressions of \citet{kip10b}, given below in
equations (5) and (6) to choose our numerical resolution such that the
maximum photometric error induced by the finite numerical resolution is
less than the observed photometric uncertainties, as seen in
Table~\ref{tab:integ}.

\begin{eqnarray}
\underset{\mathrm{Comp. Simp.}}{\sigma_{\tilde{F}}} &= \frac{\delta}{\tau} \frac{\mathcal{I}}{24 m^2} \\
\underset{\mathrm{Resampling}}{\sigma_{\tilde{F}}} &= \frac{\delta}{\tau} \frac{\mathcal{I}}{8 N^2}
\end{eqnarray},
where $\delta$ is the transit depth, $\tau$ is ingress duration, and
$\mathcal{I}$ is the integration time.

\begin{table*}
\caption{\emph{
	Integration resolutions required to match observed
	photometric performance, based upon expressions of 
	\citet{kip10b}.}
}
\centering 
\begin{tabular}{c c c c c} 
\hline\hline 
Planet & Required $m$ for & $m$ used for  & Required $N$ & $N$ used \\
 & Comp. Simp. Rule & Comp. Simp. Rule & for Resampling & for Resampling \\
\hline
Kepler-4b & 0.9434 & 2 & 1.6339 & 4 \\
Kepler-5b & 1.7108 & 2 & 2.9632 & 4 \\
Kepler-6b & 2.1848 & 2 & 3.7839 & 4 \\
Kepler-7b & 1.7199 & 2 & 2.9790 & 4 \\
Kepler-8b & 1.0810 & 2 & 1.8723 & 4 \\ [1ex]
\hline\hline 
\end{tabular}
\label{tab:integ} 
\end{table*}

\section{Kepler-4\lowercase{b}}
\label{sec:kep4}

\input{kep4.tex}

\section{Kepler-5\lowercase{b}}
\label{sec:kep5}

\input{kep5.tex}

\section{Kepler-6\lowercase{b}}
\label{sec:kep6}

\input{kep6.tex}

\section{Kepler-7\lowercase{b}}
\label{sec:kep7}

\input{kep7.tex}

\section{Kepler-8\lowercase{b}}
\label{sec:kep8}

\input{kep8.tex}


\section{Discussion}
\label{sec:discussion}

Due to the large number of results presented in this paper, we will
here summarize the key findings not available in the original discovery
papers:

\begin{itemize}

\item[{\tiny$\blacksquare$}] Secondary eclipse of \kepviib\ is detected
to 3.5-$\sigma$ confidence with depth $(F_{pd}/F_\star) = 47
\pm 14$\,ppm, indicative of a geometric albedo of $A_g = (0.38 \pm 0.12)$.

\item[{\tiny$\blacksquare$}] A marginally significant orbital eccentricity is
detected for the Neptune-mass planet \kepivb\ to 2-$\sigma$
confidence, with an eccentricity of $e = (0.25 \pm 0.12)$.

\item[{\tiny$\blacksquare$}] A marginally significant (1.8-$\sigma$) secondary 
eclipse is detected for \kepvb\ with a depth of $(26 \pm 17)$\,ppm,
for which the most plausible explanation is reflected light due to a
geometric albedo of $A_g = (0.15 \pm 0.10)$.

\item[{\tiny$\blacksquare$}] A 2.6-$\sigma$ significance peak in the
TTV periodogram of \kepvib\ is detected, which is not easily explained
as an alias frequency. Perturbing planets and exomoons are unlikely to
be responsible either and currently our favored hypothesis is one of
stellar rotation.

\item[{\tiny$\blacksquare$}] We derive significantly different impact
parameters for all of the Kepler planets except \kepviib.

\end{itemize}

\subsection{Comparison of three hot-Neptunes}

The eccentricity of \kepivb\ is marginally detected to 2-$\sigma$ 
confidence. If confirmed, this would mean that three of the four 
hot-Neptunes discovered through the transit method have 
eccentricities around 0.15-0.20, the other two being GJ 436b and 
HAT-P-11b. HAT-P-26b is the fourth transiting hot-Neptune, recently
discovered by \citet{hartman:2010} and this system also possesses
a marginal eccentricity of $e = 0.124 \pm 0.060$. With HAT-P-26b
possessing a FAP of 12\% and Kepler-4b being 11.8\%, it would
seem likely that at least one of these two systems is genuinely 
eccentric.

In all cases, the  tidal circularization time is expected to be 
much lower than the age of the system for Jovian-like tidal 
dissipation values, which raises the question as to why these 
planets still retain eccentric orbits. Generally two hypotheses 
have been put forward: i) an unseen perturbing planet pumps 
the eccentricity of the hot Neptune ii) hot Neptunes have 
larger $Q_P$ values than expected.

At the time when only one example of an eccentric Neptune was known, GJ
436b, \citet{ribas:2008} made the reasonable deduction that is was more
likely an unseen perturber was present. However, given that three such
planets are now known, Occam's razor seems to favor the alternative
hypothesis. For all three hot-Neptunes to have perturbing planets which
all remain hidden from detection and yet produce nearly identical
eccentricity pumping levels appears to be a less likely scenario than
that of a common solution due to intrinsically different tidal
dissipation values and formation history. For \kepivb, the observed
eccentricity indicates $Q_P \geq 10^6$ if no eccentricity pumping is
occurring.

We also point out that the physical properties of Kepler-4b differ
greatly between the eccentric and circular fits (see 
Table~\ref{tab:neptunes}), due to the large impact of eccentricity 
on the lightcurve derived stellar density \citep{kip10a}. 
Further RV measurements are likely required to
resolve the eccentricity of this system, since an occultation is
generally not expected to be observable given the very low $R_P/R_*$.

%

\begin{table*}
\caption{
	\emph{Comparison of two known hot-Neptunes with Kepler-4b. 
        GJ 436b values taken from \citet{tor08} except
	eccentricity which comes from \citet{gillon:2007}; HAT-P-11b values
	taken from \citet{bakos:2009}. The classification of Kepler-4b
        depends greatly upon whether the system is confirmed as maintaining
        an eccentric orbit or not.}
}
\centering 
\begin{tabular}{l l l l l l l} 
\hline\hline 
Planet & Mass/$M_J$ & Radius/$R_J$ & Density/gcm$^{-3}$ & log$(g/$cgs) & e & $T_{eq}$/K \\
\hline
GJ 436b & $0.0729_{-0.0025}^{+0.0025}$ & $0.376_{-0.009}^{+0.008}$ & $1.69_{-0.12}^{+0.14}$ & $3.107 \pm 0.040$ & $0.16 \pm 0.02$ & $650 \pm 60$ \\
HAT-P-11b & $0.081 \pm 0.009$ & $0.422 \pm 0.014$ & $1.33 \pm 0.20$ & $3.05 \pm 0.06$ & $0.198 \pm 0.046$ & $878 \pm 15$ \\
Kepler-4b (circ) & $0.081 \pm 0.031$ & $0.460_{-0.084}^{+0.272}$ & $0.86_{-0.63}^{+0.97}$ & $2.90_{-0.38}^{+0.25}$ & $0^{\mathrm{*}}$ & $1777_{-132}^{+308}$ \\
Kepler-4b (ecc) & $0.096 \pm 0.023$ & $0.79_{-0.109}^{+0.145}$ & $0.24_{-0.13}^{+0.46}$ & $2.58_{-0.20}^{+0.28}$ & $0.25 \pm 0.12$ & $2215_{-339}^{+233}$ \\
\hline
Neptune & $0.05395$ & $0.3464$ & $1.638$ & $3.047$ & $0.011$ & 9.6 \\
Uranus &  $0.0457$ & $0.3575$ & $1.27$ & $2.939$ & $0.044$ & 12.1 \\ [1ex]
Saturn & $0.299$ & $0.843$ & $0.687$ & $3.019$ & $0.056$ & 26.0 \\ [1ex]
\hline\hline 
\end{tabular}
\label{tab:neptunes} 
\end{table*}

\subsection{Secondary eclipses}

We detect a secondary eclipse for the bloated \kepviib\ of depth $(47
\pm 14)$\,ppm and 3.5-$\sigma$ confidence. This is above our formal
detection threshold of 3-$\sigma$ thus represents an unambiguous
detection. Thermal emission is an unlikely source for the eclipse since
the depth indicates a brightness temperature of $2570_{-85}^{+108}$\,K,
which is far in excess of the equilibrium temperature of the planet of
$(1554 \pm 32)$\,K. A geometric albedo of $A_g = (0.38 \pm 0.12)$ offers
a more plausible explanation for this eclipse depth. However, we do
note that a significant nightside flux is apparently present which is
not consistent with the reflection hypothesis. We believe that this
nightside flux is likely an artifact of the Kepler pipeline whose
effects may mimic a long-cut filter. Although \kepviib\ exhibits
similarities to HD 209458b in terms of its very low density, the albedo
clearly marks the planet at distinct given that $A_g < 0.17$ for 
HD 209458b \citep{row08}. The study of \citet{burrows:08} seems to 
indicate that the albedo requires the presence of reflective clouds, 
possibly composed of iron and/or silicates, as seen in L-dwarf 
atmospheres. It is interesting to note that \kepvii\ is 
$1.3 \pm 0.2$ times more metal-rich than HD 209458, and the planet
has a higher equilibrium temperature of $(1565 \pm 30)$\,K than
that of HD 209458b with $(1130 \pm 50)$\,K.

We also note that, to our knowledge and if confirmed, the above 
measurement is the first determination of a transiting planet's albedo 
at visible wavelengths, and thus the first detection the associated reflected 
light. This compliments the recent polarimetric detection of reflected light of 
HD 189733b was made by \citet{ber08}, and recently confirmed in \citet{ber11}. 
Visible secondary eclipses of other planets 
have been made for HAT-P-7b \citep{bor09} and CoRoT-1b \citep{sne09} but
in both cases the eclipse can be explained through a combination of 
reflected light plus thermal emission or even simply pure thermal
emission for models of large day-night contrasts, which is in fact
expected for these extremely hot-Jupiters \citep{fortney:2008}. In 
contrast, thermal emission cannot explain the Kepler-7b secondary eclipse
unless the planet has an internal heat source with a luminosity 
equivalent to that of a M6.5 dwarf star, which is highly improbable.

We also find weak significance of a secondary eclipse for \kepvb\ of
$\sim$2-$\sigma$ confidence. Further transits will either confirm or
reject this hypothesis but we note that the obtained depth is quite
reasonable for this planet. Whilst thermal emission again seems
unlikely based upon the required temperature of 2500\,K versus the
equilibrium temperature of 1800\,K, a geometric albedo of $A_g = (0.15
\pm 0.10)$ offers a satisfactory explanation for the eclipse.

For \kepvib\ and \kepviib, we find no evidence of a secondary
eclipse, but the measurements do constrain the geometric albedos to
$A_g \leq 0.32$ and $A_g \leq 0.63$ at 3-$\sigma$ confidence. For
\kepivb, the measurement place no constraints on the geometric albedo.

\subsection{Differences with the discovery papers}

For almost all cases, we find significantly different impact parameters
from the discovery papers, but self-consistent between methods A and B.
There are also several examples of other parameters being different.
These can be easily reviewed by consulting Tables \ref{tab:kep4tab},
\ref{tab:kep5tab}, \ref{tab:kep6tab}, \ref{tab:kep7tab} and
\ref{tab:kep8tab}.  For example, for \kepivb, we find a very low
stellar density implying a larger stellar radius and thus larger
planetary radius.  
Other examples 
of differing parameters include the eccentricity and secondary eclipse 
parameters already discussed. 

\subsection{On the error budget of planetary parameters}

The exquisite \lcs\ measured by Kepler lead to very low uncertainties
on certain parameters, even in the long-cadence mode.  One example is
the fitted ratio-of-radii $p$, where the error is about 0.5\% (e.g.,
$p=\kepviieccLCrprstar$ for \kepviib). The median error for the first 70
known transiting exoplanets (TEPs) on $\rpl/\rstar$ is 0.9\%, about
double that of \kepviib. Limiting factors on the precision of this
purely geometrical factor are the limb-darkening values for the \kept\
bandpass and the slight degeneracy with other parameters that are
affected by the long-cadence binning, even if re-sampled models are
fitted (such as the $b$ impact parameter). Using short-cadence data,
deriving precise limb-darkening coefficients, and accumulating many
transits will significantly improve errors in $\rpl/\rstar$.

The typical errors on the period for the first five \kept\ planets are
$0.00004$\,days, about 6 times greater than the median error of the
period of the first 70 published TEPs, or about 20-times the error quoted in
e.g.~\citet{hartman:09} for HAT-P-12b, an example of a TEP discovery
using photometry data spanning 2 years. This error on the \kept\ planet
periods, however, comes from a dataset with short time-span (44\,days),
while the ground-based results come from multi-year campaigns. The
median error on the ephemeris ($T_c$) for the current 5 \kept\ planets
is $0.000145$\,days, about half that of the known TEPs
($0.0003$\,days). Precision of the ephemeris data will greatly improve
during the lifetime of the \kept\ mission just by accumulating more
data, and switching to short-cadence mode.

At first glance perhaps somewhat surprisingly, the errors on the \kept\
planetary masses and radii are not different from the rest of the
transiting exoplanet population. The typical error on planetary masses
for 70 transiting exoplanets is $0.06$\,\mjup, whereas the error for
radii is $0.05\,\rjup$. In our analysis the errors for planetary masses
and radii for Kepler planets have roughly the same values.

There are multiple reasons behind this. Planetary mass and radius scale with
the respective parameters of the host star. In our analysis, \mstar\
and \rstar\ are determined from stellar isochrones, using \teff, \feh\
and \arstar. Each of these quantities have significant errors. The
stellar \teff\ and \feh\ come from SME analysis of the spectra, which,
in turn, due to the faintness of the \kept\ targets, are of low S/N,
and have larger than usual errors. The other input parameter, \arstar,
is related to the \lc\ parameters \zrstar, $b$, and the RV parameters
$k$ and $h$. While \zrstar\ is quite precise for the \kept\ transits
(median error 0.02 as compared to 0.15 for the known TEPs), the $b$
impact parameter from the current LC \kept\ data-set have errors comparable
to ground-based transits. The main limiting factor, however, are the
$k$ and $h$ Lagrangian orbital parameters from the RV data. Due to the
faintness of the targets, the orbital fits have considerable errors in
$k$ and $h$. As a result, the error on \arstar\ for the current \kept\
data in this work (median error 0.26) is similar to that of the known
TEPs (median error 0.2). This effect is well visible on the isochrone
figures (Fig.~\ref{fig:kep4iso}, \ref{fig:kep5iso}, \ref{fig:kep6iso},
\ref{fig:kep7iso}, \ref{fig:kep8iso}), where the 1-$\sigma$ and
2-$\sigma$ confidence ellipses are shown for all \kept\ planets for
both the circular and eccentric solutions, with the latter clearly
covering a much larger area on the isochrones. Finally, the error in
the planetary mass also scales with the $K$ RV semi-amplitude, which
has typical errors for the \kept\ planets similar to the median error
of the known planets ($\Delta K \approx 4$\,\ms). Given the faintness
of the \kept\ targets this is an impressive result.

A breakthrough in both accuracy and precision of \mpl\ and \rpl\ can be
expected from exploiting the extraordinary precision of \kept, and
nailing down the stellar parameters via asteroseismology 
(see \citet{mulet:2009}, \citet{christ:2010}). Alternatively (or simultaneously), if parallaxes for the
\kept\ targets become available (possibly from the \kept\ data), then
analyses similar to that performed for HAT-P-11b \citep{bakos:2009} can
be performed, where the ``luminosity-indicator'' in the isochrone
fitting will be the absolute magnitude of the stars, as opposed to the
$\arstar$ constraint. A final possibility is a dynamical measurement of
the stellar mass by measuring the transit timing variations of two TEPs
within the same system \citep{agol:2005}.

\subsection{Transit timing with LC data}

After the discovery of an exoplanet in Kepler's field, the subsequent
photometry is switched to short-cadence (SC) mode. It was generally
expected that this would be the only way to produce meaningful TTV
studies. However, we have found that by carefully correcting for the
long integration time, the long-cadence (LC) photometry yields transit
times consistent with a linear ephemeris and of r.m.s.~scatters of
10-20 seconds are possible.

Despite the evident timing precision possible with the LC photometry,
some issues require further investigation. The effect of phasing, as
defined in \S2.3.4 remains somewhat unclear but in numerous instances
we find that peaks in the TTV and TDV periodograms occur near to the
phasing period.  Subsequent aliasing and frequency mixing also seems to
lead to numerous false positives in the periodograms and these issues
may be diminished by using SC data, but are unlikely to completely
disappear.

Out of all of the studied planets, we found only one peak in the timing
periodograms which does not appear spurious. \kepvib\ exhibits a
2.6-$\sigma$ peak of period $(17.27 \pm 0.84)$\,days and amplitude $(19.7
\pm 5.0)$\,seconds. We investigated the plausibility of this signal being
a perturbing planet or exomoon but found no convincing supporting
evidence, although such hypotheses could not excluded. Our favored
hypothesis is that of stellar rotation inducing such a signal, as
reported by \citet{alo08} for CoRoT-2b. The host star has a rotational
period of $23.5_{-5.9}^{+11.7}$\,days and thus it is consistent with the
TTV period. A bisector analysis, as performed by \citet{alo08}, is not
possible for this data due to the long cadence. However, the SC data
will be able to either confirm or reject this hypothesis.

Finally, we note that the uncertainties in all of the parameters from
individual fits show evidence for being overestimates, based upon the
observed scatter of these parameters between various transits. Both
methods employed in this analysis found very similar errors and thus
the reason for consistent overestimation of the errors remains unclear.
Both methods employed Markov Chain Monte Carlo (MCMC) techniques for
obtaining these errors, which may be linked to the overestimation. If
the MCMC trials are unable to sample the true global minimum, perhaps
due to slightly erroneous limb darkening coefficients or insufficient
numerical resolution in the integration time compensation procedure,
the errors are expected to be overestimated. Nevertheless, in searching
for evidence of signals, the accurate estimation of error bars is just
as important as the best-fit value. We have proposed a F-test
periodogram which is insensitive to absolute errors, only their
relative weightings, and also penalizes overly-complex models. Further
detailed investigation with future Kepler timings will be possible and
should shed light on this issue.

\subsection{Constraints on planets, moons \& Trojans}

We find no convincing evidence for perturbing planets, moons or Trojans
in any of the systems studied here. We are able to exclude the presence of 
mean-motion resonant (MMR) planets for \kepvb, \kepvib\ and \kepviiib\ 
of masses $\geq$ $0.79 \mearth$, $0.38 \mearth$ and $0.50 \mearth$ 
respectively, to 3-$\sigma$ confidence. The corresponding resonant 
configurations are period ratios of 3:2, 4:3 and 4:3. We do not 
yet possess sufficient temporal baseline to investigate MMR planets 
for \kepivb\ and \kepviib.

Extrasolar moons at a maximum orbital separation are excluded by the 
TTV measurements for \kepivb\ through to \kepviiib\ of masses $\geq$ 
$11.0 \mearth$, $10.6 \mearth$, $4.8 \mearth$, $2.5 \mearth$ and
$2.1 \mearth$ respectively, to 3-$\sigma$ confidence. 
Extrasolar moons at a minimum orbital separation are excluded by 
the TDV measurements of masses $\geq$ $13.3 \mearth$, 
$17.3 \mearth$, $8.2 \mearth$, $5.9 \mearth$ and $21.5 \mearth$ 
respectively, to 3-$\sigma$ confidence.

For \kepivb, \kepviib\ and \kepviiib, we do not yet possess sufficient
temporal baseline to search for Trojans through TTV measurements.
However, for \kepvb\ and \kepvib\ we are able to exclude Trojans of
angular displacement $\sim 10^{\circ}$ from L4/L5 of cumulative mass
$\geq$ $3.14 \mearth$ and $0.67 \mearth$ respectively, to 3-$\sigma$ 
confidence. For all five planets, we can inspect the photometry at
$\pm P/6$ from the transit center for signs of a photometric dip due
the occultation of the host star by the Trojans. This search yielded
no detections but does exclude Trojans of effective radii $\geq$ 
$1.22 R_{\oplus}$, $1.13 R_{\oplus}$, $0.87 R_{\oplus}$, 
$1.11 R_{\oplus}$ and $1.58 R_{\oplus}$ for the five planets 
sequentially, to 3-$\sigma$ confidence.


\acknowledgements

We would like to thank the Kepler Science Team and everyone who
contributed to making the \emph{Kepler Mission} possible. We are
extremely grateful to the Kepler Science Team for making the reduced
photometry and radial velocities of the first five planets publicly
available. Thanks to J. Jenkins, D. Latham and J. Rowe for
useful comments in revising this manuscript. Special thanks to the
anonymous referee for their highly useful comments in revising this manuscript.

D.M.K.~has been supported by STFC studentships, and by the HATNet as an
SAO predoctoral fellow. We acknowledge NASA NNG04GN74G, NNX08AF23G
grants, and Postdoctoral Fellowship of the NSF Astronomy and Astrophysics
Program (AST-0702843 for G.~B.).


\input{biblio.tex}

\appendix

\begin{table*}
\caption{\emph{List of important parameters used in this paper, in order of appearance.}} 
\centering 
\begin{tabular}{l l l} 
\hline\hline 
Parameter & Name & Definition \\ [0.5ex] 
\hline 
$u_1$ & Linear limb darkening coefficient & Linear term of the quadratic description of the stellar limb darkening \\
$u_2$ & Quadratic limb darkening coefficient & Quadratic term of the quadratic description of the stellar limb darkening \\
$\varrho_c$ & Mid companion-star separation & Companion-star separation in units of stellar radii at the approximate moment of mid-transit  \\
$\varrho$ & Companion-star separation & Companion-star separation in units of the stellar radius \\
$e$ & Eccentricity & Orbital eccentricity of the companion's orbit \\
$\omega$ & Argument of pericentre & Argument of pericentre of the companion's orbit \\
$P$ & Period & Orbital period of the companion \\
$p$ & Ratio-of-radii & Ratio of the companion's radius to the stellar radius ($R_P/R_*$) \\
$R_P$ & Radius of the companion & Radius of the companion \\ 
$R_*$ & Radius of the host star & Radius of the host star \\
$b$ & Impact parameter & Approximate value of $S$ when $\mathrm{d}S/\mathrm{d}t = 0$; also given by $b = a/R_* \varrho_c \cos i$ \\
$S$ & Sky-projected separation & Sky-projected separation of the companion's centre \\
& & and the host star's centre in units of stellar radii \\
$i$ & Inclination & Orbital inclination of the companion's orbit relative to the line-of-sight of the observer \\
$a/R_*$ & Semi-major axis & Semi-major axis of the companion's orbit in units of the stellar radius \\
$a$ & Semi-major axis & Semi-major axis of the companion's orbit \\
$E$ & Epoch of mid-transit & Mid-transit time of the first transit in the sequence \\
$t_C$ & Mid-transit time & Mid-transit time for a given epoch \\
$k$ & Lagrange parameter $k$ & $k = e\cos\omega$ \\ 
$h$ & Lagrange parameter $h$ & $k = e\sin\omega$ \\
$T$ & Transit duration & Time for companion to move across the stellar disc with entry \\ 
$t_{I}$ & First contact & Instant when $S = 1+p$ and $\mathrm{d}S/\mathrm{d}t < 0$ \\
$t_{II}$ & Second contact & Instant when $S = 1-p$ and $\mathrm{d}S/\mathrm{d}t < 0$ \\
$t_{C}$ & Mid-transit time & Instant when $\mathrm{d}S/\mathrm{d}t = 0$ i.e.~inferior conjunction \\
$t_{III}$ & Third contact & Instant when $S = 1-p$ and $\mathrm{d}S/\mathrm{d}t > 0$ \\
$t_{IV}$ & Fourth contact & Instant when $S = 1+p$ and $\mathrm{d}S/\mathrm{d}t > 0$ \\
$T_{1,4}$ & Total duration & Time for companion to move between contact points I and IV \\
$T_{2,3}$ & Full duration & Time for companion to move between contact points II and III \\
$T_{1,2}$ & Ingress duration & Time for companion to move between contact points 1 and 2 \\
$T_{3,4}$ & Egress duration & Time for companion to move between contact points 3 and 4 \\
$S_{1,4}$ & Total eclipse duration & Analogous to $T_{1,4}$ but for the secondary eclipse \\
$\tau$ & Ingress/Egress duration & For circular orbits, $t_{12} = t_{34} = \tau$ \\
$T_1$ & $T_1$ duration & Approximate expression for $T_1$ from \citet{kip10a} \\
$\rho_*$ & Stellar density & Average density of the host star \\
$\zeta/R_*$ & Zeta over $R_*$ & The reciprocal of one half of the transit duration using the \citet{tin05} approximation. \\
$\Upsilon/R_*$ & Upsilon over $R_*$ & The reciprocal of one half of the transit duration using the \citet{kip10a} approximation \\ 
$F_{oot}$ & Primary flux & Constant flux term for primary-portion of phase curve \\
$F_{oos}$ & Secondary flux & Constant flux term for secondary-portion of phase curve \\
$F_{oom}$ & Middle flux & Constant flux term for middle-portion of phase curve \\
$F_{obs}(t)$ & Observed flux & Flux observed by the instrument as a function of time \\
$F_{\mathrm{model}}(t)$ & Model flux & Flux from a given model as a function of time \\
$F_*$ & Stellar flux & Flux received from the star \\
$F_{\mathrm{pd}}$ & Day-side flux & Flux received from the day-side of the planet \\
$\epsilon_{\mathrm{pd}}$ & Relative dayside flux & Ratio of flux emitted from dayside of planet to stellar flux \\
$\epsilon_{\mathrm{pn}}$ & Relative nightside flux & Ratio of flux emitted from nightside of planet to stellar flux \\
$F_t$ & Transit model flux & Model flux for the primary transit \\
$F_s$ & Secondary model flux & Model flux for secondary eclipse \\
$B$ & Blending factor & Parameter to quantify amount of blended light in the aperture \\
$\Phi$ & Phase function & Flux received from planet+star+blend without transits or eclipses \\
$\gamma_{\mathrm{rel}}$ & RV offset & Constant used for offsetting the radial velocity measurements \\
$K$ & RV semi-amplitude & Semi-amplitude of the radial velocity signal \\ 
$T_{\mathrm{eff}}$ & Effective temperature & Effective temperature of the star \\
\feh & Metallicity & Metallicity of the host star \\
$v\sin i$ & Projected rotation velocity & Projected stellar rotation velocity \\
$M_*$ & Stellar mass & Mass of the host star \\
$M_P$ & Planetary mass & Mass of the planet \\
$L_*$ & Luminosity & Stellar luminosity \\
$M_V$ & Absolute magnitude & Absolute magnitude of the host star \\
$\rho_P$ planetary density & Density of the planet \\
$g$ & Stellar surface gravity & Surface gravity of the host star \\
$g_P$ & Planetary surface gravity & Surface gravity of the planet \\
$T_{\mathrm{eq}}$ & Equilibrium temperature & Equilibrium temperature of the planet \\ 
$\tau_{\mathrm{circ}}$ & Circularization timescale & Time for eccentricity to reduce by an e-fold \\
$Q_P$ & Quality factor of planet & Quality factor of planet due to tidal dissipation \\ [1ex]
\hline\hline 
\end{tabular}
\label{table:nonlin} 
\end{table*}

\begin{table*}
\caption{\emph{Comparison of various models tried for the radial velocity data of each system, in method A. Results come from a preliminary MCMC plus simplex. Details regarding the differences between the various models is provided in \S2.1. The Bayesian Information Criterion (BIC) is used to select the preferred model for the circular and eccentric fits, which is then used in the final analyses.}} 
\centering 
\begin{tabular}{c c c} 
\hline\hline 
Model & Circular Orbit BICs & Eccentric Orbit BICs \\ [0.5ex] 
\hline
\emph{Kepler-4} & \\
\hline
Free $\dot{\gamma}$; Free $\Delta t_{\mathrm{troj}}$ & 34.395 & 36.354 \\
Free $\dot{\gamma}$; Fixed $\Delta t_{\mathrm{troj}}$ & 31.144 & 33.300 \\
Fixed $\dot{\gamma}$; Free $\Delta t_{\mathrm{troj}}$ & 31.486 & 35.732 \\
Fixed $\dot{\gamma}$; Fixed $\Delta t_{\mathrm{troj}}$ & \textbf{29.556} & \textbf{32.685} \\
\hline
\emph{Kepler-5} & \\
\hline
Free $\dot{\gamma}$; Free $\Delta t_{\mathrm{troj}}$ & 29.275 & 29.222 \\
Free $\dot{\gamma}$; Fixed $\Delta t_{\mathrm{troj}}$ & \textbf{26.768} & \textbf{28.487} \\
Fixed $\dot{\gamma}$; Free $\Delta t_{\mathrm{troj}}$ & 43.990 & 47.592 \\
Fixed $\dot{\gamma}$; Fixed $\Delta t_{\mathrm{troj}}$ & 45.321 & 45.889 \\
\hline
\emph{Kepler-6} & \\
\hline
Free $\dot{\gamma}$; Free $\Delta t_{\mathrm{troj}}$ & 17.467 & 19.790 \\
Free $\dot{\gamma}$; Fixed $\Delta t_{\mathrm{troj}}$ & 21.234 & 17.693 \\
Fixed $\dot{\gamma}$; Free $\Delta t_{\mathrm{troj}}$ & \textbf{15.592} & 17.878 \\
Fixed $\dot{\gamma}$; Fixed $\Delta t_{\mathrm{troj}}$ & 20.157 & \textbf{16.834} \\
\hline
\emph{Kepler-7} & \\
\hline
Free $\dot{\gamma}$; Free $\Delta t_{\mathrm{troj}}$ & 13.195 & 17.276 \\
Free $\dot{\gamma}$; Fixed $\Delta t_{\mathrm{troj}}$ & 13.0113 & 15.518 \\
Fixed $\dot{\gamma}$; Free $\Delta t_{\mathrm{troj}}$ & 11.508 & 15.530 \\
Fixed $\dot{\gamma}$; Fixed $\Delta t_{\mathrm{troj}}$ & \textbf{11.160} & \textbf{13.664} \\
\hline
\emph{Kepler-8} & \\
\hline
Free $\dot{\gamma}$; Free $\Delta t_{\mathrm{troj}}$ & 30.877 & 37.213 \\
Free $\dot{\gamma}$; Fixed $\Delta t_{\mathrm{troj}}$ & 28.162 & 34.311 \\
Fixed $\dot{\gamma}$; Free $\Delta t_{\mathrm{troj}}$ & 28.163 & 34.340 \\
Fixed $\dot{\gamma}$; Fixed $\Delta t_{\mathrm{troj}}$ & \textbf{27.130} & \textbf{31.675} \\[1ex]
\hline\hline
\end{tabular}
\label{tab:BIC} 
\end{table*}

\begin{table*}
\caption{\emph{Outliers identified during our analysis for each system.}} 
\centering 
\begin{tabular}{c c c c} 
\hline\hline 
(BJD-2454900) & Standard deviations & (BJD-2454900) & Standard deviations \\ [0.5ex] 
\hline
\emph{Kepler-4} & \\
\hline
73.176 & 3.5 & 93.590 & 5.3 \\
84.660 & -5.1 & 96.982 & -3.7 \\
\hline
\emph{Kepler-5} & \\
\hline
55.418 & 3.9 & 53.641 & 4.1 \\
62.795 & 5.3 & 54.437 & 9.6 \\
70.683 & 4.3 & 71.030 & -3.8 \\
77.753 & 3.7 & & \\
\hline
\emph{Kepler-6} & \\
\hline
61.733 & 4.1 & 56.236 & 4.1 \\
67.945 & -3.5 & 66.024 & 3.6 \\
86.070 & 3.9 & 88.011 & 4.2 \\
55.929 & 3.5 & 94.775 & 4.0 \\
\hline
\emph{Kepler-7} & \\
\hline
72.113 & -3.6 & 84.415 & -3.5 \\
96.287 & -3.9 & 94.571 & 3.6 \\
\hline
\emph{Kepler-8} & \\
\hline
78.192 & 10.1 & 56.656 & 3.7 \\
58.883 & 17.3 & 58.781 & 4.2 \\
60.681 & 4.1 & 73.349 & 3.7 \\
53.039 & 4.7 & & \\ [1ex]
\hline\hline
\end{tabular}
\label{tab:outliers} 
\end{table*}

\end{document}

%% file: newcommand_gen.tex
\newcommand{\ctbd}[1]{}

\newcommand{\lc}{lightcurve}
\newcommand{\lcs}{lightcurves}
\newcommand{\Lc}{Lightcurve}

\newcommand{\chisq}{\ensuremath{\chi^2}}

\newcommand{\kms}{\ensuremath{\rm km\,s^{-1}}}
\newcommand{\ms}{\ensuremath{\rm m\,s^{-1}}}

\newcommand{\teff}{\ensuremath{T_{\rm eff}}}

\newcommand{\vsini}{\ensuremath{v \sin{i}}}
\newcommand{\feh}{\ensuremath{\rm [Fe/H]}}

\newcommand{\rsun}{\ensuremath{R_\sun}}
\newcommand{\msun}{\ensuremath{M_\sun}}
\newcommand{\lsun}{\ensuremath{L_\sun}}

\newcommand{\rstar}{\ensuremath{R_\star}}
\newcommand{\mstar}{\ensuremath{M_\star}}
\newcommand{\lstar}{\ensuremath{L_\star}}

\newcommand{\teffstar}{\ensuremath{T_{\rm eff\star}}}
\newcommand{\rhostar}{\ensuremath{\rho_\star}}

\newcommand{\rearth}{\ensuremath{R_\earth}}
\newcommand{\mearth}{\ensuremath{M_\earth}}

\newcommand{\rpl}{\ensuremath{R_{p}}}
\newcommand{\mpl}{\ensuremath{M_{p}}}

\newcommand{\arstar}{\ensuremath{a/\rstar}}
\newcommand{\zrstar}{\ensuremath{\zeta/\rstar}}
\newcommand{\fpdfs}{\ensuremath{F_{p,d}/F_\star}}

\newcommand{\rjup}{\ensuremath{R_{\rm J}}}
\newcommand{\mjup}{\ensuremath{M_{\rm J}}}

\newcommand{\reffig}[1]{Fig.~\ref{fig:#1}}

\newcommand{\reffigl}[1]{Figure~\ref{fig:#1}}

\newcommand{\kept}{{\em Kepler}}



%% file: newcommand_spec_4.tex
\newcommand{\kepiv}{Kepler-4}
\newcommand{\kepivb}{Kepler-4b}

\input{Kep4-info-cir.tex}

\input{Kep4-info-ecc.tex}

\newcommand{\kepivcirRVgammarel}{\kepivcirRVgamma}                           
\newcommand{\kepivcirSMEversion}{ii}                                       
\newcommand{\kepivcirSMEteff}{\ifthenelse{\equal{\kepivcirSMEversion}{i}}{\kepivcirSMEiteff}{\kepivcirSMEiiteff}}
\newcommand{\kepivcirSMEzfeh}{\ifthenelse{\equal{\kepivcirSMEversion}{i}}{\kepivcirSMEizfeh}{\kepivcirSMEiizfeh}}
\newcommand{\kepivcirSMEzfehshort}{\ifthenelse{\equal{\kepivcirSMEversion}{i}}{\kepivcirSMEizfehshort}{\kepivcirSMEiizfehshort}}
\newcommand{\kepivcirSMElogg}{\ifthenelse{\equal{\kepivcirSMEversion}{i}}{\kepivcirSMEilogg}{\kepivcirSMEiilogg}}
\newcommand{\kepivcirSMEvsin}{\ifthenelse{\equal{\kepivcirSMEversion}{i}}{\kepivcirSMEivsin}{\kepivcirSMEiivsin}}
\newcommand{\kepivcirSMEvmac}{\ifthenelse{\equal{\kepivcirSMEversion}{i}}{\kepivcirSMEivmac}{\kepivcirSMEiivmac}}
\newcommand{\kepivcirSMEvmic}{\ifthenelse{\equal{\kepivcirSMEversion}{i}}{\kepivcirSMEivmic}{\kepivcirSMEiivmic}}

\newcommand{\kepiveccRVgammarel}{\kepiveccRVgamma}                           
\newcommand{\kepiveccSMEversion}{ii}                                       
\newcommand{\kepiveccSMEteff}{\ifthenelse{\equal{\kepiveccSMEversion}{i}}{\kepiveccSMEiteff}{\kepiveccSMEiiteff}}
\newcommand{\kepiveccSMEzfeh}{\ifthenelse{\equal{\kepiveccSMEversion}{i}}{\kepiveccSMEizfeh}{\kepiveccSMEiizfeh}}
\newcommand{\kepiveccSMEzfehshort}{\ifthenelse{\equal{\kepiveccSMEversion}{i}}{\kepiveccSMEizfehshort}{\kepiveccSMEiizfehshort}}
\newcommand{\kepiveccSMElogg}{\ifthenelse{\equal{\kepiveccSMEversion}{i}}{\kepiveccSMEilogg}{\kepiveccSMEiilogg}}
\newcommand{\kepiveccSMEvsin}{\ifthenelse{\equal{\kepiveccSMEversion}{i}}{\kepiveccSMEivsin}{\kepiveccSMEiivsin}}
\newcommand{\kepiveccSMEvmac}{\ifthenelse{\equal{\kepiveccSMEversion}{i}}{\kepiveccSMEivmac}{\kepiveccSMEiivmac}}
\newcommand{\kepiveccSMEvmic}{\ifthenelse{\equal{\kepiveccSMEversion}{i}}{\kepiveccSMEivmic}{\kepiveccSMEiivmic}}

%% file: Kep4-info-cir.tex
\newcommand{\kepivcirCCtwomassJmag}{\ensuremath{11.122\pm0.023}}       
\newcommand{\kepivcirCCtwomassHmag}{\ensuremath{10.836\pm0.031}}       
\newcommand{\kepivcirCCtwomassKmag}{\ensuremath{10.805\pm0.023}}       
\newcommand{\kepivcirCCesoJKmag}{\ensuremath{0.340\pm0.009}}           
\newcommand{\kepivcirLCrprstar}{\ensuremath{0.0257\pm0.0009}}          
\newcommand{\kepivcirLCbsq}{\ensuremath{0.290_{-0.160}^{+0.199}}}      
\newcommand{\kepivcirLCimp}{\ensuremath{0.539_{-0.205}^{+0.155}}}      
\newcommand{\kepivcirLCzeta}{\ensuremath{12.58\pm0.17}}                
\newcommand{\kepivcirLCP}{\ensuremath{3.21344\pm0.00024}}            
\newcommand{\kepivcirSMEiteff}{\ensuremath{NULL\pmNULL}}               
\newcommand{\kepivcirSMEizfeh}{\ensuremath{NULL\pmNULL}}               
\newcommand{\kepivcirSMEizfehshort}{\ensuremath{NULL}}                 
\newcommand{\kepivcirSMEilogg}{\ensuremath{NULL\pmNULL}}               
\newcommand{\kepivcirSMEivsin}{\ensuremath{NULL\pmNULL}}               
\newcommand{\kepivcirSMEivmac}{\ensuremath{NULL}}                      
\newcommand{\kepivcirSMEivmic}{\ensuremath{NULL}}                      
\newcommand{\kepivcirSMEiiteff}{\ensuremath{5857\pm120}}               
\newcommand{\kepivcirSMEiizfeh}{\ensuremath{+0.17\pm0.06}}             
\newcommand{\kepivcirSMEiizfehshort}{\ensuremath{+0.17}}               
\newcommand{\kepivcirSMEiilogg}{\ensuremath{4.25\pm0.10}}              
\newcommand{\kepivcirSMEiivsin}{\ensuremath{2.2\pm1.0}}                
\newcommand{\kepivcirSMEiivmac}{\ensuremath{NULL}}                     
\newcommand{\kepivcirSMEiivmic}{\ensuremath{NULL}}                     
\newcommand{\kepivcirLBikep}{\ensuremath{0.4086}}                      
\newcommand{\kepivcirLBiikep}{\ensuremath{0.2633}}                     
\newcommand{\kepivcirISOmlong}{\ensuremath{1.275\pm0.109}}             
\newcommand{\kepivcirISOrlong}{\ensuremath{1.829_{-0.190}^{+0.426}}}   
\newcommand{\kepivcirISOrho}{\ensuremath{0.29\pm0.10}}                 
\newcommand{\kepivcirISOlogg}{\ensuremath{4.01\pm0.11}}                
\newcommand{\kepivcirISOlum}{\ensuremath{3.56_{-0.77}^{+2.03}}}        
\newcommand{\kepivcirISOmv}{\ensuremath{3.43\pm0.37}}                  
\newcommand{\kepivcirISOage}{\ensuremath{4.6_{-1.0}^{+1.7}}}           
\newcommand{\kepivcirISOMK}{\ensuremath{1.95\pm0.35}}                  
\newcommand{\kepivcirISOJK}{\ensuremath{0.38\pm0.02}}                  
\newcommand{\kepivcirRVK}{\ensuremath{9.3\pm1.4}}                      
\newcommand{\kepivcirRVgamma}{\ensuremath{-0.9\pm0.9}}                 
\newcommand{\kepivcirRVjitter}{\ensuremath{2.0}}                       
\newcommand{\kepivcirRVfitrms}{\ensuremath{3.8}}                       
\newcommand{\kepivcirPPi}{\ensuremath{84.3_{-3.2}^{+2.4}}}             
\newcommand{\kepivcirPPlogg}{\ensuremath{2.96_{-0.19}^{+0.13}}}        
\newcommand{\kepivcirPPar}{\ensuremath{5.41_{-0.82}^{+0.56}}}          
\newcommand{\kepivcirPParel}{\ensuremath{0.0462\pm0.0013}}             
\newcommand{\kepivcirPPrho}{\ensuremath{1.00\pm0.49}}                  
\newcommand{\kepivcirPPmlong}{\ensuremath{0.079\pm0.013}}              
\newcommand{\kepivcirPPrlong}{\ensuremath{0.457_{-0.056}^{+0.134}}}    
\newcommand{\kepivcirPPteff}{\ensuremath{1783_{-92}^{+166}}}           
\newcommand{\kepivcirXdist}{\ensuremath{600_{-65}^{+140}}}             

%% file: Kep4-info-ecc.tex
\newcommand{\kepiveccLCrprstar}{\ensuremath{0.0256\pm0.0006}}          
\newcommand{\kepiveccLCbsq}{\ensuremath{0.259_{-0.143}^{+0.132}}}      
\newcommand{\kepiveccLCimp}{\ensuremath{0.509_{-0.196}^{+0.113}}}      
\newcommand{\kepiveccLCzeta}{\ensuremath{12.61\pm0.20}}                
\newcommand{\kepiveccLCP}{\ensuremath{3.21340\pm0.00027}}            
\newcommand{\kepiveccSMEiteff}{\ensuremath{NULL\pmNULL}}               
\newcommand{\kepiveccSMEizfeh}{\ensuremath{NULL\pmNULL}}               
\newcommand{\kepiveccSMEizfehshort}{\ensuremath{NULL}}                 
\newcommand{\kepiveccSMEilogg}{\ensuremath{NULL\pmNULL}}               
\newcommand{\kepiveccSMEivsin}{\ensuremath{NULL\pmNULL}}               
\newcommand{\kepiveccSMEivmac}{\ensuremath{NULL}}                      
\newcommand{\kepiveccSMEivmic}{\ensuremath{NULL}}                      
\newcommand{\kepiveccSMEiiteff}{\ensuremath{5857\pm120}}               
\newcommand{\kepiveccSMEiizfeh}{\ensuremath{+0.17\pm0.06}}             
\newcommand{\kepiveccSMEiizfehshort}{\ensuremath{+0.17}}               
\newcommand{\kepiveccSMEiilogg}{\ensuremath{4.25\pm0.10}}              
\newcommand{\kepiveccSMEiivsin}{\ensuremath{2.2\pm1.0}}                
\newcommand{\kepiveccSMEiivmac}{\ensuremath{NULL}}                     
\newcommand{\kepiveccSMEiivmic}{\ensuremath{NULL}}                     
\newcommand{\kepiveccISOmlong}{\ensuremath{1.282_{-0.123}^{+0.160}}}   
\newcommand{\kepiveccISOrlong}{\ensuremath{1.922_{-0.430}^{+0.572}}}   
\newcommand{\kepiveccISOrho}{\ensuremath{0.25_{-0.11}^{+0.34}}}        
\newcommand{\kepiveccISOlogg}{\ensuremath{3.98\pm0.18}}                
\newcommand{\kepiveccISOlum}{\ensuremath{3.91_{-1.52}^{+2.98}}}        
\newcommand{\kepiveccISOmv}{\ensuremath{3.33\pm0.58}}                  
\newcommand{\kepiveccISOage}{\ensuremath{4.3_{-1.0}^{+1.8}}}           
\newcommand{\kepiveccRVK}{\ensuremath{9.1\pm1.5}}                      
\newcommand{\kepiveccRVk}{\ensuremath{0.083_{-0.036}^{+0.114}}}        
\newcommand{\kepiveccRVh}{\ensuremath{0.062_{-0.219}^{+0.166}}}        
\newcommand{\kepiveccRVgamma}{\ensuremath{-0.8\pm1.0}}                 
\newcommand{\kepiveccRVjitter}{\ensuremath{1.6}}                       
\newcommand{\kepiveccRVeccen}{\ensuremath{0.208\pm0.104}}              
\newcommand{\kepiveccRVomega}{\ensuremath{76\pm145}}                   
\newcommand{\kepiveccRVfitrms}{\ensuremath{3.5}}                       
\newcommand{\kepiveccPPi}{\ensuremath{84.2_{-4.3}^{+2.6}}}             
\newcommand{\kepiveccPPlogg}{\ensuremath{2.92\pm0.19}}                 
\newcommand{\kepiveccPPar}{\ensuremath{5.17_{-0.94}^{+1.50}}}          
\newcommand{\kepiveccPParel}{\ensuremath{0.0463\pm0.0017}}             
\newcommand{\kepiveccPPrho}{\ensuremath{0.87_{-0.38}^{+1.16}}}         
\newcommand{\kepiveccPPmlong}{\ensuremath{0.077\pm0.015}}              
\newcommand{\kepiveccPPrlong}{\ensuremath{0.480_{-0.109}^{+0.145}}}    
\newcommand{\kepiveccPPteff}{\ensuremath{1832\pm212}}                  
\newcommand{\kepiveccXdist}{\ensuremath{630_{-142}^{+188}}}            

%% file: newcommand_spec_5.tex
\newcommand{\kepv}{Kepler-5}
\newcommand{\kepvb}{Kepler-5b}

\input{Kep5-info-cir.tex}

\input{Kep5-info-ecc.tex}

\newcommand{\kepvcirRVgammarel}{\kepvcirRVgamma}                           
\newcommand{\kepvcirSMEversion}{ii}                                       
\newcommand{\kepvcirSMEteff}{\ifthenelse{\equal{\kepvcirSMEversion}{i}}{\kepvcirSMEiteff}{\kepvcirSMEiiteff}}
\newcommand{\kepvcirSMEzfeh}{\ifthenelse{\equal{\kepvcirSMEversion}{i}}{\kepvcirSMEizfeh}{\kepvcirSMEiizfeh}}
\newcommand{\kepvcirSMEzfehshort}{\ifthenelse{\equal{\kepvcirSMEversion}{i}}{\kepvcirSMEizfehshort}{\kepvcirSMEiizfehshort}}
\newcommand{\kepvcirSMElogg}{\ifthenelse{\equal{\kepvcirSMEversion}{i}}{\kepvcirSMEilogg}{\kepvcirSMEiilogg}}
\newcommand{\kepvcirSMEvsin}{\ifthenelse{\equal{\kepvcirSMEversion}{i}}{\kepvcirSMEivsin}{\kepvcirSMEiivsin}}
\newcommand{\kepvcirSMEvmac}{\ifthenelse{\equal{\kepvcirSMEversion}{i}}{\kepvcirSMEivmac}{\kepvcirSMEiivmac}}
\newcommand{\kepvcirSMEvmic}{\ifthenelse{\equal{\kepvcirSMEversion}{i}}{\kepvcirSMEivmic}{\kepvcirSMEiivmic}}

\newcommand{\kepveccRVgammarel}{\kepveccRVgamma}                           
\newcommand{\kepveccSMEversion}{ii}                                       
\newcommand{\kepveccSMEteff}{\ifthenelse{\equal{\kepveccSMEversion}{i}}{\kepveccSMEiteff}{\kepveccSMEiiteff}}
\newcommand{\kepveccSMEzfeh}{\ifthenelse{\equal{\kepveccSMEversion}{i}}{\kepveccSMEizfeh}{\kepveccSMEiizfeh}}
\newcommand{\kepveccSMEzfehshort}{\ifthenelse{\equal{\kepveccSMEversion}{i}}{\kepveccSMEizfehshort}{\kepveccSMEiizfehshort}}
\newcommand{\kepveccSMElogg}{\ifthenelse{\equal{\kepveccSMEversion}{i}}{\kepveccSMEilogg}{\kepveccSMEiilogg}}
\newcommand{\kepveccSMEvsin}{\ifthenelse{\equal{\kepveccSMEversion}{i}}{\kepveccSMEivsin}{\kepveccSMEiivsin}}
\newcommand{\kepveccSMEvmac}{\ifthenelse{\equal{\kepveccSMEversion}{i}}{\kepveccSMEivmac}{\kepveccSMEiivmac}}
\newcommand{\kepveccSMEvmic}{\ifthenelse{\equal{\kepveccSMEversion}{i}}{\kepveccSMEivmic}{\kepveccSMEiivmic}}

%% file: Kep5-info-cir.tex
\newcommand{\kepvcirLCrprstar}{\ensuremath{0.0789\pm0.0002}}           
\newcommand{\kepvcirLCbsq}{\ensuremath{0.021_{-0.012}^{+0.027}}}       
\newcommand{\kepvcirLCimp}{\ensuremath{0.143_{-0.057}^{+0.067}}}       
\newcommand{\kepvcirLCzeta}{\ensuremath{11.34\pm0.02}}                 
\newcommand{\kepvcirLCP}{\ensuremath{3.548459\pm0.000035}}             
\newcommand{\kepvcirSMEiteff}{\ensuremath{NULL\pmNULL}}                
\newcommand{\kepvcirSMEizfeh}{\ensuremath{NULL\pmNULL}}                
\newcommand{\kepvcirSMEizfehshort}{\ensuremath{NULL}}                  
\newcommand{\kepvcirSMEilogg}{\ensuremath{NULL\pmNULL}}                
\newcommand{\kepvcirSMEivsin}{\ensuremath{NULL\pmNULL}}                
\newcommand{\kepvcirSMEivmac}{\ensuremath{NULL}}                       
\newcommand{\kepvcirSMEivmic}{\ensuremath{NULL}}                       
\newcommand{\kepvcirSMEiiteff}{\ensuremath{6297\pm60}}                 
\newcommand{\kepvcirSMEiizfeh}{\ensuremath{+0.04\pm0.06}}              
\newcommand{\kepvcirSMEiizfehshort}{\ensuremath{+0.04}}                
\newcommand{\kepvcirSMEiilogg}{\ensuremath{3.96\pm0.10}}               
\newcommand{\kepvcirSMEiivsin}{\ensuremath{4.8\pm1.0}}                 
\newcommand{\kepvcirSMEiivmac}{\ensuremath{NULL}}                      
\newcommand{\kepvcirSMEiivmic}{\ensuremath{NULL}}                      
\newcommand{\kepvcirLBikep}{\ensuremath{0.3462}}                       
\newcommand{\kepvcirLBiikep}{\ensuremath{0.2972}}                      
\newcommand{\kepvcirISOmlong}{\ensuremath{1.352\pm0.026}}              
\newcommand{\kepvcirISOrlong}{\ensuremath{1.709_{-0.017}^{+0.029}}}    
\newcommand{\kepvcirISOrho}{\ensuremath{0.38_{-0.02}^{+0.01}}}         
\newcommand{\kepvcirISOlogg}{\ensuremath{4.10\pm0.01}}                 
\newcommand{\kepvcirISOlum}{\ensuremath{4.12\pm0.22}}                  
\newcommand{\kepvcirISOmv}{\ensuremath{3.22\pm0.07}}                   
\newcommand{\kepvcirISOage}{\ensuremath{2.9\pm0.2}}                    
\newcommand{\kepvcirRVK}{\ensuremath{228.1\pm8.1}}                     
\newcommand{\kepvcirRVgamma}{\ensuremath{-44.8\pm6.5}}                 
\newcommand{\kepvcirRVjitter}{\ensuremath{15.5}}                       
\newcommand{\kepvcirRVfitrms}{\ensuremath{16.8}}                       
\newcommand{\kepvcirPPi}{\ensuremath{88.7\pm0.6}}                      
\newcommand{\kepvcirPPlogg}{\ensuremath{3.48\pm0.02}}                  
\newcommand{\kepvcirPPar}{\ensuremath{6.34_{-0.09}^{+0.04}}}           
\newcommand{\kepvcirPParel}{\ensuremath{0.0503\pm0.0003}}              
\newcommand{\kepvcirPPrho}{\ensuremath{1.15\pm0.06}}                   
\newcommand{\kepvcirPPmlong}{\ensuremath{2.094\pm0.079}}               
\newcommand{\kepvcirPPrlong}{\ensuremath{1.312_{-0.014}^{+0.026}}}     
\newcommand{\kepvcirPPteff}{\ensuremath{1770\pm19}}                    
\newcommand{\kepvcirXdist}{\ensuremath{906\pm17}}                      

%% file: Kep5-info-ecc.tex
\newcommand{\kepveccLCrprstar}{\ensuremath{0.0789\pm0.0002}}           
\newcommand{\kepveccLCbsq}{\ensuremath{0.022_{-0.013}^{+0.030}}}       
\newcommand{\kepveccLCimp}{\ensuremath{0.149_{-0.060}^{+0.072}}}       
\newcommand{\kepveccLCzeta}{\ensuremath{11.35\pm0.02}}                 
\newcommand{\kepveccLCP}{\ensuremath{3.548459\pm0.000035}}             
\newcommand{\kepveccSMEiteff}{\ensuremath{NULL\pmNULL}}                
\newcommand{\kepveccSMEizfeh}{\ensuremath{NULL\pmNULL}}                
\newcommand{\kepveccSMEizfehshort}{\ensuremath{NULL}}                  
\newcommand{\kepveccSMEilogg}{\ensuremath{NULL\pmNULL}}                
\newcommand{\kepveccSMEivsin}{\ensuremath{NULL\pmNULL}}                
\newcommand{\kepveccSMEivmac}{\ensuremath{NULL}}                       
\newcommand{\kepveccSMEivmic}{\ensuremath{NULL}}                       
\newcommand{\kepveccSMEiiteff}{\ensuremath{6297\pm60}}                 
\newcommand{\kepveccSMEiizfeh}{\ensuremath{+0.04\pm0.06}}              
\newcommand{\kepveccSMEiizfehshort}{\ensuremath{+0.04}}                
\newcommand{\kepveccSMEiilogg}{\ensuremath{3.96\pm0.10}}               
\newcommand{\kepveccSMEiivsin}{\ensuremath{4.8\pm1.0}}                 
\newcommand{\kepveccSMEiivmac}{\ensuremath{NULL}}                      
\newcommand{\kepveccSMEiivmic}{\ensuremath{NULL}}                      
\newcommand{\kepveccISOmlong}{\ensuremath{1.358\pm0.039}}              
\newcommand{\kepveccISOrlong}{\ensuremath{1.730\pm0.091}}              
\newcommand{\kepveccISOrho}{\ensuremath{0.37_{-0.04}^{+0.06}}}         
\newcommand{\kepveccISOlogg}{\ensuremath{4.09\pm0.04}}                 
\newcommand{\kepveccISOlum}{\ensuremath{4.21\pm0.48}}                  
\newcommand{\kepveccISOmv}{\ensuremath{3.20\pm0.13}}                   
\newcommand{\kepveccISOage}{\ensuremath{2.9\pm0.2}}                    
\newcommand{\kepveccRVK}{\ensuremath{228.5\pm8.1}}                     
\newcommand{\kepveccRVk}{\ensuremath{-0.009\pm0.023}}                  
\newcommand{\kepveccRVh}{\ensuremath{0.008\pm0.044}}                   
\newcommand{\kepveccRVgamma}{\ensuremath{-43.9\pm6.8}}                 
\newcommand{\kepveccRVjitter}{\ensuremath{15.0}}                       
\newcommand{\kepveccRVeccen}{\ensuremath{0.038\pm0.028}}               
\newcommand{\kepveccRVomega}{\ensuremath{150\pm83}}                    
\newcommand{\kepveccRVfitrms}{\ensuremath{16.3}}                       
\newcommand{\kepveccPPi}{\ensuremath{88.6_{-0.7}^{+0.6}}}              
\newcommand{\kepveccPPlogg}{\ensuremath{3.47\pm0.04}}                  
\newcommand{\kepveccPPar}{\ensuremath{6.26\pm0.29}}                    
\newcommand{\kepveccPParel}{\ensuremath{0.0504\pm0.0005}}              
\newcommand{\kepveccPPrho}{\ensuremath{1.11_{-0.14}^{+0.19}}}          
\newcommand{\kepveccPPmlong}{\ensuremath{2.101\pm0.087}}               
\newcommand{\kepveccPPrlong}{\ensuremath{1.330\pm0.071}}               
\newcommand{\kepveccPPteff}{\ensuremath{1779\pm43}}                    
\newcommand{\kepveccXdist}{\ensuremath{916\pm49}}                      

%% file: newcommand_spec_6.tex
\newcommand{\kepvi}{Kepler-6}
\newcommand{\kepvib}{Kepler-6b}

\input{Kep6-info-cir.tex}

\input{Kep6-info-ecc.tex}

\newcommand{\kepvicirRVgammarel}{\kepvicirRVgamma}                           
\newcommand{\kepvicirSMEversion}{ii}                                       
\newcommand{\kepvicirSMEteff}{\ifthenelse{\equal{\kepvicirSMEversion}{i}}{\kepvicirSMEiteff}{\kepvicirSMEiiteff}}
\newcommand{\kepvicirSMEzfeh}{\ifthenelse{\equal{\kepvicirSMEversion}{i}}{\kepvicirSMEizfeh}{\kepvicirSMEiizfeh}}
\newcommand{\kepvicirSMEzfehshort}{\ifthenelse{\equal{\kepvicirSMEversion}{i}}{\kepvicirSMEizfehshort}{\kepvicirSMEiizfehshort}}
\newcommand{\kepvicirSMElogg}{\ifthenelse{\equal{\kepvicirSMEversion}{i}}{\kepvicirSMEilogg}{\kepvicirSMEiilogg}}
\newcommand{\kepvicirSMEvsin}{\ifthenelse{\equal{\kepvicirSMEversion}{i}}{\kepvicirSMEivsin}{\kepvicirSMEiivsin}}
\newcommand{\kepvicirSMEvmac}{\ifthenelse{\equal{\kepvicirSMEversion}{i}}{\kepvicirSMEivmac}{\kepvicirSMEiivmac}}
\newcommand{\kepvicirSMEvmic}{\ifthenelse{\equal{\kepvicirSMEversion}{i}}{\kepvicirSMEivmic}{\kepvicirSMEiivmic}}

\newcommand{\kepvieccRVgammarel}{\kepvieccRVgamma}                           
\newcommand{\kepvieccSMEversion}{ii}                                       
\newcommand{\kepvieccSMEteff}{\ifthenelse{\equal{\kepvieccSMEversion}{i}}{\kepvieccSMEiteff}{\kepvieccSMEiiteff}}
\newcommand{\kepvieccSMEzfeh}{\ifthenelse{\equal{\kepvieccSMEversion}{i}}{\kepvieccSMEizfeh}{\kepvieccSMEiizfeh}}
\newcommand{\kepvieccSMEzfehshort}{\ifthenelse{\equal{\kepvieccSMEversion}{i}}{\kepvieccSMEizfehshort}{\kepvieccSMEiizfehshort}}
\newcommand{\kepvieccSMElogg}{\ifthenelse{\equal{\kepvieccSMEversion}{i}}{\kepvieccSMEilogg}{\kepvieccSMEiilogg}}
\newcommand{\kepvieccSMEvsin}{\ifthenelse{\equal{\kepvieccSMEversion}{i}}{\kepvieccSMEivsin}{\kepvieccSMEiivsin}}
\newcommand{\kepvieccSMEvmac}{\ifthenelse{\equal{\kepvieccSMEversion}{i}}{\kepvieccSMEivmac}{\kepvieccSMEiivmac}}
\newcommand{\kepvieccSMEvmic}{\ifthenelse{\equal{\kepvieccSMEversion}{i}}{\kepvieccSMEivmic}{\kepvieccSMEiivmic}}

%% file: Kep6-info-cir.tex
\newcommand{\kepvicirLCbsq}{\ensuremath{0.088_{-0.043}^{+0.047}}}      
\newcommand{\kepvicirLCimp}{\ensuremath{0.296_{-0.099}^{+0.068}}}      
\newcommand{\kepvicirLCzeta}{\ensuremath{14.67\pm0.02}}                
\newcommand{\kepvicirLCP}{\ensuremath{3.234720\pm0.000018}}            
\newcommand{\kepvicirSMEiteff}{\ensuremath{NULL\pmNULL}}               
\newcommand{\kepvicirSMEizfeh}{\ensuremath{NULL\pmNULL}}               
\newcommand{\kepvicirSMEizfehshort}{\ensuremath{NULL}}                 
\newcommand{\kepvicirSMEilogg}{\ensuremath{NULL\pmNULL}}               
\newcommand{\kepvicirSMEivsin}{\ensuremath{NULL\pmNULL}}               
\newcommand{\kepvicirSMEivmac}{\ensuremath{NULL}}                      
\newcommand{\kepvicirSMEivmic}{\ensuremath{NULL}}                      
\newcommand{\kepvicirSMEiiteff}{\ensuremath{5647\pm88}}                
\newcommand{\kepvicirSMEiizfeh}{\ensuremath{+0.34\pm0.08}}             
\newcommand{\kepvicirSMEiizfehshort}{\ensuremath{+0.34}}               
\newcommand{\kepvicirSMEiilogg}{\ensuremath{4.59\pm0.10}}              
\newcommand{\kepvicirSMEiivsin}{\ensuremath{3.0\pm1.0}}                
\newcommand{\kepvicirSMEiivmac}{\ensuremath{NULL}}                     
\newcommand{\kepvicirSMEiivmic}{\ensuremath{NULL}}                     
\newcommand{\kepvicirLBikep}{\ensuremath{0.4471}}                      
\newcommand{\kepvicirLBiikep}{\ensuremath{0.2397}}                     
\newcommand{\kepvicirISOmlong}{\ensuremath{1.136_{-0.072}^{+0.034}}}   
\newcommand{\kepvicirISOrlong}{\ensuremath{1.325\pm0.042}}             
\newcommand{\kepvicirISOrho}{\ensuremath{0.68\pm0.05}}                 
\newcommand{\kepvicirISOlogg}{\ensuremath{4.24\pm0.02}}                
\newcommand{\kepvicirISOlum}{\ensuremath{1.61\pm0.17}}                 
\newcommand{\kepvicirISOmv}{\ensuremath{4.33\pm0.13}}                  
\newcommand{\kepvicirISOage}{\ensuremath{5.6_{-0.9}^{+2.5}}}           
\newcommand{\kepvicirRVK}{\ensuremath{78.2\pm3.7}}                     
\newcommand{\kepvicirRVgamma}{\ensuremath{-15.9\pm2.9}}                
\newcommand{\kepvicirRVjitter}{\ensuremath{3.2}}                       
\newcommand{\kepvicirRVfitrms}{\ensuremath{5.7}}                       
\newcommand{\kepvicirPPi}{\ensuremath{87.7_{-0.6}^{+0.8}}}             
\newcommand{\kepvicirPPlogg}{\ensuremath{3.01\pm0.03}}                 
\newcommand{\kepvicirPPar}{\ensuremath{7.21\pm0.18}}                   
\newcommand{\kepvicirPParel}{\ensuremath{0.0447_{-0.0010}^{+0.0004}}}  
\newcommand{\kepvicirPPrho}{\ensuremath{0.41\pm0.04}}                  
\newcommand{\kepvicirPPmlong}{\ensuremath{0.616\pm0.035}}              
\newcommand{\kepvicirPPrlong}{\ensuremath{1.224\pm0.043}}              
\newcommand{\kepvicirPPteff}{\ensuremath{1488\pm30}}                   
\newcommand{\kepvicirXdist}{\ensuremath{621\pm22}}                     

%% file: Kep6-info-ecc.tex
\newcommand{\kepvieccLCbsq}{\ensuremath{0.085_{-0.044}^{+0.049}}}      
\newcommand{\kepvieccLCimp}{\ensuremath{0.292_{-0.102}^{+0.071}}}      
\newcommand{\kepvieccLCzeta}{\ensuremath{14.69\pm0.02}}                
\newcommand{\kepvieccLCP}{\ensuremath{3.234721\pm0.000018}}            
\newcommand{\kepvieccSMEiteff}{\ensuremath{NULL\pmNULL}}               
\newcommand{\kepvieccSMEizfeh}{\ensuremath{NULL\pmNULL}}               
\newcommand{\kepvieccSMEizfehshort}{\ensuremath{NULL}}                 
\newcommand{\kepvieccSMEilogg}{\ensuremath{NULL\pmNULL}}               
\newcommand{\kepvieccSMEivsin}{\ensuremath{NULL\pmNULL}}               
\newcommand{\kepvieccSMEivmac}{\ensuremath{NULL}}                      
\newcommand{\kepvieccSMEivmic}{\ensuremath{NULL}}                      
\newcommand{\kepvieccSMEiiteff}{\ensuremath{5647\pm88}}                
\newcommand{\kepvieccSMEiizfeh}{\ensuremath{+0.34\pm0.08}}             
\newcommand{\kepvieccSMEiizfehshort}{\ensuremath{+0.34}}               
\newcommand{\kepvieccSMEiilogg}{\ensuremath{4.59\pm0.10}}              
\newcommand{\kepvieccSMEiivsin}{\ensuremath{3.0\pm1.0}}                
\newcommand{\kepvieccSMEiivmac}{\ensuremath{NULL}}                     
\newcommand{\kepvieccSMEiivmic}{\ensuremath{NULL}}                     
\newcommand{\kepvieccISOmlong}{\ensuremath{1.170_{-0.087}^{+0.064}}}   
\newcommand{\kepvieccISOrlong}{\ensuremath{1.503\pm0.073}}             
\newcommand{\kepvieccISOrho}{\ensuremath{0.48\pm0.06}}                 
\newcommand{\kepvieccISOlogg}{\ensuremath{4.15\pm0.03}}                
\newcommand{\kepvieccISOlum}{\ensuremath{2.06\pm0.28}}                 
\newcommand{\kepvieccISOmv}{\ensuremath{4.05\pm0.16}}                  
\newcommand{\kepvieccISOage}{\ensuremath{5.7_{-1.0}^{+2.6}}}           
\newcommand{\kepvieccRVK}{\ensuremath{86.6\pm1.6}}                     
\newcommand{\kepvieccRVk}{\ensuremath{0.012_{-0.015}^{+0.029}}}        
\newcommand{\kepvieccRVh}{\ensuremath{0.117\pm0.036}}                  
\newcommand{\kepvieccRVgamma}{\ensuremath{-13.6\pm2.7}}                
\newcommand{\kepvieccRVjitter}{\ensuremath{2.7}}                       
\newcommand{\kepvieccRVeccen}{\ensuremath{0.120\pm0.036}}              
\newcommand{\kepvieccRVomega}{\ensuremath{83\pm12}}                    
\newcommand{\kepvieccRVfitrms}{\ensuremath{5.5}}                       
\newcommand{\kepvieccPPi}{\ensuremath{87.0\pm0.9}}                     
\newcommand{\kepvieccPPlogg}{\ensuremath{2.95\pm0.04}}                 
\newcommand{\kepvieccPPar}{\ensuremath{6.42\pm0.26}}                   
\newcommand{\kepvieccPParel}{\ensuremath{0.0451_{-0.0012}^{+0.0008}}}  
\newcommand{\kepvieccPPrho}{\ensuremath{0.32\pm0.04}}                  
\newcommand{\kepvieccPPmlong}{\ensuremath{0.692\pm0.032}}              
\newcommand{\kepvieccPPrlong}{\ensuremath{1.389\pm0.072}}              
\newcommand{\kepvieccPPteff}{\ensuremath{1579\pm41}}                   
\newcommand{\kepvieccXdist}{\ensuremath{704\pm37}}                     

%% file: newcommand_spec_7.tex
\newcommand{\kepvii}{Kepler-7}
\newcommand{\kepviib}{Kepler-7b}

\input{Kep7-info-cir.tex}

\input{Kep7-info-ecc.tex}

\newcommand{\kepviicirRVgammarel}{\kepviicirRVgamma}                           
\newcommand{\kepviicirSMEversion}{ii}                                       
\newcommand{\kepviicirSMEteff}{\ifthenelse{\equal{\kepviicirSMEversion}{i}}{\kepviicirSMEiteff}{\kepviicirSMEiiteff}}
\newcommand{\kepviicirSMEzfeh}{\ifthenelse{\equal{\kepviicirSMEversion}{i}}{\kepviicirSMEizfeh}{\kepviicirSMEiizfeh}}
\newcommand{\kepviicirSMEzfehshort}{\ifthenelse{\equal{\kepviicirSMEversion}{i}}{\kepviicirSMEizfehshort}{\kepviicirSMEiizfehshort}}
\newcommand{\kepviicirSMElogg}{\ifthenelse{\equal{\kepviicirSMEversion}{i}}{\kepviicirSMEilogg}{\kepviicirSMEiilogg}}
\newcommand{\kepviicirSMEvsin}{\ifthenelse{\equal{\kepviicirSMEversion}{i}}{\kepviicirSMEivsin}{\kepviicirSMEiivsin}}
\newcommand{\kepviicirSMEvmac}{\ifthenelse{\equal{\kepviicirSMEversion}{i}}{\kepviicirSMEivmac}{\kepviicirSMEiivmac}}
\newcommand{\kepviicirSMEvmic}{\ifthenelse{\equal{\kepviicirSMEversion}{i}}{\kepviicirSMEivmic}{\kepviicirSMEiivmic}}

\newcommand{\kepviieccRVgammarel}{\kepviieccRVgamma}                           
\newcommand{\kepviieccSMEversion}{ii}                                       
\newcommand{\kepviieccSMEteff}{\ifthenelse{\equal{\kepviieccSMEversion}{i}}{\kepviieccSMEiteff}{\kepviieccSMEiiteff}}
\newcommand{\kepviieccSMEzfeh}{\ifthenelse{\equal{\kepviieccSMEversion}{i}}{\kepviieccSMEizfeh}{\kepviieccSMEiizfeh}}
\newcommand{\kepviieccSMEzfehshort}{\ifthenelse{\equal{\kepviieccSMEversion}{i}}{\kepviieccSMEizfehshort}{\kepviieccSMEiizfehshort}}
\newcommand{\kepviieccSMElogg}{\ifthenelse{\equal{\kepviieccSMEversion}{i}}{\kepviieccSMEilogg}{\kepviieccSMEiilogg}}
\newcommand{\kepviieccSMEvsin}{\ifthenelse{\equal{\kepviieccSMEversion}{i}}{\kepviieccSMEivsin}{\kepviieccSMEiivsin}}
\newcommand{\kepviieccSMEvmac}{\ifthenelse{\equal{\kepviieccSMEversion}{i}}{\kepviieccSMEivmac}{\kepviieccSMEiivmac}}
\newcommand{\kepviieccSMEvmic}{\ifthenelse{\equal{\kepviieccSMEversion}{i}}{\kepviieccSMEivmic}{\kepviieccSMEiivmic}}

%% file: Kep7-info-cir.tex
\newcommand{\kepviicirLCrprstar}{\ensuremath{0.0808\pm0.0005}}         
\newcommand{\kepviicirLCbsq}{\ensuremath{0.155_{-0.051}^{+0.045}}}     
\newcommand{\kepviicirLCimp}{\ensuremath{0.394_{-0.079}^{+0.052}}}     
\newcommand{\kepviicirLCzeta}{\ensuremath{10.20\pm0.01}}               
\newcommand{\kepviicirLCP}{\ensuremath{4.885535\pm0.000041}}           
\newcommand{\kepviicirSMEiteff}{\ensuremath{NULL\pmNULL}}              
\newcommand{\kepviicirSMEizfeh}{\ensuremath{NULL\pmNULL}}              
\newcommand{\kepviicirSMEizfehshort}{\ensuremath{NULL}}                
\newcommand{\kepviicirSMEilogg}{\ensuremath{NULL\pmNULL}}              
\newcommand{\kepviicirSMEivsin}{\ensuremath{NULL\pmNULL}}              
\newcommand{\kepviicirSMEivmac}{\ensuremath{NULL}}                     
\newcommand{\kepviicirSMEivmic}{\ensuremath{NULL}}                     
\newcommand{\kepviicirSMEiiteff}{\ensuremath{5933\pm88}}               
\newcommand{\kepviicirSMEiizfeh}{\ensuremath{+0.11\pm0.06}}            
\newcommand{\kepviicirSMEiizfehshort}{\ensuremath{+0.11}}              
\newcommand{\kepviicirSMEiilogg}{\ensuremath{3.98\pm0.10}}             
\newcommand{\kepviicirSMEiivsin}{\ensuremath{4.2\pm0.5}}               
\newcommand{\kepviicirSMEiivmac}{\ensuremath{NULL}}                    
\newcommand{\kepviicirSMEiivmic}{\ensuremath{NULL}}                    
\newcommand{\kepviicirLBikep}{\ensuremath{0.3948}}                     
\newcommand{\kepviicirLBiikep}{\ensuremath{0.2711}}                    
\newcommand{\kepviicirISOmlong}{\ensuremath{1.226_{-0.045}^{+0.108}}}  
\newcommand{\kepviicirISOrlong}{\ensuremath{1.791\pm0.067}}            
\newcommand{\kepviicirISOrho}{\ensuremath{0.31\pm0.03}}                
\newcommand{\kepviicirISOlogg}{\ensuremath{4.03\pm0.02}}               
\newcommand{\kepviicirISOlum}{\ensuremath{3.55\pm0.41}}                
\newcommand{\kepviicirISOmv}{\ensuremath{3.43\pm0.14}}                 
\newcommand{\kepviicirISOage}{\ensuremath{5.1_{-1.3}^{+0.7}}}          
\newcommand{\kepviicirRVK}{\ensuremath{42.6\pm4.3}}                    
\newcommand{\kepviicirRVgamma}{\ensuremath{0.0\pm2.9}}                 
\newcommand{\kepviicirRVjitter}{\ensuremath{0.0}}                      
\newcommand{\kepviicirRVfitrms}{\ensuremath{8.1}}                      
\newcommand{\kepviicirPPi}{\ensuremath{86.9_{-0.5}^{+0.7}}}            
\newcommand{\kepviicirPPlogg}{\ensuremath{2.71\pm0.05}}                
\newcommand{\kepviicirPPar}{\ensuremath{7.29\pm0.20}}                  
\newcommand{\kepviicirPParel}{\ensuremath{0.0603_{-0.0007}^{+0.0017}}} 
\newcommand{\kepviicirPPrho}{\ensuremath{0.18\pm0.03}}                 
\newcommand{\kepviicirPPmlong}{\ensuremath{0.414\pm0.045}}             
\newcommand{\kepviicirPPrlong}{\ensuremath{1.409\pm0.058}}             
\newcommand{\kepviicirPPteff}{\ensuremath{1554\pm32}}                  
\newcommand{\kepviicirXdist}{\ensuremath{829\pm36}}                    

%% file: Kep7-info-ecc.tex
\newcommand{\kepviieccLCrprstar}{\ensuremath{0.0808\pm0.0005}}         
\newcommand{\kepviieccLCbsq}{\ensuremath{0.152_{-0.051}^{+0.047}}}     
\newcommand{\kepviieccLCimp}{\ensuremath{0.391_{-0.079}^{+0.054}}}     
\newcommand{\kepviieccLCzeta}{\ensuremath{10.21_{-0.02}^{+0.02}}}      
\newcommand{\kepviieccLCP}{\ensuremath{4.885537\pm0.000042}}           
\newcommand{\kepviieccSMEiteff}{\ensuremath{NULL\pmNULL}}              
\newcommand{\kepviieccSMEizfeh}{\ensuremath{NULL\pmNULL}}              
\newcommand{\kepviieccSMEizfehshort}{\ensuremath{NULL}}                
\newcommand{\kepviieccSMEilogg}{\ensuremath{NULL\pmNULL}}              
\newcommand{\kepviieccSMEivsin}{\ensuremath{NULL\pmNULL}}              
\newcommand{\kepviieccSMEivmac}{\ensuremath{NULL}}                     
\newcommand{\kepviieccSMEivmic}{\ensuremath{NULL}}                     
\newcommand{\kepviieccSMEiiteff}{\ensuremath{5933\pm88}}               
\newcommand{\kepviieccSMEiizfeh}{\ensuremath{+0.11\pm0.06}}            
\newcommand{\kepviieccSMEiizfehshort}{\ensuremath{+0.11}}              
\newcommand{\kepviieccSMEiilogg}{\ensuremath{3.98\pm0.10}}             
\newcommand{\kepviieccSMEiivsin}{\ensuremath{4.2\pm0.5}}               
\newcommand{\kepviieccSMEiivmac}{\ensuremath{NULL}}                    
\newcommand{\kepviieccSMEiivmic}{\ensuremath{NULL}}                    
\newcommand{\kepviieccLBikep}{\ensuremath{0.3948}}                     
\newcommand{\kepviieccLBiikep}{\ensuremath{0.2711}}                    
\newcommand{\kepviieccISOmlong}{\ensuremath{1.272_{-0.054}^{+0.102}}}  
\newcommand{\kepviieccISOrlong}{\ensuremath{1.941_{-0.118}^{+0.178}}}  
\newcommand{\kepviieccISOrho}{\ensuremath{0.25\pm0.05}}                
\newcommand{\kepviieccISOlogg}{\ensuremath{3.97\pm0.05}}               
\newcommand{\kepviieccISOlum}{\ensuremath{4.18_{-0.56}^{+0.94}}}       
\newcommand{\kepviieccISOmv}{\ensuremath{3.25\pm0.19}}                 
\newcommand{\kepviieccISOage}{\ensuremath{4.7_{-1.0}^{+0.7}}}          
\newcommand{\kepviieccRVK}{\ensuremath{43.1\pm4.3}}                    
\newcommand{\kepviieccRVk}{\ensuremath{-0.003_{-0.025}^{+0.005}}}      
\newcommand{\kepviieccRVh}{\ensuremath{0.074\pm0.063}}                 
\newcommand{\kepviieccRVgamma}{\ensuremath{-0.6\pm2.9}}                
\newcommand{\kepviieccRVjitter}{\ensuremath{0.0}}                      
\newcommand{\kepviieccRVeccen}{\ensuremath{0.076\pm0.057}}             
\newcommand{\kepviieccRVomega}{\ensuremath{93\pm52}}                   
\newcommand{\kepviieccRVfitrms}{\ensuremath{6.9}}                      
\newcommand{\kepviieccPPi}{\ensuremath{86.4\pm0.8}}                    
\newcommand{\kepviieccPPlogg}{\ensuremath{2.65\pm0.07}}                
\newcommand{\kepviieccPPar}{\ensuremath{6.78\pm0.43}}                  
\newcommand{\kepviieccPParel}{\ensuremath{0.0611_{-0.0009}^{+0.0016}}} 
\newcommand{\kepviieccPPrho}{\ensuremath{0.15\pm0.03}}                 
\newcommand{\kepviieccPPmlong}{\ensuremath{0.424\pm0.046}}             
\newcommand{\kepviieccPPrlong}{\ensuremath{1.527_{-0.097}^{+0.144}}}   
\newcommand{\kepviieccPPteff}{\ensuremath{1612_{-50}^{+68}}}           
\newcommand{\kepviieccXdist}{\ensuremath{898_{-57}^{+84}}}             

%% file: newcommand_spec_8.tex
\newcommand{\kepviii}{Kepler-8}
\newcommand{\kepviiib}{Kepler-8b}

\input{Kep8-info-cir.tex}

\input{Kep8-info-ecc.tex}

\newcommand{\kepviiicirRVgammarel}{\kepviiicirRVgamma}                           
\newcommand{\kepviiicirSMEversion}{ii}                                       
\newcommand{\kepviiicirSMEteff}{\ifthenelse{\equal{\kepviiicirSMEversion}{i}}{\kepviiicirSMEiteff}{\kepviiicirSMEiiteff}}
\newcommand{\kepviiicirSMEzfeh}{\ifthenelse{\equal{\kepviiicirSMEversion}{i}}{\kepviiicirSMEizfeh}{\kepviiicirSMEiizfeh}}
\newcommand{\kepviiicirSMEzfehshort}{\ifthenelse{\equal{\kepviiicirSMEversion}{i}}{\kepviiicirSMEizfehshort}{\kepviiicirSMEiizfehshort}}
\newcommand{\kepviiicirSMElogg}{\ifthenelse{\equal{\kepviiicirSMEversion}{i}}{\kepviiicirSMEilogg}{\kepviiicirSMEiilogg}}
\newcommand{\kepviiicirSMEvsin}{\ifthenelse{\equal{\kepviiicirSMEversion}{i}}{\kepviiicirSMEivsin}{\kepviiicirSMEiivsin}}
\newcommand{\kepviiicirSMEvmac}{\ifthenelse{\equal{\kepviiicirSMEversion}{i}}{\kepviiicirSMEivmac}{\kepviiicirSMEiivmac}}
\newcommand{\kepviiicirSMEvmic}{\ifthenelse{\equal{\kepviiicirSMEversion}{i}}{\kepviiicirSMEivmic}{\kepviiicirSMEiivmic}}

\newcommand{\kepviiieccRVgammarel}{\kepviiieccRVgamma}                           
\newcommand{\kepviiieccSMEversion}{ii}                                       
\newcommand{\kepviiieccSMEteff}{\ifthenelse{\equal{\kepviiieccSMEversion}{i}}{\kepviiieccSMEiteff}{\kepviiieccSMEiiteff}}
\newcommand{\kepviiieccSMEzfeh}{\ifthenelse{\equal{\kepviiieccSMEversion}{i}}{\kepviiieccSMEizfeh}{\kepviiieccSMEiizfeh}}
\newcommand{\kepviiieccSMEzfehshort}{\ifthenelse{\equal{\kepviiieccSMEversion}{i}}{\kepviiieccSMEizfehshort}{\kepviiieccSMEiizfehshort}}
\newcommand{\kepviiieccSMElogg}{\ifthenelse{\equal{\kepviiieccSMEversion}{i}}{\kepviiieccSMEilogg}{\kepviiieccSMEiilogg}}
\newcommand{\kepviiieccSMEvsin}{\ifthenelse{\equal{\kepviiieccSMEversion}{i}}{\kepviiieccSMEivsin}{\kepviiieccSMEiivsin}}
\newcommand{\kepviiieccSMEvmac}{\ifthenelse{\equal{\kepviiieccSMEversion}{i}}{\kepviiieccSMEivmac}{\kepviiieccSMEiivmac}}
\newcommand{\kepviiieccSMEvmic}{\ifthenelse{\equal{\kepviiieccSMEversion}{i}}{\kepviiieccSMEivmic}{\kepviiieccSMEiivmic}}

%% file: Kep8-info-cir.tex
\newcommand{\kepviiicirLCrprstar}{\ensuremath{0.0933\pm0.0012}}        
\newcommand{\kepviiicirLCbsq}{\ensuremath{0.460_{-0.067}^{+0.051}}}    
\newcommand{\kepviiicirLCimp}{\ensuremath{0.678_{-0.053}^{+0.036}}}    
\newcommand{\kepviiicirLCzeta}{\ensuremath{17.29\pm0.08}}              
\newcommand{\kepviiicirLCP}{\ensuremath{3.522418\pm0.000057}}          
\newcommand{\kepviiicirSMEiteff}{\ensuremath{NULL\pmNULL}}             
\newcommand{\kepviiicirSMEizfeh}{\ensuremath{NULL\pmNULL}}             
\newcommand{\kepviiicirSMEizfehshort}{\ensuremath{NULL}}               
\newcommand{\kepviiicirSMEilogg}{\ensuremath{NULL\pmNULL}}             
\newcommand{\kepviiicirSMEivsin}{\ensuremath{NULL\pmNULL}}             
\newcommand{\kepviiicirSMEivmac}{\ensuremath{NULL}}                    
\newcommand{\kepviiicirSMEivmic}{\ensuremath{NULL}}                    
\newcommand{\kepviiicirSMEiiteff}{\ensuremath{6213\pm150}}             
\newcommand{\kepviiicirSMEiizfeh}{\ensuremath{-0.055\pm0.06}}          
\newcommand{\kepviiicirSMEiizfehshort}{\ensuremath{-0.055}}            
\newcommand{\kepviiicirSMEiilogg}{\ensuremath{4.28\pm0.10}}            
\newcommand{\kepviiicirSMEiivsin}{\ensuremath{10.5\pm0.7}}             
\newcommand{\kepviiicirSMEiivmac}{\ensuremath{NULL}}                   
\newcommand{\kepviiicirSMEiivmic}{\ensuremath{NULL}}                   
\newcommand{\kepviiicirLBikep}{\ensuremath{0.3585}}                    
\newcommand{\kepviiicirLBiikep}{\ensuremath{0.2917}}                   
\newcommand{\kepviiicirISOmlong}{\ensuremath{1.228_{-0.082}^{+0.052}}} 
\newcommand{\kepviiicirISOrlong}{\ensuremath{1.460\pm0.088}}           
\newcommand{\kepviiicirISOrho}{\ensuremath{0.55_{-0.07}^{+0.10}}}      
\newcommand{\kepviiicirISOlogg}{\ensuremath{4.19\pm0.04}}              
\newcommand{\kepviiicirISOlum}{\ensuremath{2.84\pm0.50}}               
\newcommand{\kepviiicirISOmv}{\ensuremath{3.64\pm0.20}}                
\newcommand{\kepviiicirISOage}{\ensuremath{3.4_{-0.6}^{+1.6}}}         
\newcommand{\kepviiicirRVK}{\ensuremath{68.9\pm18.4}}                  
\newcommand{\kepviiicirRVgamma}{\ensuremath{-7.4\pm12.4}}              
\newcommand{\kepviiicirRVjitter}{\ensuremath{31.5}}                    
\newcommand{\kepviiicirRVfitrms}{\ensuremath{45.7}}                    
\newcommand{\kepviiicirPPi}{\ensuremath{84.5\pm0.6}}                   
\newcommand{\kepviiicirPPlogg}{\ensuremath{2.92_{-0.17}^{+0.11}}}      
\newcommand{\kepviiicirPPar}{\ensuremath{7.12\pm0.37}}                 
\newcommand{\kepviiicirPParel}{\ensuremath{0.0485_{-0.0011}^{+0.0007}}} 
\newcommand{\kepviiicirPPrho}{\ensuremath{0.31_{-0.09}^{+0.12}}}       
\newcommand{\kepviiicirPPmlong}{\ensuremath{0.589\pm0.160}}            
\newcommand{\kepviiicirPPrlong}{\ensuremath{1.327\pm0.092}}            
\newcommand{\kepviiicirPPteff}{\ensuremath{1646\pm58}}                 
\newcommand{\kepviiicirXdist}{\ensuremath{978\pm62}}                   

%% file: Kep8-info-ecc.tex
\newcommand{\kepviiieccLCrprstar}{\ensuremath{0.0933\pm0.0012}}        
\newcommand{\kepviiieccLCbsq}{\ensuremath{0.460_{-0.068}^{+0.049}}}    
\newcommand{\kepviiieccLCimp}{\ensuremath{0.678_{-0.054}^{+0.035}}}    
\newcommand{\kepviiieccLCzeta}{\ensuremath{17.28\pm0.07}}              
\newcommand{\kepviiieccLCP}{\ensuremath{3.522416\pm0.000057}}          
\newcommand{\kepviiieccSMEiteff}{\ensuremath{NULL\pmNULL}}             
\newcommand{\kepviiieccSMEizfeh}{\ensuremath{NULL\pmNULL}}             
\newcommand{\kepviiieccSMEizfehshort}{\ensuremath{NULL}}               
\newcommand{\kepviiieccSMEilogg}{\ensuremath{NULL\pmNULL}}             
\newcommand{\kepviiieccSMEivsin}{\ensuremath{NULL\pmNULL}}             
\newcommand{\kepviiieccSMEivmac}{\ensuremath{NULL}}                    
\newcommand{\kepviiieccSMEivmic}{\ensuremath{NULL}}                    
\newcommand{\kepviiieccSMEiiteff}{\ensuremath{6213\pm150}}             
\newcommand{\kepviiieccSMEiizfeh}{\ensuremath{-0.055\pm0.06}}          
\newcommand{\kepviiieccSMEiizfehshort}{\ensuremath{-0.055}}            
\newcommand{\kepviiieccSMEiilogg}{\ensuremath{4.28\pm0.10}}            
\newcommand{\kepviiieccSMEiivsin}{\ensuremath{10.5\pm0.7}}             
\newcommand{\kepviiieccSMEiivmac}{\ensuremath{NULL}}                   
\newcommand{\kepviiieccSMEiivmic}{\ensuremath{NULL}}                   
\newcommand{\kepviiieccLBikep}{\ensuremath{0.3585}}                    
\newcommand{\kepviiieccLBiikep}{\ensuremath{0.2917}}                   
\newcommand{\kepviiieccISOmlong}{\ensuremath{1.211_{-0.076}^{+0.058}}} 
\newcommand{\kepviiieccISOrlong}{\ensuremath{1.408\pm0.107}}           
\newcommand{\kepviiieccISOrho}{\ensuremath{0.61_{-0.10}^{+0.15}}}      
\newcommand{\kepviiieccISOlogg}{\ensuremath{4.22\pm0.05}}              
\newcommand{\kepviiieccISOlum}{\ensuremath{2.64\pm0.53}}               
\newcommand{\kepviiieccISOmv}{\ensuremath{3.72\pm0.23}}                
\newcommand{\kepviiieccISOage}{\ensuremath{3.4_{-0.7}^{+1.5}}}         
\newcommand{\kepviiieccRVK}{\ensuremath{68.1\pm14.0}}                  
\newcommand{\kepviiieccRVk}{\ensuremath{0.030_{-0.108}^{+0.081}}}      
\newcommand{\kepviiieccRVh}{\ensuremath{-0.036\pm0.062}}               
\newcommand{\kepviiieccRVgamma}{\ensuremath{-6.0\pm12.1}}              
\newcommand{\kepviiieccRVjitter}{\ensuremath{27.1}}                    
\newcommand{\kepviiieccRVeccen}{\ensuremath{0.099\pm0.056}}            
\newcommand{\kepviiieccRVomega}{\ensuremath{253\pm111}}                
\newcommand{\kepviiieccRVfitrms}{\ensuremath{42.8}}                    
\newcommand{\kepviiieccPPi}{\ensuremath{84.9\pm0.6}}                   
\newcommand{\kepviiieccPPlogg}{\ensuremath{2.94\pm0.13}}               
\newcommand{\kepviiieccPPar}{\ensuremath{7.35\pm0.49}}                 
\newcommand{\kepviiieccPParel}{\ensuremath{0.0483_{-0.0010}^{+0.0007}}} 
\newcommand{\kepviiieccPPrho}{\ensuremath{0.34_{-0.10}^{+0.16}}}       
\newcommand{\kepviiieccPPmlong}{\ensuremath{0.575\pm0.116}}            
\newcommand{\kepviiieccPPrlong}{\ensuremath{1.278\pm0.102}}            
\newcommand{\kepviiieccPPteff}{\ensuremath{1623\pm67}}                 
\newcommand{\kepviiieccXdist}{\ensuremath{943\pm74}}                   

%% file: kep4.tex

\subsection{Global Fits}

The discovery of \kepivb\ was made by \citet{bor10b}. 
The planet is particularly interesting for joining the
club of HAT-P-11b and GJ 436b as a Neptune-mass transiting exoplanet.
\kepivb\ exhibits a sub-mmag transit around a $12^{\mathrm{th}}$
magnitude star, which leads to relatively large uncertainties on the
system parameters but demonstrates the impressive performance of the
Kepler photometry. 

However, the combination of the course sampling (e.g.~the ingress/egress
duration is 3-4 times shorter than the cadence of the observations), the
very low RV amplitude and a sub-mmag transit
make \kepivb\ the most challenging object to determine reliable
system parameters for in this paper. The BIC test for the approriate RV 
model prefers the simplest description possible, reflecting the low 
signal-to-noise levels encountered for the radial velocities.

The largest difference between our own fits and that
of \citet{bor10b} is the retrieved $\arstar$ and impact parameter, $b$.
\citet{bor10b} find an equatorial transit solution whereas our method A
circular fit, and both modes of method B, place \kepivb\ at an impact 
parameter of 0.5-0.6. Curiously, the eccentric fit of method A prefers
an even larger impact parameter than this, as a result of letting the
limb darkening be fitted.

\begin{figure}[!ht]
\plotone{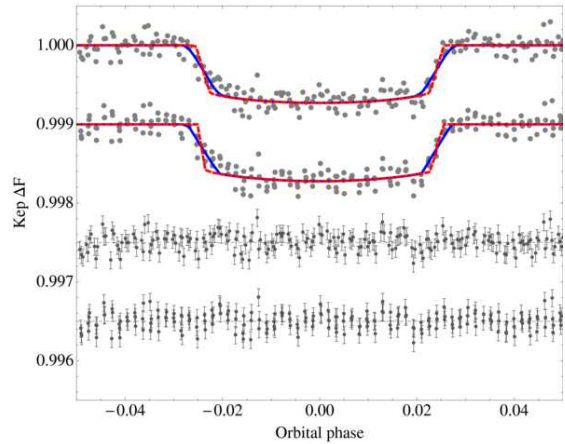}
\caption{
    Phase-folded primary transit \lc\ of \kepiv\ for method
    B. The upper curve shows the circular fit, the bottom curve the
    eccentric fit.  Solid (blue) lines indicate the best fit resampled
    model (with bin-number 4). The dashed (red) lines show the
    corresponding unbinned model, which one would get if the transit
    was observed with infinitely fine time resolution.
	Residuals from the best fit resampled model for the
	circular and eccentric solutions are shown below in
        respective order.
\label{fig:kep4prim}}
\end{figure}

\begin{figure}[ht]
\plotone{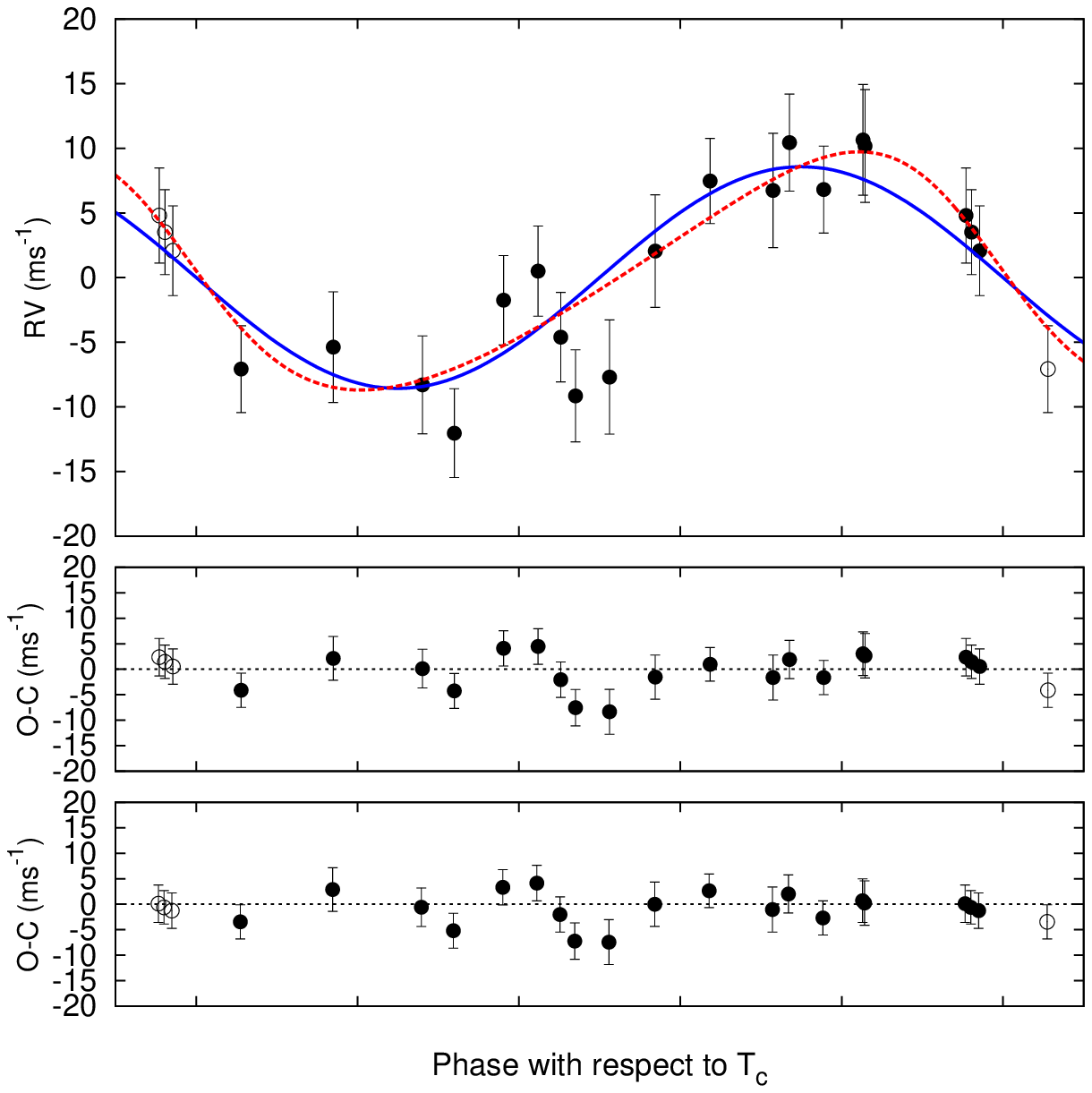}
\caption{
    {\bf Top:} RV measurements from Keck for \kepiv{}, along with an
    orbital fit, shown as a function of orbital phase, using our
    best-fit period. Solid (blue) line shows the circular orbital fit
    with binned RV model (3 bins, separated by 600\,seconds). The
    dashed (red) line shows the eccentric orbital fit (with the same
    bin parameters). Zero phase is defined by the transit midpoint. The
    center-of-mass velocity has been subtracted.  Note that the error
    bars include the stellar jitter (taken for the circular solution),
    added in quadrature to the formal errors given by the spectrum
    reduction pipeline. Fits from method B.
    {\bf Middle:} phased residuals after subtracting the orbital fit
    for the circular solution. The r.m.s.~variation of the residuals is
    \kepivcirRVfitrms\,\ms, and the stellar jitter that was added in
    quadrature to the formal RV errors is $\kepivcirRVjitter$\,\ms.
	{\bf Bottom:} phased residuals after subtracting the orbital fit
    for the eccentric solution.  Here the r.m.s.~variation of the
    residuals is \kepiveccRVfitrms\,\ms, and the stellar jitter is
    $\kepiveccRVjitter$\,\ms.
\label{fig:kep4rv}}
\end{figure}

Due to the well-known negative correlation between $b$ and $\arstar$
\citep{car08}, these larger $b$ values lead to lower $\arstar$ values and
consequently a significantly lower \lc\ derived stellar density, $\rho_*$.
Indeed, the lower stellar density results in one of the largest stellar
radii out of all the known transiting systems. The fitted models on the
phase-folded \lcs\ are shown on \reffigl{kep4prim} using method B. Correlated
noise was checked for in the residuals using the Durbin-Watson statistic, which
finds $d = 2.082$, well inside the 1\% critical boundary of $2.135$. The
orbital fits to the RV points are shown in \reffig{kep4rv} (method B),
depicting both the circular and eccentric fits. The final table of all
results are shown in Table~\ref{tab:kep4tab}.

\subsubsection{Eccentricity}

From Table~\ref{tab:BIC} of the Appendix, the circular fit with no other 
free parameters is the preferred model to describe the radial velocities 
of \kepivb. We may perform other tests aside from the BIC model selection 
though.
The global circular fit, using method A, yields a $\chi^2 = 28.54$ and
the eccentric fit obtains $\chi^2 = 20.56$. These values correspond to
the lowest $\chi^2$ solution of the simplex global fit, but for the RV
points only, which dominate the eccentricity determination. Assuming the 
period and epoch are essentially completely driven by the photometry, the
number of free parameters are 2 and 4 for the circular and eccentric fits
respectively. Therefore, based upon an F-test, the
inclusion of two new degrees of freedom to describe the eccentricity is
justified at the 1.7-$\sigma$ level (91.5\% confidence).
Another test we can implement is the \citet{lucy71} test, where the
significance of the eccentric fit is given by:

\begin{equation}
\mathrm{P}(e>0) = 1 - \mathrm{exp}\Big[-\frac{\hat{e}^2}{2 \sigma_e^2}\Big]
\end{equation}

Where $\hat{e}$ is the modal value of $e$ and $\sigma_e$ is the error (in
the negative direction). We find that this gives a significance of 88.2\%
or 1.6-$\sigma$, in close agreement with the F-test result. We also
computed the posterior distribution of the distance of the pericentre
passage, in units of the stellar radius, and found 
$(a/R_*) (1-e) = 2.6_{-0.6}^{+1.5}$ for method A. Therefore, if the eccentric
fit is the true solution, \kepivb\ would make an extremely close 
pericentre passage.

We note that \citet{bor10b} found a best-fit eccentricity of $e = 0.22$ 
(no quoted uncertainty) but the authors concluded the eccentric fit was
not statistically significant, but provided no quantification. Based upon
the current observations, it is not possible to make a reliable conclusion
as to whether \kepivb\ is genuinely eccentric or not. At best, it can
be considered a $\sim2$-$\sigma$ marginal detection of eccentricity. The
only assured thing we can say is that $e<0.43$ to 95\% confidence.
Despite the ambiguity of the eccentricity, it is interesting to consider
the consequences of this object possessing $e>0$.

For the eccentric fit, the extreme proximity of this planet to the star 
raises the issue of tidal circularization. Let us continue under the 
assumption that no third body is responsible for
pumping the eccentricity of \kepivb. For a planet initially with $e\sim1$,
the eccentricity decreases to $1/\exp(1)^n$ after $n$ circularization 
timescales. This means the maximum number of circularization timescales 
which have transpired is given by $n\leq-\log(e)$. Since 
$n=\tau_{\mathrm{age}}/\tau_{\mathrm{circ}}$ (age of the system divided by
circularization timescale) then we may write
$\tau_{\mathrm{circ}} \geq \tau_{\mathrm{age}}/n$. Using
the method of \citet{ada06}, the tidal circularization time may be
written as:

\begin{equation}
\tau_{\mathrm{circ}} = Q_P \frac{4}{63} \frac{P}{2\pi}
\frac{\mpl}{\mstar} \Big(\frac{a}{\rstar} \frac{1}{p} \Big)^5 (1-e^2)^{13/2}
\end{equation}

Given that $\tau_{\mathrm{circ}} \geq \tau_{\mathrm{age}}/n$, we may compute
the minimum possible value of $Q_P$ through re-arrangement:

\begin{equation}
Q_P \geq \frac{63}{4} \frac{2\pi}{P}
\frac{\mstar}{\mpl} \Big(\frac{p}{(a/\rstar)}\Big)^5 \frac{\tau_{\mathrm{age}}}{-\log(e)} \frac{1}{(1-e^2)^{13/2}}
\end{equation}

The final term, $(1-e^2)^{13/2}$ may be neglected since we are interested
in the minimum limit of $Q_P$ and this term only acts to further increase
the $Q_P$ for non-zero $e$. The advantage of doing this is that $e$ is a 
function of time and thus we can eliminate a term which would otherwise
have to be integrated over time. Taking the posterior distribution of
the $Q_P$ equation for method A yields 
$Q_P \geq (1.1_{-0.9}^{+1.5})\times10^7$. We find that 99.9\% of the
MCMC trials correspond to $Q_P \geq 10^5$ for the eccentric fit. Thus,
if the eccentric fit was genuine, \kepivb\ would exhibit very poor
tidal dissipation. This is somewhat reminiscent of GJ 436b \citep{deming:2007}
and HAT-P-11b \citep{bakos:2009} for which a large $Q_P$ value is also predicted 
and raises the interesting question as to whether hot-Neptune generally possess 
larger $Q_P$ values than their Jovian counterparts. We also note that WASP-18b
has also been proposed to possess a greater-than-Jovian-$Q_P$ 
\citep{hellier:2009}.

This discussion highlights the importance of obtaining more statistics
regarding the eccentricity of transiting hot-Neptunes, which may reveal
some fundamental insights into the origin of these objects.

\subsubsection{Secondary eclipse}

We here consider the circular model only since the eccentric fit
is not unambiguously accepted. Global fits do not find any
significant detection of the secondary eclipse. Secondary depths $\geq
104$\,ppm are excluded to 3-$\sigma$ confidence, which corresponds to
geometric albedos greater than unity and thus this places no
constraints on the detectability of reflected light from the planet.
The thermal emission is limited by this constraint such that
$T_{P,\mathrm{brightness}} \leq 3988$\,K, which places no constraints on
redistribution.

\subsubsection{Properties of the parent star \kepiv}
\label{sec:stelparam}

\begin{figure}[!ht]
\plotone{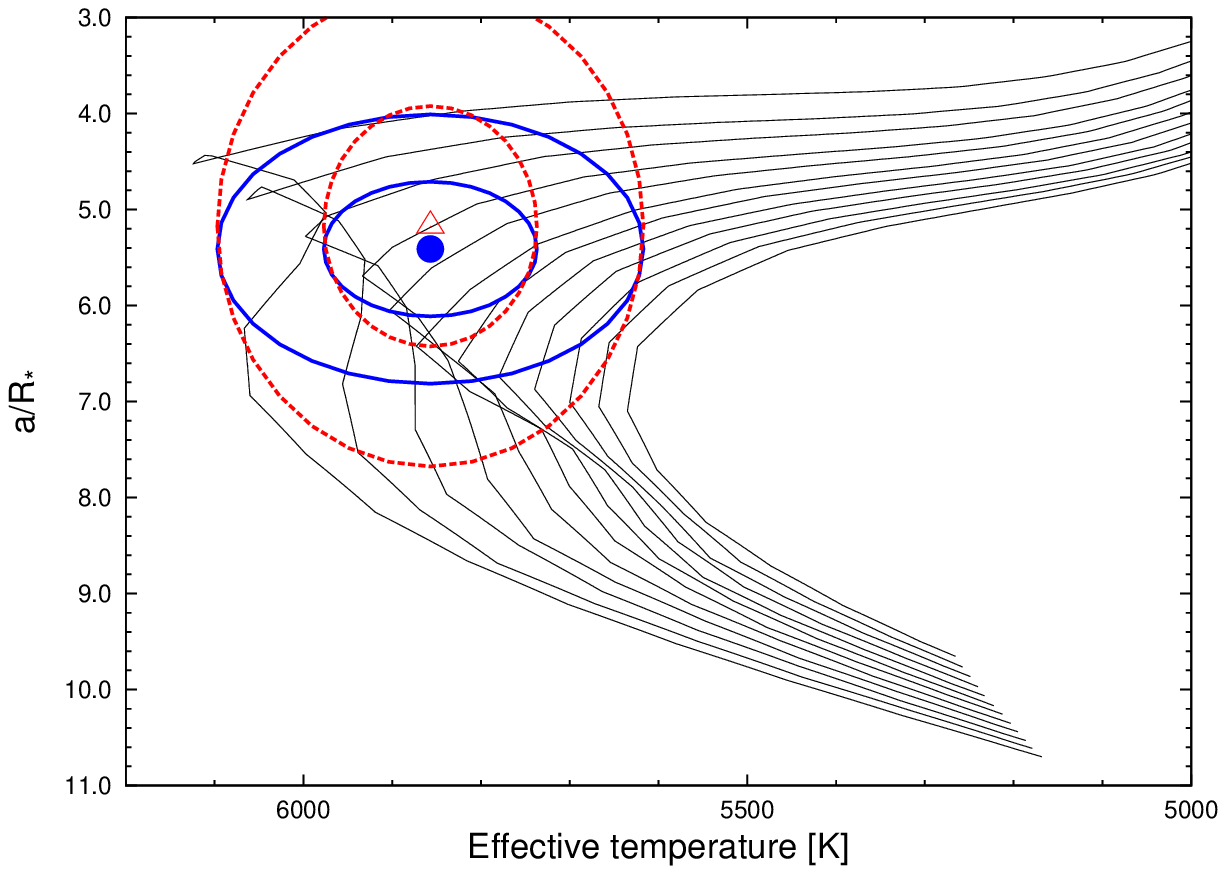}
\caption{
    Stellar isochrones from \citet{yi:2001} for metallicity
    \feh=\kepivcirSMEiizfehshort\ and ages 
	3.5, 4.0, 4.5, 5.0, 5.5, 6.0, 6.5, 7.0, 7.5, 8.0, 8.5 and
	9.0\,Gyrs.
	The final choice of $\teffstar$ and \arstar\ for the circular
    solution are marked by a filled circle, and is encircled by the
    1$\sigma$ and 2$\sigma$ confidence ellipsoids (solid, blue lines). 
    Corresponding values and confidence ellipsoids for the eccentric
    solution are shown with a triangle and dashed (red) lines. Fits
    from method B.
\label{fig:kep4iso}}
\end{figure}

The Yonsei-Yale model isochrones from \citet{yi:2001} for metallicity
\feh=\kepivcirSMEzfehshort\ are plotted in \reffigl{kep4iso}, with the
final choice of effective temperature $\teffstar$ and \arstar\ marked,
and encircled by the 1$\sigma$ and 2$\sigma$ confidence ellipsoids,
both for the circular and the eccentric orbital solution from method B.
Here the MCMC distribution of \arstar\ comes from the global modeling
of the data. Naturally, errors of the stellar parameters for the
eccentric solution are larger, due to the larger error in \arstar,
which, in turn, is due to the uncertainty in the Lagrangian orbital
parameters $k$ and $h$ (as opposed to fixing these to zero in the
circular solution).

The stellar evolution modeling provides color indices that may be
compared against the measured values as a sanity check. The best
available measurements are the near-infrared magnitudes from the 2MASS
Catalogue \citep{skrutskie:2006}. For \kepiv, these are:
$J_{\rm 2MASS}=\kepivcirCCtwomassJmag$, 
$H_{\rm 2MASS}=\kepivcirCCtwomassHmag$ and 
$K_{\rm 2MASS}=\kepivcirCCtwomassKmag$;
which we have converted to the photometric system of the models (ESO
system) using the transformations by \citet{carpenter:2001}. The
resulting measured color index is $J-K = \kepivcirCCesoJKmag$. This is
within 2-$\sigma$ of the predicted value from the isochrones of $J-K =
\kepivcirISOJK$ (see \ref{tab:kep4tab}).
The distance to the object may be computed from the absolute $K$
magnitude from the models ($M_{\rm K}=\kepivcirISOMK$) and the 2MASS
$K_s$ magnitude, which has the advantage of being less affected by
extinction than optical magnitudes. The result is $\kepivcirXdist$\,pc,
where the uncertainty excludes possible systematics in the model
isochrones that are difficult to quantify.

\begin{table*}
\caption{\emph{Global fits for Kepler-4b.  Quoted values are medians of
MCMC trials with errors given by 1-$\sigma$ quantiles. Two independent
methods are used to fit the data (A\&B) with both circular and eccentric modes
(c\&e). * = fixed parameter; $\dagger$ = parameter was floated but not fitted.}} 
\centering 
\begin{tabular}{c c c c c c} 
\hline\hline 
Parameter & Discovery & Method A.c & Method A.e & Model B.c & Model B.e \\ [0.5ex] 
\hline
\emph{Fitted params.} & & \\
\hline 
$P/$days & $3.21346 \pm 0.00022$ & $3.21345_{-0.00058}^{+0.00059}$ & $3.21343_{-0.00030}^{+ 0.00030}$ & $\kepivcirLCP$ & $\kepiveccLCP$ \\
$E$ (BJD-2,454,000) & $956.6127 \pm 0.0015$ & $956.6125_{-0.0041}^{+0.0041}$ & $956.61322_{-0.0026}^{+0.0027}$ & $956.6124 \pm 0.0017$ & $956.6132 \pm 0.0020$ \\
$T_{1,4}$/s & - & $14259_{-564}^{+701}$ & $14222_{-400}^{+535}$ & $14282 \pm 190$ & $14187 \pm 207$ \\
$T$/s & - & $13578_{-536}^{+593}$ & $13550_{-340}^{+396}$ & $13733 \pm 182$ & $13706 \pm 204$ \\ 
$(T_{1,2} \simeq T_{3,4})$/s & - & $513_{-149}^{+614}$ & $809_{-338}^{+600}$ & $501 \pm 207$ & $475 \pm 104$ \\
$(\rpl/\rstar)^2$/\% & - & $0.0691_{-0.0070}^{+0.0119}$ & $0.0740_{-0.0063}^{+0.0094}$ & $0.0660 \pm 0.0046$ & $0.0656 \pm 0.0031$ \\
$b^2$ & - & $0.30_{-0.26}^{+0.36}$ & $0.54_{-0.30}^{+0.19}$ & $\kepivcirLCbsq$ & $\kepiveccLCbsq$ \\
$(\zeta/\rstar)$/days$^{-1}$ & - & $12.88_{-0.57}^{+0.62}$ & $13.50_{-0.56}^{+0.65}$ & $\kepivcirLCzeta$ & $\kepiveccLCzeta$ \\
$(F_{P,\mathrm{d}}/F_\star)$/ppm & - & $10_{-29}^{+29}$ & $8_{-355}^{+67}$ & $-6 \pm 13$ & $5 \pm 29$ \\
$e\sin \omega$ & 0$^{*}$ & 0$^{*}$ & $0.24_{-0.16}^{+0.12}$ & 0$^{*}$ & $\kepiveccRVh$ \\
$e\cos \omega$ & 0$^{*}$ & 0$^{*}$ & $0.033_{-0.061}^{+0.054}$ & 0$^{*}$ & $\kepiveccRVk$ \\
$K$/ms$^{-1}$ & $9.3_{-1.9}^{+1.1}$ & $9.2_{-3.3}^{+3.3}$ & - & $\kepivcirRVK$ & $\kepiveccRVK$ \\
$\gamma$/ms$^{-1}$ & $-1.27 \pm 1.1$ & $-0.9_{-2.2}^{+2.2}$ & $-1.2_{-1.1}^{+1.1}$ & $\kepivcirRVgammarel$ & $\kepiveccRVgammarel$ \\
$\dot{\gamma}$/ms$^{-1}$day$^{-1}$ & 0$^{*}$ & 0$^{*}$ & 0$^{*}$ & 0$^{*}$ & 0$^{*}$ \\
$t_{\mathrm{troj}}$/days & 0$^{*}$ & 0$^{*}$ & 0$^{*}$ & 0$^{*}$ & 0$^{*}$ \\
$B$ & $1.02 \pm 0.02$ & $1.02 \pm 0.02$ $\dagger$ & $1.02 \pm 0.02$ $\dagger$ & $1.02 \pm 0.02$ $\dagger$ & $1.02 \pm 0.02$ $\dagger$ \\
\hline
\emph{SME derived params.} & & \\
\hline
$T_{\mathrm{eff}}$/K & \kepivcirSMEteff & \kepivcirSMEteff & \kepivcirSMEteff & \kepivcirSMEteff & \kepivcirSMEteff \\
log$(g/\mathrm{cgs})$ & $4.25 \pm 0.10$ & $4.25 \pm 0.10$ & $4.25 \pm 0.10$ & $4.25 \pm 0.10$ & $4.25 \pm 0.10$ \\
$\feh$ (dex) & \kepivcirSMEzfeh & \kepivcirSMEzfeh & \kepivcirSMEzfeh & \kepivcirSMEzfeh & \kepivcirSMEzfeh \\
$\vsini$ (\kms) & \kepivcirSMEvsin & \kepivcirSMEvsin & \kepivcirSMEvsin & \kepivcirSMEvsin & \kepivcirSMEvsin \\
$u_1$ & - & $0.61_{-0.39}^{+0.59}$ & $0.54_{-0.35}^{+0.63}$ & \kepivcirLBikep$^{*}$ & \kepivcirLBikep$^{*}$ \\
$u_2$ & - & $-0.21_{-0.68}^{+0.52}$ & $-0.23_{-0.67}^{+0.46}$ & \kepivcirLBiikep$^{*}$ & \kepivcirLBiikep$^{*}$ \\
\hline
\emph{Model indep.~params.} & & \\
\hline
$\rpl/\rstar$ & $0.02470_{-0.00030}^{+0.00031}$ & $0.0263_{-0.0014}^{+0.0022}$ & $0.0272_{-0.0012}^{+0.0017}$ & $\kepivcirLCrprstar$ & $\kepiveccLCrprstar$ \\
$\arstar$ & $6.47_{-0.28}^{+0.26}$ & $5.45_{-1.49}^{+0.87}$ & $3.57_{-0.63}^{+1.34}$ & $\kepivcirPPar$ & $\kepiveccPPar$ \\
$b$ & $0.022_{-0.022}^{+0.234}$ & $0.55_{-0.35}^{+0.26}$ & $0.73_{-0.25}^{+0.12}$ & $\kepivcirLCimp$ & $\kepiveccLCimp$ \\
$i$/$^{\circ}$ & $89.76_{-2.05}^{+0.24}$ & $84.3_{-6.0}^{+4.0}$ & $74.1_{-7.4}^{+9.5}$ & $\kepivcirPPi$ & $\kepiveccPPi$ \\
$e$ & 0$^{*}$ & 0$^{*}$ & $0.25_{-0.12}^{+0.11}$ & 0$^{*}$ & $\kepiveccRVeccen$ \\
$\omega$/$^{\circ}$ & - & - & $84.5_{-18.1}^{+18.0}$ & - & $\kepiveccRVomega$ \\
RV jitter (\ms) & - & 1.7 & 1.0 & \kepivcirRVjitter & \kepiveccRVjitter \\
$\rho_\star$/gcm$^{-3}$ & $0.525$ & $0.30_{-0.18}^{+0.17}$ & $0.084_{-0.037}^{+0.134}$ & \kepivcirISOrho & \kepiveccISOrho \\
log$(g_P/\mathrm{cgs})$ & $3.16_{-0.10}^{+0.06}$ & $2.90_{-0.38}^{+0.25}$ & $2.58_{-0.20}^{+0.28}$ & $\kepivcirPPlogg$ & $\kepiveccPPlogg$ \\
$S_{1,4}$/s & - & $14259_{-564}^{+701}$ & $0_{0}^{+14946}$ & $14282 \pm 190$ & $11975 \pm 544$ \\
$t_{sec}$/(BJD-2,454,000) & - & $961.4325_{-0.0058}^{+0.0059}$ & $961.28_{-1.51}^{+0.10}$ & $980.713 \pm 0.001$ & $980.89 \pm 0.17$ \\
\hline
\emph{Derived stellar params.} & & \\
\hline
$\mstar/\msun$ & $1.223_{-0.091}^{+0.053}$ & $1.28_{-0.13}^{+0.20}$ & $1.56_{-0.24}^{+0.20}$ & \kepivcirISOmlong & \kepiveccISOmlong \\ 
$\rstar/\rsun$ & $1.487_{-0.084}^{+0.071}$ & $1.82_{-0.28}^{+0.81}$ & $2.97_{-0.93}^{+0.79}$ & \kepivcirISOrlong & \kepiveccISOrlong \\
log$(g/\mathrm{cgs})$ & $4.17 \pm 0.04$ & $4.02_{-0.26}^{+0.12}$ & $3.68_{-0.15}^{+0.25}$ & \kepivcirISOlogg & \kepiveccISOlogg \\ 
$\lstar/\lsun$ & $2.26_{-0.48}^{+0.66}$ & $3.5_{-1.1}^{+3.8}$ & $9.3_{-4.9}^{+5.7}$ & \kepivcirISOlum & \kepiveccISOlum \\
$M_{V}$/mag & $4.00 \pm 0.28$ & $3.45_{-0.81}^{+0.40}$ & $2.39_{-0.52}^{+0.82}$ & \kepivcirISOmv & \kepiveccISOmv \\
Age/Gyr & $4.5 \pm 1.5$ & $4.2_{-1.2}^{+2.1}$ & $2.74_{-0.86}^{+1.31}$ & \kepivcirISOage & \kepiveccISOage \\
Distance/pc & $550 \pm 80$ & $566_{-94}^{+254}$ & $922_{-289}^{+252}$ & \kepivcirXdist & \kepiveccXdist \\
\hline
\emph{Derived planetary params.} & & \\
\hline
$\mpl/\mjup$ & $0.077 \pm 0.012$ & $0.081_{-0.030}^{+0.031}$ & $0.096_{-0.021}^{+0.023}$ & $\kepivcirPPmlong$ & $\kepiveccPPmlong$ \\ 
$\rpl/\rjup$ & $0.357 \pm 0.019$ & $0.460_{-0.084}^{+0.272}$ & $0.79_{-0.27}^{+0.26}$ & $\kepivcirPPrlong$ & $\kepiveccPPrlong$ \\ 
$\rho_P$/gcm$^{-3}$ & $1.91_{-0.47}^{+0.36}$ & $0.86_{-0.63}^{+0.97}$ & $0.24_{-0.13}^{+0.46}$ & $\kepivcirPPrho$ & $\kepiveccPPrho$ \\ 
$a$/AU & $0.0456 \pm 0.0009$ & $0.0463_{-0.0016}^{+0.0023}$ & $0.0494_{-0.0026}^{+0.0020}$ & $\kepivcirPParel$ & $\kepiveccPParel$ \\ 
$T_{\mathrm{eq}}/K$ & $1650 \pm 200$ & $1777_{-132}^{+308}$ & $2215_{-339}^{+233}$ & $\kepivcirPPteff$ & $\kepiveccPPteff$ \\ [1ex]
\hline\hline 
\end{tabular}
\label{tab:kep4tab} 
\end{table*}

\subsection{Transit timing analysis}
\subsubsection{Analysis of variance}

We find TTV residuals of 207.8 seconds and TDV residuals of 217.0
seconds using method A. Repeating the TTV analysis for method B finds
best-fit times of consistently less than 0.25-$\sigma$ deviance (we
note that similar results were found for all planets).  The TTV
indicates timings consistent with a linear ephemeris, producing a
$\chi^2$ of 5.7 for 11 degrees of freedom.  This is somewhat low with a
probability of 10.9\% of occurring by coincidence. This may be an
indication that the MCMC methods adopted here yield artificially large
error bars, perhaps due to our choice of free-fitting every \lc\ 
rather than fixing certain parameters.  However, we prefer to remain on
the side prudence by overestimating such errors rather than
underestimating them. The TDV too is consistent with a non-variable
system exhibiting a $\chi^2$ of 6.4 for 12 degrees of freedom (10.6\%
probability of random occurrence).

Under the assumption that the timing uncertainties are accurate or
overestimated, it is possible for us to determine what signal
amplitudes are excluded for waveforms of a period less than the
temporal baseline. For the TTV, we note that
for 11 degrees of freedom, excess scatter producing a $\chi^2$ of 28.5
would be detected to 3-$\sigma$ confidence. This therefore excludes a 
short-period signal of r.m.s.~amplitude $\geq 341.6$\,seconds. Similarly, 
for the TDV, we exclude scatter producing a $\chi^2$ of 30.1, or a 
short-period r.m.s.~amplitude of $\geq 339.3$\,seconds, to 3-$\sigma$ 
confidence. This constitutes a 4.9\% variation in the transit duration.

These limits place constraints on the presence of perturbing planets, moons
and Trojans. For the case a 2:1 mean-motion resonance perturbing planet,
the libration period would be $\sim 328$\,cycles (2.9 years) and thus we
do not possess sufficient baseline to look for such perturbers yet.
However, this libration period is less than the mission lifetime of
Kepler (3.5 years) and so assuming the same timing precision for all
measurements, the TTV will be sensitive to a 0.14\,$\mearth$
perturber.

The current TTV data rules out an exomoon at the maximum orbital radius
(for retrograde revolution) of mass $\geq 11.0 \mearth$.  The TDV data rules
an exomoon of $\geq 13.3\,\mearth$ at a minimum orbital distance.

Using the expressions of \citet{for07} and assuming $\mpl \gg 
M_{\mathrm{Trojan}}$, the expected libration period of \kepivb\
induced by Trojans at L4/L5 is 50.5 cycles and therefore we do not yet
possess sufficient temporal baseline to look for the TTVs. 
However, inspection of the folded photometry at $\pm P/6$ from 
the transit centre excludes Trojans of effective 
radius $\geq 1.22 R_{\oplus}$ to 3-$\sigma$ confidence.

\subsubsection{Periodograms}

As discussed in section \S2.3.2, we may use an F-test periodogram to search
for periodic signals, as this negates any problems introduced by
underestimating or overestimating timing uncertainties. Whilst a search
for excess variance is sensitive to any period below the temporal
baseline, periodograms are limited to a range of $2 P_P <
P_{\mathrm{signal}} < 2 P_{\mathrm{baseline}}$.

There are no peaks in the TTV periodogram below 5\% FAP\@. The TDV
does seem to exhibit a significant peak, which can also be seen by eye
in the TDV plot itself, at a period of $(15.5 \pm 3.1)$\,cycles, amplitude
$(204.9 \pm 61.2)$\,seconds and FAP 2.1\% (2.3-$\sigma$). A linear trend
in the durations could be confused with a long period signal. Fitting a
simple $a + b t$ model yields $a = (170.9 \pm 95.2)$\,seconds and $b =
(-31.6 \pm 13.8)$\,seconds/cycle with FAP 2.6\% (2.2-$\sigma$). 
Therefore, the two models yield very similar significances and it is
difficult to distinguish between these two scenarios. The linear drift
model seems improbable due to the very rapid change in duration the
trend implies. The gradient equates to a change of around one hour
per year in the half-duration, i.e.~two hours per year in
$T$. The period of the signal is too large to be caused by an
exomoon and TDV is generally insensitive to perturbing planets, meaning
that any planet which could produce such a large effect would dominate
the radial velocity. Other TDV effects, for example apsidal precession,
are not expected to occur on timescales of 10-20 days which puts the
reality of the signal in question.

The phasing of the measurements, as defined in \S2.3.4, yields a clear
periodic waveform of period 4 cycles. Therefore one possible
explanation for the TDV signal is a mixing of this phasing frequency
with the Nyquist frequency to produce a signal of period 16 cycles,
which is consistent with the value determined earlier. The analysis of
further measurements is evidently required to assess the reality of the
putative signal.

\begin{figure*}
\begin{center}
\includegraphics[width=16.8 cm]{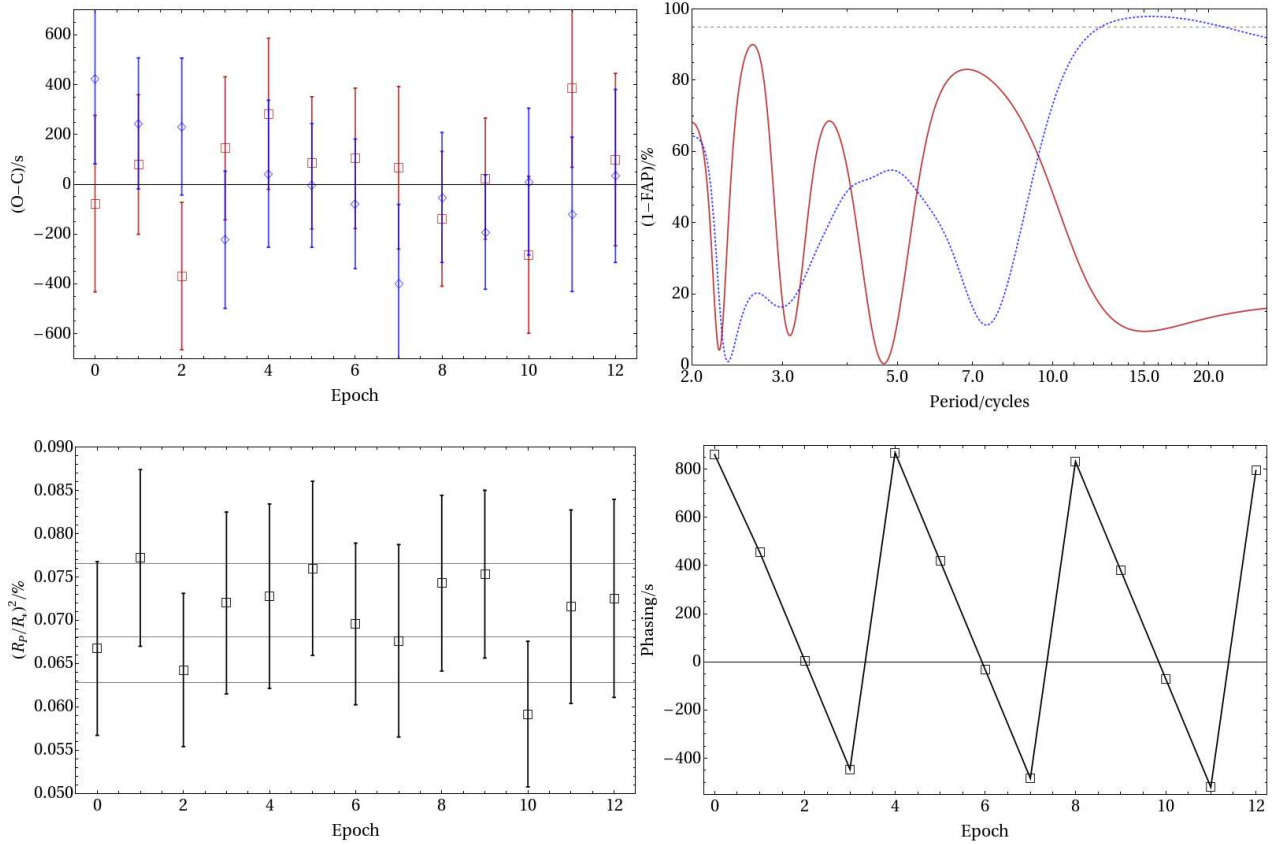}
\caption{\emph{
{\bf Upper Left:} TTV (squares) and TDV (diamonds) for \kepivb\ (see \S2.3.1 for details).
{\bf Upper Right:} TTV periodogram (solid) and TDV periodogram (dashed) for \kepivb, calculated using a sequential F-test (see \S2.3.2 for description).
{\bf Lower Left:} Transit depths from individual fits of \kepivb\ (see \S2.3.1 for details).
{\bf Lower Right:} ``Phasing'' of Kepler long cadence photometry for \kepivb\ (see \S2.3.4 for description).
}} \label{fig:kep4ttv}
\end{center}
\end{figure*}

\subsection{Depth and OOT Variations}

The transit depth is consistent with the globally fitted value yielding
a $\chi^2$ of 4.3 for 12 degrees of freedom. Similarly, the OOT levels
are flat yielding a $\chi^2$ of 6.1 for 12 degrees of freedom.

\begin{table*}
\caption{\emph{Mid-transit times, transit durations, transit depths and
out-of-transit (normalized) fluxes for Kepler-4b.}}
\centering 
\begin{tabular}{c c c c c} 
\hline\hline 
Epoch & $t_c$/(BJD-2,454,000) & $T$/s & $(\rpl/\rstar)^2/\%$ & $F_{oot}$ \\ [0.5ex] 
\hline
0 & $956.6126_{-0.0038}^{+0.0044}$ & $14684.0_{-720.4}^{+653.8}$ & $0.067_{-0.008}^{+0.012}$ & $1.000006_{-0.000013}^{+0.000013}$ \\
1 & $959.8278_{-0.0032}^{+0.0033}$ & $14321.2_{-512.9}^{+539.4}$ & $0.077_{-0.008}^{+0.012}$ & $0.999998_{-0.000013}^{+0.000013}$ \\
2 & $963.0361_{-0.0033}^{+0.0036}$ & $14296.0_{-548.9}^{+551.0}$ & $0.064_{-0.007}^{+0.010}$ & $0.999978_{-0.000016}^{+0.000016}$ \\
3 & $966.2554_{-0.0032}^{+0.0034}$ & $13387.4_{-545.8}^{+556.2}$ & $0.072_{-0.008}^{+0.013}$ & $1.000015_{-0.000013}^{+0.000013}$ \\
4 & $969.4704_{-0.0033}^{+0.0037}$ & $13917.1_{-585.2}^{+596.0}$ & $0.073_{-0.008}^{+0.013}$ & $0.999995_{-0.000013}^{+0.000013}$ \\
5 & $972.6816_{-0.0029}^{+0.0032}$ & $13823.7_{-486.1}^{+509.0}$ & $0.076_{-0.008}^{+0.012}$ & $1.000005_{-0.000013}^{+0.000013}$ \\
6 & $975.8952_{-0.0031}^{+0.0034}$ & $13675.3_{-518.5}^{+522.8}$ & $0.070_{-0.008}^{+0.011}$ & $0.999998_{-0.000013}^{+0.000013}$ \\
7 & $979.1082_{-0.0037}^{+0.0039}$ & $13038.3_{-631.6}^{+633.0}$ & $0.069_{-0.009}^{+0.014}$ & $0.999996_{-0.000013}^{+0.000013}$ \\
8 & $982.3193_{-0.0030}^{+0.0032}$ & $13726.6_{-512.5}^{+531.7}$ & $0.074_{-0.008}^{+0.012}$ & $1.000000_{-0.000013}^{+0.000013}$ \\
9 & $985.5345_{-0.0027}^{+0.0029}$ & $13448.9_{-450.2}^{+468.2}$ & $0.075_{-0.008}^{+0.012}$ & $1.000003_{-0.000013}^{+0.000013}$ \\
10 & $988.7444_{-0.0035}^{+0.0038}$ & $13854.2_{-573.2}^{+605.5}$ & $0.059_{-0.007}^{+0.010}$ & $1.000004_{-0.000013}^{+0.000013}$ \\
11 & $991.9656_{-0.0035}^{+0.0039}$ & $13590.9_{-605.8}^{+634.7}$ & $0.072_{-0.009}^{+0.014}$ & $1.000008_{-0.000013}^{+0.000013}$ \\
12 & $995.1757_{-0.0040}^{+0.0040}$ & $13899.6_{-650.2}^{+739.9}$ & $0.073_{-0.009}^{+0.014}$ & $1.000013_{-0.000013}^{+0.000013}$ \\ [1ex]
\hline
\end{tabular}
\label{tab:kep4ttv} 
\end{table*}

%% file: kep5.tex

\subsection{Global fits}
\subsubsection{Comparison to discovery paper}

\begin{figure}[!ht]
\plotone{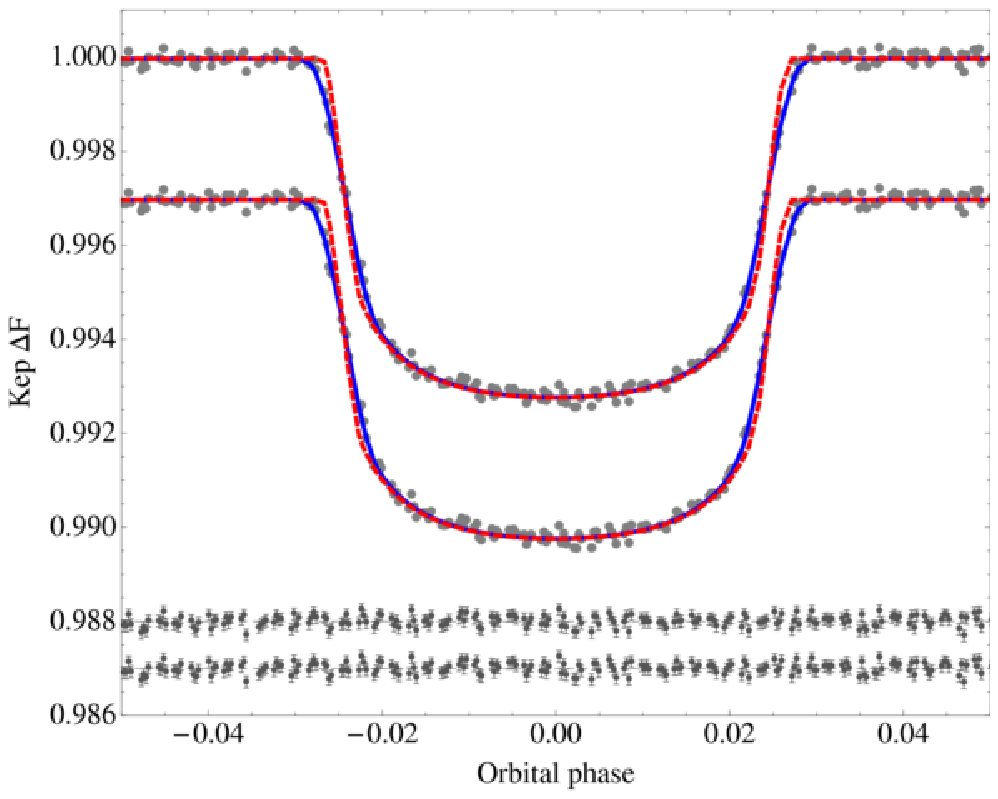}
\caption{
    {\bf Top:} Phase-folded primary transit \lc\ of \kepv. The upper
    curve shows the circular fit, the bottom curve the eccentric fit. 
    Solid (blue) lines indicate the best fit resampled
    model (with bin-number 4). The dashed (red) lines show the
    corresponding unbinned model, which one would get if the transit
    was observed with infinitely fine time resolution.
	Residuals from the best fit resampled model for the
	circular and eccentric solutions are shown below in
        respective order.
\label{fig:kep5prim}}
\end{figure}

\begin{figure}[ht]
\plotone{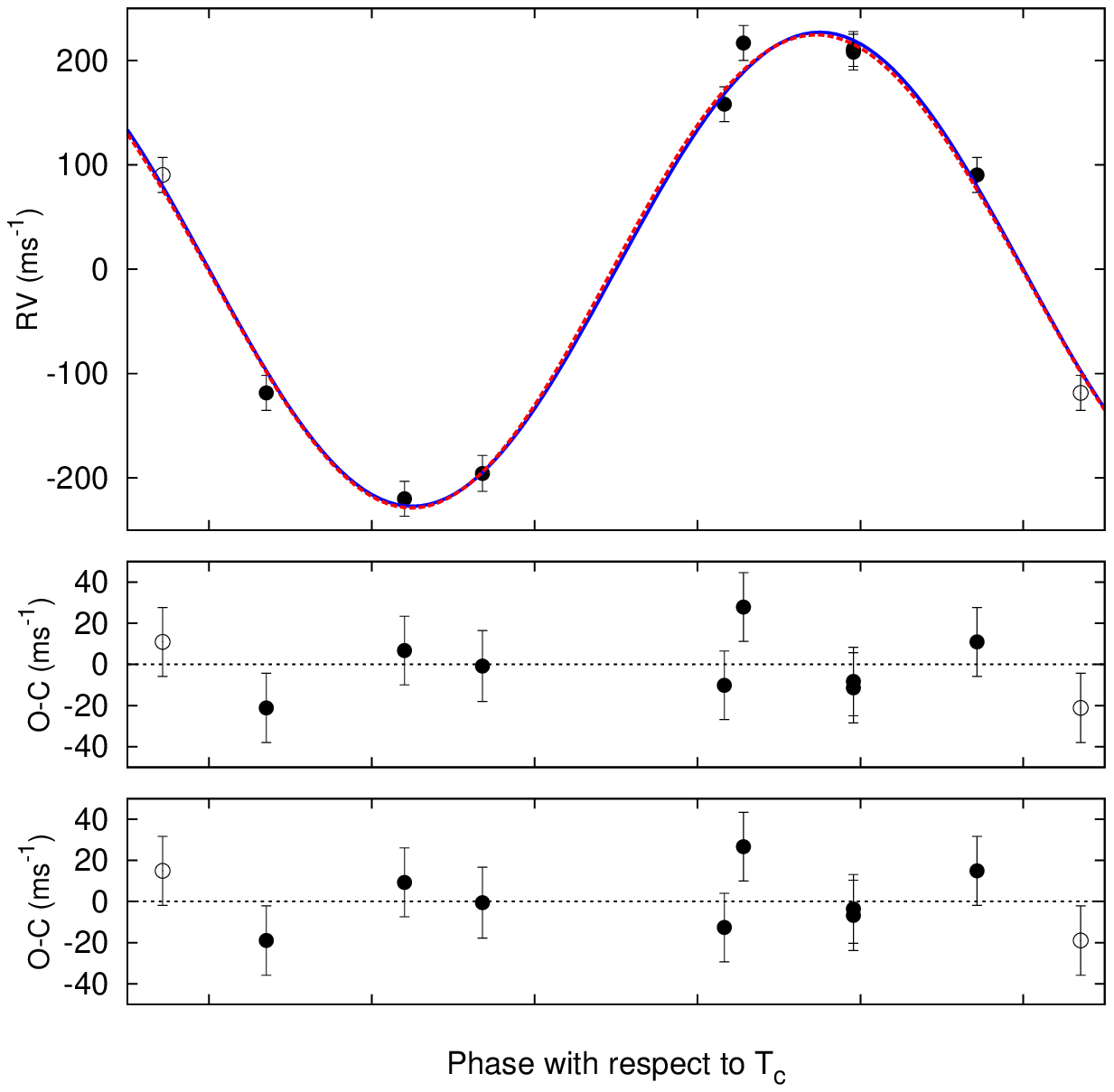}
\caption{
    {\bf Top:} RV measurements from Keck for \kepv{}, along with an
    orbital fit, shown as a function of orbital phase, using our
    best-fit period. Solid (blue) line shows the circular orbital fit
    with binned RV model (3 bins, separated by 600\,seconds). The
    dashed (red) line shows the eccentric orbital fit (with the same
    bin parameters). Zero phase is defined by the transit midpoint. The
    center-of-mass velocity has been subtracted.  Note that the error
    bars include the stellar jitter (taken for the circular solution),
    added in quadrature to the formal errors given by the spectrum
    reduction pipeline. Fits from method B.
    {\bf Middle:} phased residuals after subtracting the orbital fit
    for the circular solution. The r.m.s.~variation of the residuals is
    \kepvcirRVfitrms\,\ms, and the stellar jitter that was added in
    quadrature to the formal RV errors is $\kepvcirRVjitter$\,\ms.
	{\bf Bottom:} phased residuals after subtracting the orbital fit
    for the eccentric solution.  Here the r.m.s.~variation of the
    residuals is \kepveccRVfitrms\,\ms, and the stellar jitter is
    $\kepveccRVjitter$\,\ms.
\label{fig:kep5rv}}
\end{figure}

\kepvb\ was discovered by \citet{koc10}. Our global fits reveal
\kepvb\ to be a $\sim1.35 \rjup$ planet on an orbit consistent with that 
of a circular orbit, transiting the host star at a near-equatorial impact
parameter. The fitted models for method B of the phase-folded \lcs\ are
shown on \reffigl{kep5prim}. Correlated noise was checked for in the residuals
using the Durbin-Watson statistic, which finds $d=1.991$, which is well within
the 1\% critical level of $1.867$. The orbital fits to the RV points are
shown in \reffig{kep5rv} for method B, depicting both the circular and
eccentric solutions. We obtain transit parameters broadly consistent with that
of the discovery paper, except for $b$ where method A favors an equatorial
transit and method B favors a near-equatorial transit, whereas \citet{koc10}
found $b \simeq 0.4$. The final parameters are summarized in
Table~\ref{tab:kep5tab}. \Lc\ fits reveal that the theoretical limb 
darkening values differ from the fitted values by a noticeable amount
and the residuals of method B do reveal some structure as a consequence.
Future SC mode observations will pin down the limb darkening to a much
higher precision.

\subsubsection{Eccentricity}

The global circular fit of method A yields a $\chi^2 = 20.77$ for the RVs
and the eccentric fit obtains $\chi^2 = 19.77$. Based on an F-test, the
inclusion of two new degrees of freedom to describe the eccentricity is
justified below the 1-$\sigma$ level and is therefore not
significant. We therefore conclude that the system is consistent with a
circular orbit and constrain $e<0.086$ to 95\% confidence.

\subsubsection{Secondary eclipse}

We here only consider the circular fits since the eccentric solution is
not statistically significant. Method B provides the more accurate 
constraints due the use of multiple baseline scalings. Proceeding with 
method B, the secondary eclipse is weakly detected in our analysis to 
93.4\% confidence, or 1.8-$\sigma$, with a depth of $26_{-17}^{+17}$\,ppm. 

\begin{figure}[!ht]
\plotone{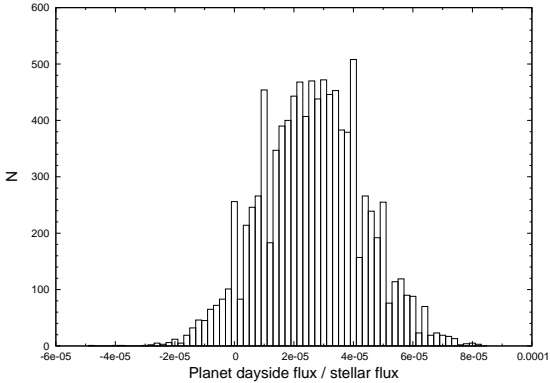}
\caption{
    Distribution of \fpdfs\ for \kepv from the global modeling. 
	\fpdfs\ is primarily constrained by the depth of the secondary
	eclipse. Only results for the circular orbital solution of method B
	are shown.
\label{fig:kep5sampid}}
\end{figure}

The MCMC distribution of the \fpdfs\
values is shown in \reffigl{kep5sampid}. The discovery paper reported a
weak 2.0-$\sigma$ detection of the eclipse but did not provide an
amplitude for the signal.  An eclipse of depth $\geq 76$\,ppm is
excluded at the 3-$\sigma$ level, which excludes a geometric albedo
$\geq 0.47$ to the same level.

If due to thermal emission, we find that by integrating a Planck
function with the Kepler response function, the planetary brightness
temperature would have to be $2480_{-284}^{+159}$\,K, which is
inconsistent with that expected for the equilibrium temperature of
$(1770 \pm 19)$\,K. Assuming negligible albedo and solving for the
redistribution factor implies a factor of 3.7, which surpasses the
maximal physically permitted value of 8/3 (corresponding to 2300K).
Internal heating also seems to be unlikely as the heat source would
have to be able to generate a surface temperature of 2300\,K alone, or
have an intrinsic luminosity of $4.6 \times 10^{-4} L_{\odot}$ which is
approximately that of a M7V star. Tidal heating from eccentricity seems
improbable given that the orbit is consistent with a circular orbit.
However, using our best-fit eccentricity and the equations of \citet{pea78},
 we estimate that tidal heating could induce $8.3 \times
10^{-8} \lsun$ for $Q_P = 10^5$ and $k_{2p} = 0.5$. In order to get
enough tidal heating, we require $Q' = Q_P/k_{2p} = 35$, which is
much less than the typical values expected. Furthermore, such a low
$Q_P$ leads to very rapid circularization which somewhat nullifies the
argument.

If the eclipse was due to reflection, we estimate a required geometric
albedo of $A_g = 0.15 \pm 0.10$. This latter option does
offer a plausible physical origin for the secondary eclipse and
therefore should be considered as a possible explanation for the
observation. Although there has been a general impression that
hot-Jupiters must have low albedos after the remarkable MOST
constraints of $A_g < 0.17$ for HD 209458b by \citet{row08}, a recent
study by \citet{cow10} finds evidence for hot-Jupiters having much
higher albedos ($>0.4$) in a statistical sample of 20 exoplanets.
Therefore, a geometric albedo of 0.15 is not exceptional in such a
sample and offers much more reasonable explanation that thermal
emission. However, at this stage, the $\sim$2-$\sigma$ significance is too
low to meet the requirements for formal detection.

\begin{figure}[!ht]
\plotone{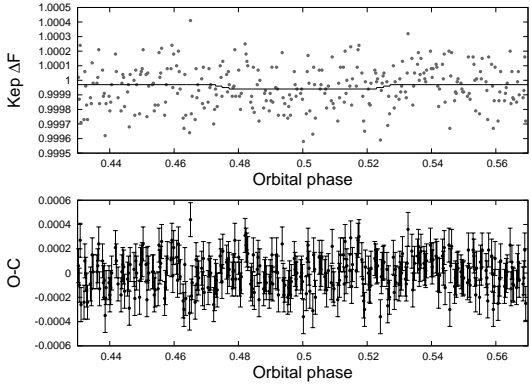}
\caption{
    Phase-folded secondary transit \lc\ of \kepv\ (top), and residuals
    from the best fit (bottom). Only the fit for the circular orbital
    solution of method B is shown.
\label{fig:kep5sec}}
\end{figure}

\subsubsection{Properties of the parent star \kepv}
\label{sec:stelparam}

\begin{figure}[!ht]
\plotone{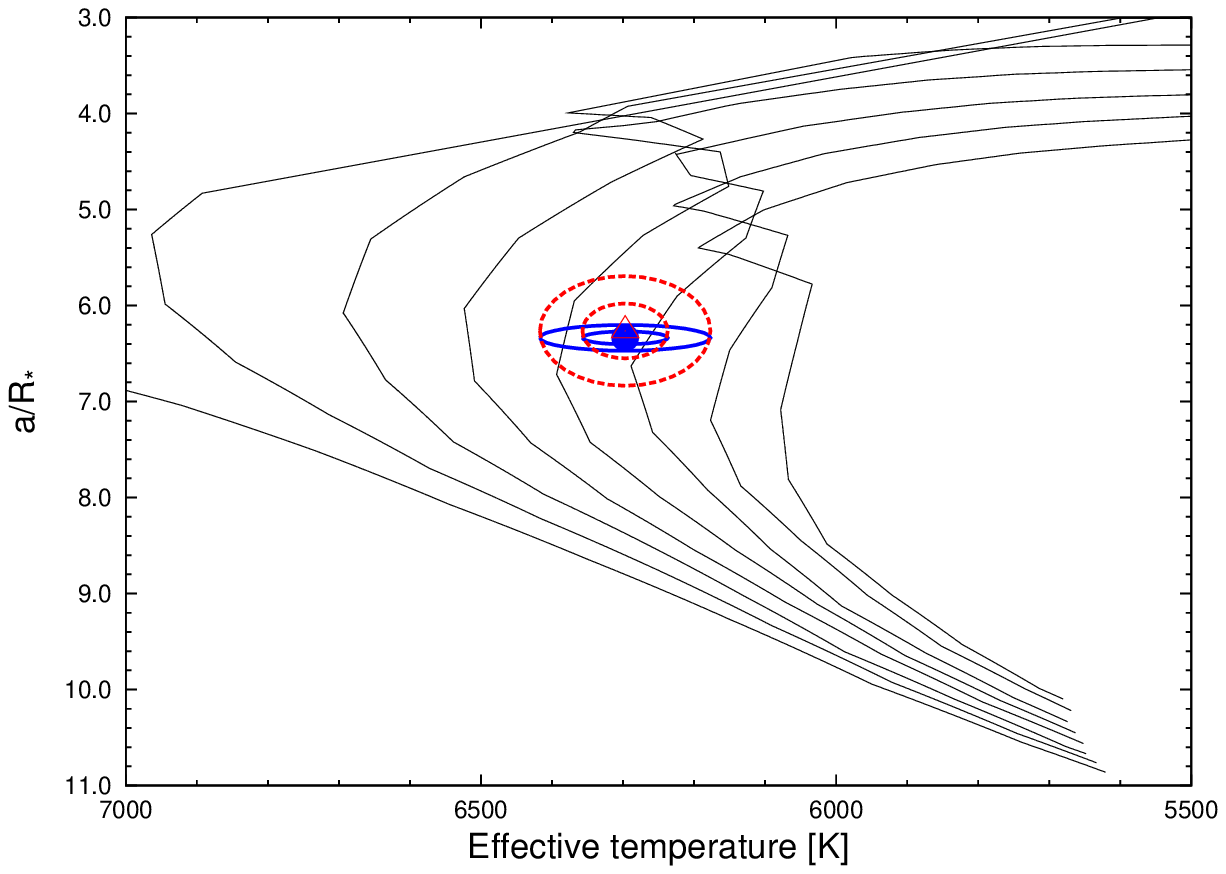}
\caption{
    Stellar isochrones from \citet{yi:2001} for metallicity
    \feh=\kepvcirSMEiizfehshort\ and ages 
	1.0, 1.4, 1.8, 2.2, 2.6, 3.0, 3.4 and 3.8\,Gyrs.
	The final choice of $\teffstar$ and \arstar\ for the circular
    solution are marked by a filled circle, and is encircled by the
    1$\sigma$ and 2$\sigma$ confidence ellipsoids (solid, blue lines). 
    Corresponding values and confidence ellipsoids for the eccentric
    solution are shown with a triangle and dashed (red) lines. Fits
    from method B.
\label{fig:kep5iso}}
\end{figure}

The Yonsei-Yale model isochrones from \citet{yi:2001} for metallicity
\feh=\kepvcirSMEzfehshort\ are plotted in \reffigl{kep5iso}, with the
final choice of effective temperature $\teffstar$ and \arstar\ marked,
and encircled by the 1$\sigma$ and 2$\sigma$ confidence ellipsoids,
both for the circular and the eccentric orbital solution. Here the MCMC
distribution of \arstar\ comes from the global modeling of the data.

\begin{table*}

\caption{
	\emph{Global fits for Kepler-5b.  Quoted values are medians of
	MCMC trials with errors given by 1-$\sigma$ quantiles. Two independent
	methods are used to fit the data (A\&B) with both circular and eccentric 
	modes (c\&e).* = fixed parameter; $\dagger$ = parameter was floated but 
	not fitted.}
}
\centering 
\begin{tabular}{c c c c c c} 
\hline\hline 
Parameter & Discovery & Method A.c & Method A.e & Model B.c & Model B.e \\ [0.5ex] 
\hline
\emph{Fitted params.} & & \\
\hline 
$P/$days & $3.548460 \pm 0.000032$ & $3.548460_{-0.000075}^{+0.000074}$ & $3.548460_{-0.000051}^{+0.000051}$ & $\kepvcirLCP$ & $\kepveccLCP$ \\
$E$ (BJD-2,454,000) & $955.90122 \pm 0.00021$ &  $955.90126_{-0.00049}^{+0.00049}$ & $955.90124_{-0.00033}^{+0.00033}$ & $955.90124 \pm 0.00023$ & $955.90124 \pm 0.00023$ \\
$T_{1,4}$/s & - & $16580_{-165}^{+184}$ & $16547_{-117}^{+137}$ & $16459 \pm 35$ & $16468 \pm 35$ \\
$T$/s & - & $15209_{-169}^{+171}$ & $15232_{-118}^{+122}$ & $15233 \pm 23$ & $15230 \pm 25$ \\ 
$(T_{1,2} \simeq T_{3,4})$/s & - & $1309_{-84}^{+257}$ & $1275_{-52}^{+173}$ & $1227 \pm 26$ & $1227 \pm 35$ \\
$(\rpl/\rstar)^2$/\% & - & $0.637_{-0.017}^{+0.026}$ & $0.632_{-0.012}^{+0.018}$ & $0.6225 \pm 0.0032$ & $0.6225 \pm 0.0032$ \\
$b^2$ & - & $0.066_{-0.059}^{+0.143}$ & $0.042_{-0.038}^{+0.106}$ & $\kepvcirLCbsq$ & $\kepveccLCbsq$ \\
$(\zeta/\rstar)$/days$^{-1}$ & - & $11.42_{-0.13}^{+0.14}$ & $11.403_{-0.093}^{+0.098}$ & $\kepvcirLCzeta$ & $\kepveccLCzeta$ \\
$(F_{P,\mathrm{d}}/F_\star)$/ppm & - & $29_{-35}^{+35}$ & $29_{-24}^{+24}$ & $26 \pm 17$ & $25 \pm 25$ \\
$e\sin \omega$ & 0$^{*}$ & 0$^{*}$ & $0.012_{-0.038}^{+0.041}$ & 0$^{*}$ & $\kepveccRVh$ \\
$e\cos \omega$ & 0$^{*}$ & 0$^{*}$ & $-0.008_{-0.016}^{+0.017}$ & 0$^{*}$ & $\kepveccRVk$ \\
$K$/ms$^{-1}$ & $231.3 \pm 2.8$ & $220.6_{-14.2}^{+14.2}$ & $221.1_{-9.4}^{+9.3}$ & $\kepvcirRVK$ & $\kepveccRVK$ \\
$\gamma$/ms$^{-1}$ & $-46.7 \pm 4.1$ & $-43.1_{-10.0}^{+10.0}$ & $-42.1_{-7.0}^{+6.8}$ & $\kepvcirRVgammarel$ & $\kepveccRVgammarel$ \\
$\dot{\gamma}$/ms$^{-1}$day$^{-1}$ & 0$^{*}$ & $0.28_{-0.26}^{+0.26}$ & $0.28_{-0.17}^{+0.17}$ & 0$^{*}$ & 0$^{*}$ \\
$t_{\mathrm{troj}}$/days & 0$^{*}$ & 0$^{*}$ & 0$^{*}$ & 0$^{*}$ & 0$^{*}$ \\
$B$ & $1.020 \pm 0.002$ & $1.020 \pm 0.002$ $\dagger$ & $1.020 \pm 0.002$ $\dagger$ & $1.020 \pm 0.002$ $\dagger$ & $1.020 \pm 0.002$ $\dagger$ \\
\hline
\emph{SME derived params.} & & \\
\hline
$T_{\mathrm{eff}}$/K & \kepvcirSMEteff & \kepvcirSMEteff & \kepvcirSMEteff & \kepvcirSMEteff & \kepvcirSMEteff \\
log$(g/\mathrm{cgs})$ & $3.96 \pm 0.10$ & $3.96 \pm 0.10$ & $3.96 \pm 0.10$ & $3.96 \pm 0.10$ & $3.96 \pm 0.10$ \\
$\feh$ (dex) & \kepvcirSMEzfeh & \kepvcirSMEzfeh & \kepvcirSMEzfeh & \kepvcirSMEzfeh & \kepvcirSMEzfeh \\
$\vsini$ (\kms) & \kepvcirSMEvsin & \kepvcirSMEvsin & \kepvcirSMEvsin & \kepvcirSMEvsin & \kepvcirSMEvsin \\
$u_1$ & - & $0.25_{-0.12}^{+0.13}$ & $0.244_{-0.085}^{+0.083}$ & \kepvcirLBikep$^{*}$ & \kepvcirLBikep$^{*}$ \\
$u_2$ & - & $0.37_{-0.27}^{+0.25}$ & $0.39_{-0.18}^{+0.19}$ & \kepvcirLBiikep$^{*}$ & \kepvcirLBiikep$^{*}$ \\
\hline
\emph{Model indep.~params.} & & \\
\hline
$\rpl/\rstar$ & $0.08195_{-0.00047}^{+0.00030}$ & $0.0798_{-0.0011}^{+0.0016}$ & $0.07952_{-0.00074}^{+0.00111}$ & $\kepvcirLCrprstar$ & $\kepveccLCrprstar$ \\
$\arstar$ & $6.06 \pm 0.14$ & $6.21_{-0.43}^{+0.18}$ & $6.15_{-0.35}^{+0.30}$ & $\kepvcirPPar$ & $\kepveccPPar$ \\
$b$ & $0.393_{-0.043}^{+0.051}$ & $0.00_{-0.00}^{+0.35}$ & $0.01_{-0.01}^{+0.29}$ & $\kepvcirLCimp$ & $\kepveccLCimp$ \\
$i$/$^{\circ}$ & $86.3 \pm 0.5$ & $87.6_{-2.2}^{+1.7}$ & $88.1_{-1.9}^{+1.3}$ & $\kepvcirPPi$ & $\kepveccPPi$ \\
$e$ & 0$^{*}$ & 0$^{*}$ & $0.034_{-0.018}^{+0.029}$ & 0$^{*}$ & $\kepveccRVeccen$ \\
$\omega$/$^{\circ}$ & - & - & $132_{-43}^{+140}$ & - & $\kepveccRVomega$ \\
RV jitter (\ms) & - & 8.2 & 7.7 & \kepvcirRVjitter & \kepveccRVjitter \\
$\rho_\star$/gcm$^{-3}$ & $0.336$ & $0.360_{-0.070}^{+0.033}$ & $0.350_{-0.057}^{+0.053}$ & \kepvcirISOrho & \kepveccISOrho \\
log$(g_P/\mathrm{cgs})$ & $3.41 \pm 0.03$ & $3.433_{-0.078}^{+0.046}$ & $3.433_{-0.062}^{+0.048}$ & $\kepvcirPPlogg$ & $\kepveccPPlogg$ \\
$S_{1,4}$/s & - & $16580_{-165}^{+184}$ & $20006_{-4329}^{+761}$ & $16459 \pm 35$ & $16718 \pm 1417$ \\
$t_{sec}$/(BJD-2,454,000) & - & $961.22372_{-0.00049}^{+0.00049}$ & $959.43_{-0.04}^{+1.77}$ & $978.9660 \pm 0.0002$ & $975.397 \pm 0.052$ \\
\hline
\emph{Derived stellar params.} & & \\
\hline
$\mstar/\msun$ & $1.374_{-0.059}^{+0.040}$ & $1.370_{-0.036}^{+0.050}$ & $1.372_{-0.041}^{+0.048}$ & \kepvcirISOmlong & \kepveccISOmlong \\ 
$\rstar/\rsun$ & $1.793_{-0.062}^{+0.043}$ & $1.749_{-0.059}^{+0.151}$ & $1.768_{-0.094}^{+0.127}$ & \kepvcirISOrlong & \kepveccISOrlong \\
log$(g/\mathrm{cgs})$ & $4.07 \pm 0.02$ & $4.087_{-0.057}^{+0.023}$ & $4.080_{-0.048}^{+0.037}$ & \kepvcirISOlogg & \kepveccISOlogg \\ 
$\lstar/\lsun$ & $4.67_{-0.59}^{+0.63}$ & $4.34_{-0.37}^{+0.77}$ & $4.41_{-0.50}^{+0.69}$ & \kepvcirISOlum & \kepveccISOlum \\
$M_{V}$/mag & - & $3.16_{-0.18}^{+0.10}$ & $3.14_{-0.16}^{+0.14}$ & \kepvcirISOmv & \kepveccISOmv \\
Age/Gyr & $3.0 \pm 0.6$ & $2.83_{-0.25}^{+0.23}$ & $2.82_{-0.24}^{+0.24}$ & \kepvcirISOage & \kepveccISOage \\
Distance/pc & - & $1101_{-51}^{+94}$ & $1109_{-67}^{+85}$ & \kepvcirXdist & \kepveccXdist \\
\hline
\emph{Derived planetary params.} & & \\
\hline
$\mpl/\mjup$ & $2.114_{-0.059}^{+0.056}$ & $2.05_{-0.14}^{+0.14}$ & $2.054_{-0.098}^{+0.099}$ & $\kepvcirPPmlong$ & $\kepveccPPmlong$ \\ 
$\rpl/\rjup$ & $1.431_{-0.052}^{+0.041}$ & $1.352_{-0.052}^{+0.149}$ & $1.366_{-0.076}^{+0.112}$ & $\kepvcirPPrlong$ & $\kepveccPPrlong$ \\ 
$\rho_P$/gcm$^{-3}$ & $0.894 \pm 0.079$ & $1.01_{-0.25}^{+0.15}$ & $1.00_{-0.20}^{+0.18}$ & $\kepvcirPPrho$ & $\kepveccPPrho$ \\ 
$a$/AU & $0.05064 \pm 0.00070$ & $0.05056_{-0.00045}^{+0.00061}$ & $0.05059_{-0.00051}^{+0.00058}$ & $\kepvcirPParel$ & $\kepveccPParel$ \\ 
$T_{\mathrm{eq}}/K$ & $1868 \pm 284$ & $1790_{-33}^{+64}$ & $1796_{-45}^{+57}$ & $\kepvcirPPteff$ & $\kepveccPPteff$ \\ [1ex]
\hline\hline 
\end{tabular}
\label{tab:kep5tab} 
\end{table*}

\subsection{Transit timing analysis}
\subsubsection{Analysis of variance}

We find TTV residuals of 30.1 seconds and TDV residuals of 35.6
seconds.  The TTV indicates timings consistent with a linear ephemeris,
producing a $\chi^2$ of 8.7 for 10 degrees of freedom.  The TDV is also
consistent with a non-variable system exhibiting a $\chi^2$ of 10.5 for
11 degrees of freedom.

For the TTV, we note that for 10 degrees of freedom, excess scatter
producing a $\chi^2$ of 26.9 would be detected to 3-$\sigma$
confidence. This therefore excludes a short-period signal of
r.m.s.~amplitude $\geq 39.8$\,seconds. Similarly, for the TDV, we exclude
scatter producing a $\chi^2$ of 28.5, or a short-period
r.m.s.~amplitude of $\geq 45.9$\,seconds, to 3-$\sigma$ confidence. This
constitutes a 0.60\% variation in the transit duration.

These limits place constraints on the presence of perturbing planets, moons
and Trojans. For the case a 2:1 mean-motion resonance perturbing planet,
the libration period would be $\sim 38.7$\,cycles and thus we do not
possess sufficient temporal baseline to look for such perturbers yet.
However, a 3:2 resonance would produce a libration period of 15.4
cycles and thus we would expect the effects to be visible over the 12
observed cycles so far. This therefore excludes a 3:2 resonant
perturber of mass $\geq 0.79 \mearth$ to 3-$\sigma$ confidence.

The current TTV data rules out an exomoon at the maximum orbital radius
(for retrograde revolution) of $\geq 10.6 \mearth$.  The current TDV data
rules an exomoon of $\geq 17.3 \mearth$ at a minimum orbital
distance.

Using the expressions of \citet{for07} and assuming $\mpl \gg 
M_{\mathrm{Trojan}}$, the expected libration period of \kepvb\
induced by Trojans at L4/L5 is 10.0 cycles and therefore we can 
search for such TTVs. We find such Trojans of cumulative mass 
$\geq 3.14 M_{\oplus}$ at angular displacement $10^{\circ}$ are excluded
by our timings. Inspection of the folded photometry at $\pm P/6$ 
from the transit centre excludes Trojans of 
effective radius $\geq 1.13 R_{\oplus}$ to 3-$\sigma$ confidence.

\subsubsection{Periodograms}

An F-test periodogram of the TTV reveals only one significant peaks
which is very close to Nyquist frequency. We therefore disregard this
frequency as being significant.

The TDV F-test periodogram reveals a significant peak at $(4.0 \pm 0.2)$
\,cycles with amplitude $(37 \pm 11)$\,seconds with significance
2.2-$\sigma$. We again note that the period is not in the range which
would be expected due to the known physical sources. The phasing effect
has a period of 3 cycles and thus a mixing between the phasing and
Nyquist frequency again offers a reasonable explanation for this
signal, especially since peaks occur at 2 and 3 cycles in the TDV
periodogram as well.

\begin{figure*}
\begin{center}
\includegraphics[width=16.8 cm]{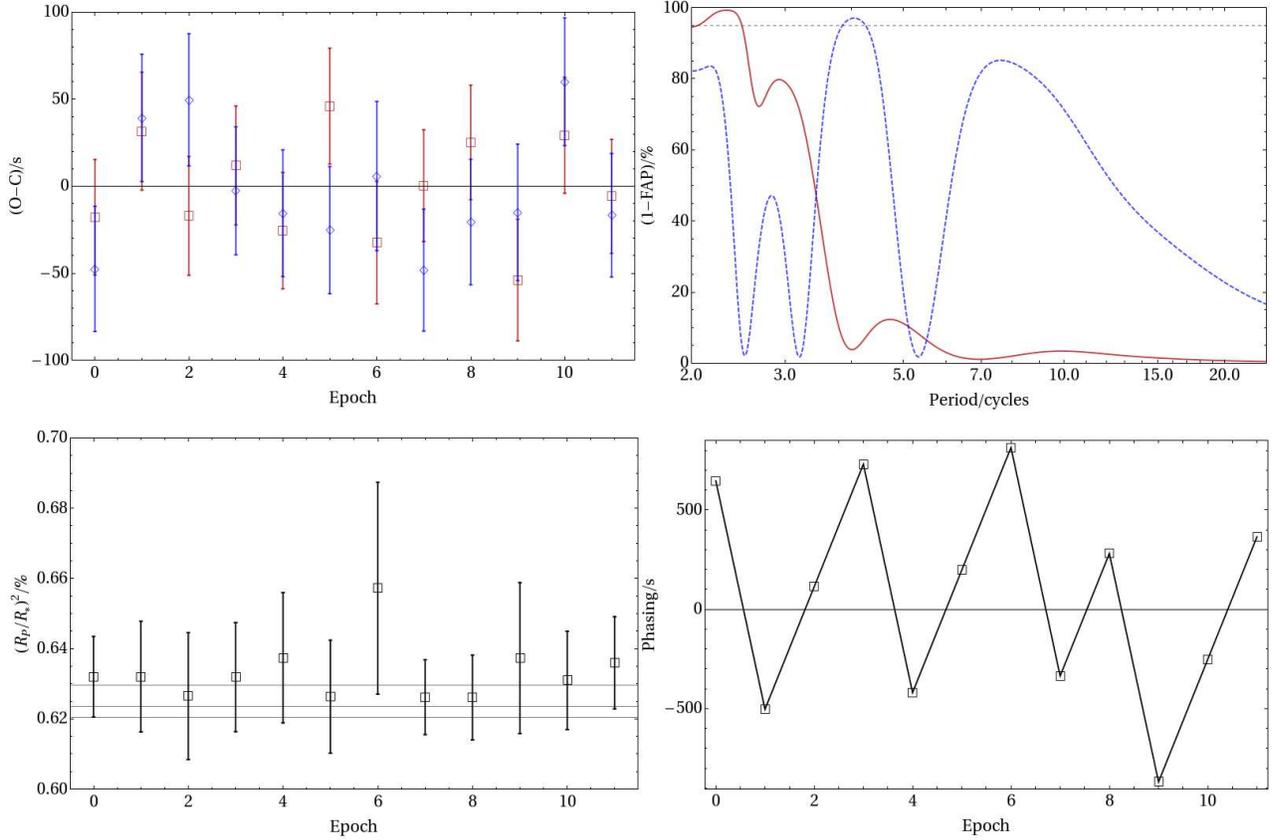}
\caption{\emph{
	{\bf Upper Left:} TTV (squares) and TDV (diamonds) for \kepvb\ (see
	\S2.3.1 for details).
	{\bf Upper Right:} TTV periodogram (solid) and TDV periodogram
	(dashed) for \kepvb, calculated using a sequential F-test (see
	\S2.3.2 for description).
	{\bf Lower Left:} Transit depths from individual fits of \kepvb\
	(see \S2.3.1 for details).
	{\bf Lower Right:} ``Phasing'' of Kepler long cadence photometry
	for \kepvb\ (see \S2.3.4 for description).
}}
\label{fig:kep5ttv}
\end{center}
\end{figure*}

\subsection{Depth and OOT Variations}

The transit depth is consistent with the globally fitted value yielding
a $\chi^2$ of 7.9 for 11 degrees of freedom.  Similarly, the OOT levels
are flat yielding a $\chi^2$ of 4.0 for 11 degrees of freedom.

\begin{table*}
\caption{
	\emph{Mid-transit times, transit durations, transit depths and
	out-of-transit (normalized) fluxes for Kepler-5b.}
}
\centering 
\begin{tabular}{c c c c c} 
\hline\hline 
Epoch & $t_c$/(BJD-2,454,000) & $T$/s & $(\rpl/\rstar)^2/\%$ & $F_{oot}$ \\ [0.5ex] 
\hline
0 & $955.90106_{-0.00039}^{+0.00038}$ & $15177.8_{-71.7}^{+72.0}$ & $0.632_{-0.008}^{+0.015}$ & $0.999952_{-0.000015}^{+0.000015}$ \\
1 & $959.45010_{-0.00040}^{+0.00039}$ & $15351.4_{-73.5}^{+72.8}$ & $0.632_{-0.011}^{+0.020}$ & $0.999958_{-0.000015}^{+0.000015}$ \\
2 & $962.99799_{-0.00040}^{+0.00039}$ & $15372.1_{-75.6}^{+76.1}$ & $0.626_{-0.015}^{+0.022}$ & $0.999943_{-0.000019}^{+0.000019}$ \\
3 & $966.54678_{-0.00040}^{+0.00039}$ & $15267.6_{-74.0}^{+72.9}$ & $0.632_{-0.011}^{+0.020}$ & $0.999965_{-0.000015}^{+0.000015}$ \\
4 & $970.09481_{-0.00039}^{+0.00039}$ & $15241.9_{-71.9}^{+73.6}$ & $0.637_{-0.015}^{+0.022}$ & $0.999970_{-0.000015}^{+0.000015}$ \\
5 & $973.64409_{-0.00038}^{+0.00038}$ & $15222.4_{-72.5}^{+73.5}$ & $0.626_{-0.011}^{+0.021}$ & $0.999970_{-0.000015}^{+0.000015}$ \\
6 & $977.19164_{-0.00041}^{+0.00041}$ & $15284.6_{-87.6}^{+83.8}$ & $0.657_{-0.028}^{+0.033}$ & $0.999964_{-0.000015}^{+0.000015}$ \\
7 & $980.74048_{-0.00037}^{+0.00037}$ & $15176.5_{-69.5}^{+70.6}$ & $0.626_{-0.008}^{+0.014}$ & $0.999965_{-0.000015}^{+0.000015}$ \\
8 & $984.28922_{-0.00038}^{+0.00038}$ & $15231.7_{-71.1}^{+72.8}$ & $0.626_{-0.009}^{+0.016}$ & $0.999966_{-0.000015}^{+0.000015}$ \\
9 & $987.83677_{-0.00040}^{+0.00041}$ & $15242.8_{-79.0}^{+77.7}$ & $0.637_{-0.016}^{+0.027}$ & $0.999975_{-0.000015}^{+0.000015}$ \\
10 & $991.38619_{-0.00038}^{+0.00039}$ & $15392.9_{-72.9}^{+73.7}$ & $0.631_{-0.010}^{+0.018}$ & $0.999977_{-0.000015}^{+0.000015}$ \\
11 & $994.93424_{-0.00038}^{+0.00038}$ & $15239.6_{-70.3}^{+71.5}$ & $0.636_{-0.009}^{+0.017}$ & $0.999974_{-0.000015}^{+0.000015}$ \\ [1ex]
\hline
\end{tabular}
\label{tab:kep5ttv} 
\end{table*}

%% file: kep6.tex

\subsection{Global Fits}
\subsubsection{Comparison to the discovery paper}

\begin{figure}[!ht]
\plotone{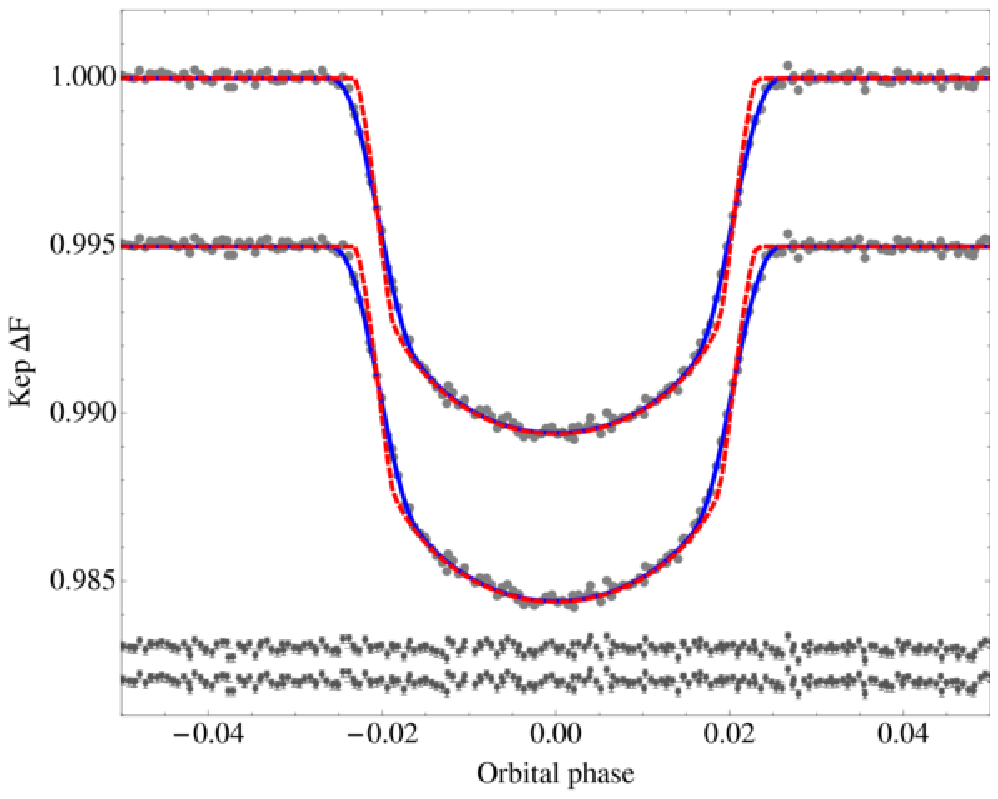}
\caption{
    {\bf Top:} Phase-folded primary transit \lc\ of \kepvi for method
    B. The upper curve shows the circular fit, the bottom curve the
    eccentric fit.  Solid (blue) lines indicate the best fit resampled
    model (with bin-number 4). The dashed (red) lines show the
    corresponding unbinned model, which one would get if the transit
    was observed with infinitely fine time resolution.
	Residuals from the best fit resampled model for the
	circular and eccentric solutions are shown below in
        respective order..
\label{fig:kep6prim}}
\end{figure}

\begin{figure}[ht]
\plotone{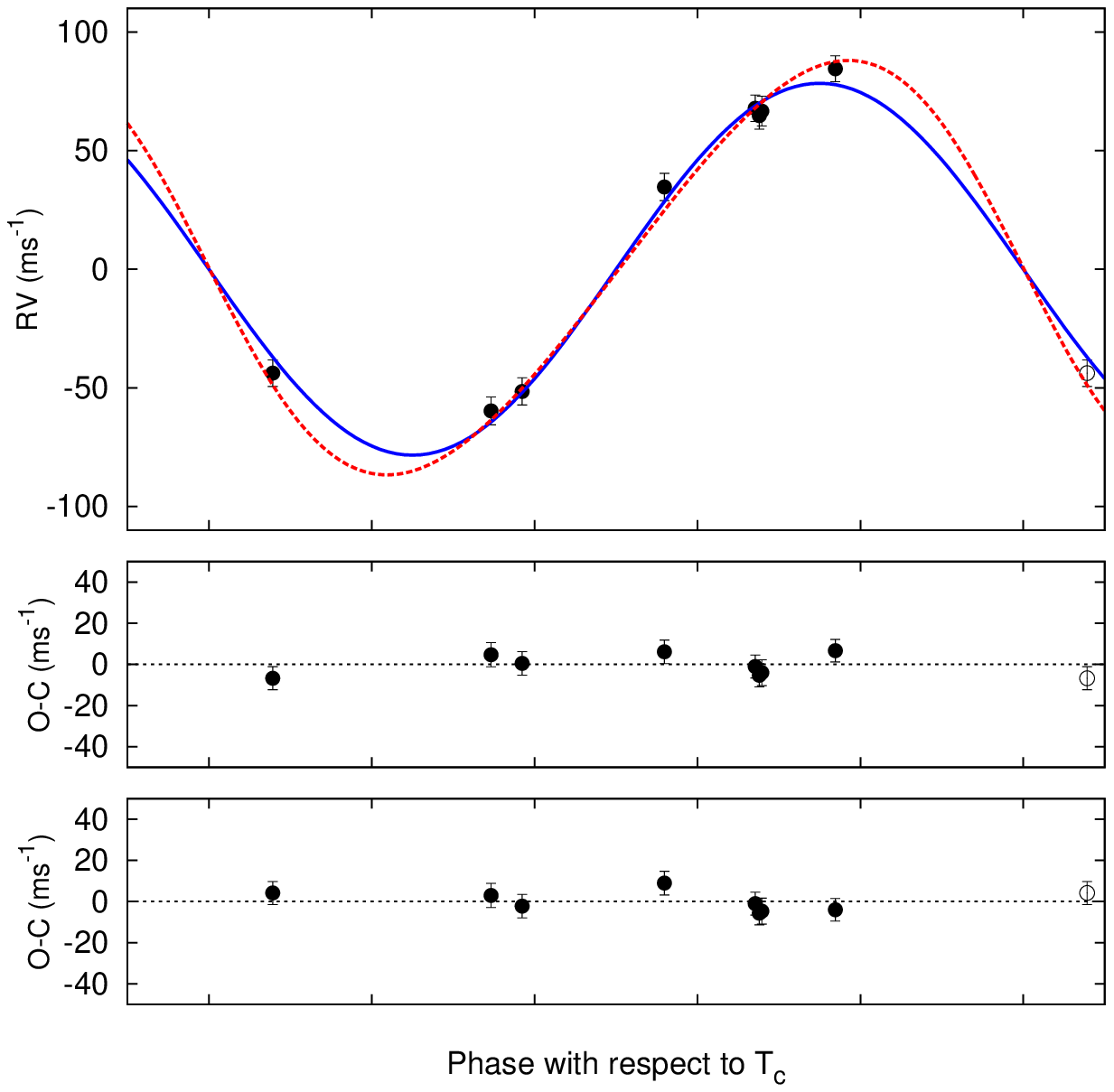}
\caption{
    {\bf Top:} RV measurements from Keck for \kepvi{}, along with an
    orbital fit, shown as a function of orbital phase, using our
    best-fit period. Solid (blue) line shows the
    circular orbital fit with binned RV model (3 bins, separated by
    600\,seconds). The dashed (red) line shows the eccentric orbital
    fit (with the same bin parameters). Zero phase is defined by the
    transit midpoint. The center-of-mass velocity has been subtracted. 
    Note that the error bars include the stellar jitter (taken for the
    circular solution), added in quadrature to the formal errors given
    by the spectrum reduction pipeline. Fits from method B.
    {\bf Middle:} phased residuals after subtracting the orbital fit
    for the circular solution. The r.m.s.~variation of the residuals is
    \kepvicirRVfitrms\,\ms, and the stellar jitter that was added in
    quadrature to the formal RV errors is $\kepvicirRVjitter$\,\ms.
	{\bf Bottom:} phased residuals after subtracting the orbital fit
    for the eccentric solution.  Here the r.m.s.~variation of the
    residuals is \kepvieccRVfitrms\,\ms, and the stellar jitter is
    $\kepvieccRVjitter$\,\ms.
\label{fig:kep6rv}}
\end{figure}

\kepvib\ was discovered by \citet{dun10}. In a similar manner to
the other planets, we find consistent values for most parameters when 
compared to the \citet{dun10} analysis. The fitted models on the
phase-folded \lcs\ for method B are shown on \reffigl{kep6prim}. Correlated
noise was checked for in the residuals using the Durbin-Watson statistic, which
finds $d=1.943$ which is well within the 1\% critical level of $1.870$. The
orbital fits to the RV points using method B are shown in
\reffig{kep6rv}, depicting both the circular and eccentric fits. The
final planetary parameters are summarized at the bottom of
Table~\ref{tab:kep6tab}.

\subsubsection{Eccentricity}
The global circular fit yields a $\chi^2 = 9.01$ and the eccentric
fit obtains $\chi^2 = 7.75$ for the RVs. Based on an F-test, the inclusion
of two new degrees of freedom to describe the eccentricity is justified 
with a significance of $<1$-$\sigma$ The system is therefore best described 
with a circular orbit with the current observations. We constrain $e<0.13$ to
95\% confidence.

\subsubsection{Secondary eclipse}

We here only consider the circular fit since the eccentric solution is
not statistically significant. We find no evidence for a secondary
eclipse detection for \kepvib.  However, our analysis does exclude
secondary eclipses of depth $51.5$\,ppm or greater to 3-$\sigma$
confidence, which excludes a geometric albedo $\geq 0.32$ to the same
level. This therefore excludes the possibility of a planet covered 
in reflective clouds. A brightness temperature of $2445$\,K is ruled 
out to the same level, which places no constraints on the redistribution. 

\subsubsection{Properties of the parent star \kepvi}
\label{sec:stelparam}

\begin{figure}[!ht]
\plotone{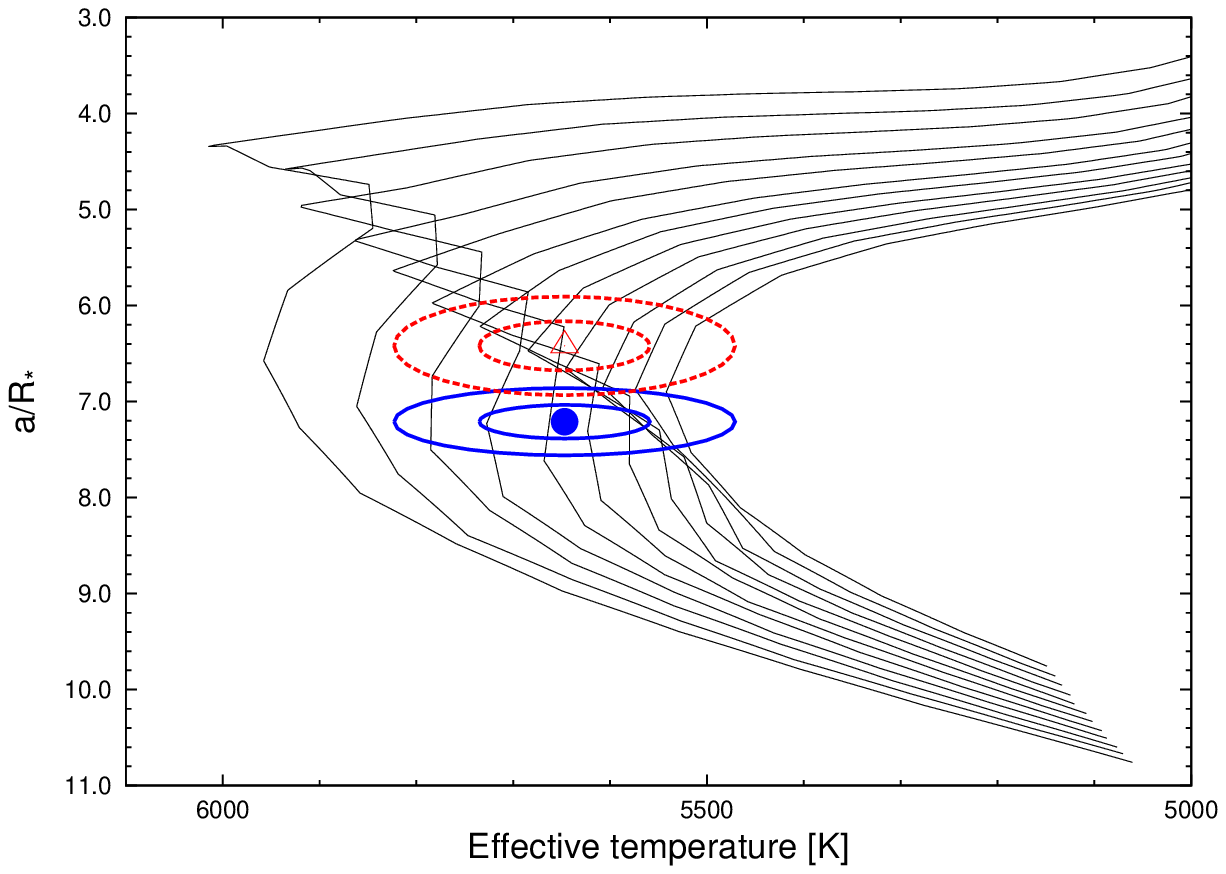}
\caption{
    Stellar isochrones from \citet{yi:2001} for metallicity
    \feh=\kepvicirSMEiizfehshort\ and ages 
	1.0, 1.4, 1.8, 2.2, 2.6, 3.0, 3.4 and 3.8\,Gyrs.
	The final choice of $\teffstar$ and \arstar\ for the circular
    solution are marked by a filled circle, and is encircled by the
    1$\sigma$ and 2$\sigma$ confidence ellipsoids (solid, blue lines). 
    Corresponding values and confidence ellipsoids for the eccentric
    solution are shown with a triangle and dashed (red) lines. Fits
    from method B.
\label{fig:kep6iso}}
\end{figure}

The Yonsei-Yale model isochrones from \citet{yi:2001} for metallicity
\feh=\kepvicirSMEzfehshort\ are plotted in \reffigl{kep6iso}, with the
final choice of effective temperature $\teffstar$ and \arstar\ marked,
and encircled by the 1$\sigma$ and 2$\sigma$ confidence ellipsoids,
both for the circular and the eccentric orbital solution. Here the MCMC
distribution of \arstar\ comes from the global modeling of the data.

\begin{table*}
\caption{
	\emph{Global fits for Kepler-6b.  Quoted values are medians of
	MCMC trials with errors given by 1-$\sigma$ quantiles. Two independent
	methods are used to fit the data (A\&B) with both circular and eccentric 
	modes (c\&e).* = fixed parameter; $\dagger$ = parameter was floated but 
	not fitted.}}
\centering 
\begin{tabular}{c c c c c c} 
\hline\hline 
Parameter & Discovery & Method A.c & Method A.e & Model B.c & Model B.e \\ [0.5ex] 
\hline
\emph{Fitted params.} & & \\
\hline 
$P$/days & $3.234723 \pm 0.000017$ & $3.234721_{-0.000043}^{+0.000043}$ & $3.234721_{-0.000028}^{+0.000028}$ & $\kepvicirLCP$ & $\kepvieccLCP$ \\
$E$ (BJD-2,454,000) & $954.48636 \pm 0.00014$ & $954.48639_{-0.00034}^{+0.00034}$ & $954.48637_{-0.00022}^{+0.00022}$ & $954.48640 \pm 0.00014$ & $954.48641 \pm 0.00015$ \\
$T_{1,4}$/s & - & $12982_{-140}^{+160}$ & $12956_{-98}^{+111}$ & $13003 \pm 60$ & $12977 \pm 60$ \\
$T$/s & - & $11720_{-142}^{+149}$ & $11728_{-98}^{+100}$ & $11871 \pm 16$ & $11762 \pm 18$ \\ 
$(T_{1,2} \simeq T_{3,4})$/s & - & $1211_{-84}^{+234}$ & $1200_{-73}^{+173}$ & $1227 \pm 69$ & $1218 \pm 69$ \\
$(\rpl/\rstar)^2$/\% & - & $0.913_{-0.028}^{+0.047}$ & $0.910_{-0.021}^{+0.036}$ & $0.901 \pm 0.011$ & $0.901 \pm 0.011$ \\
$b^2$ & - & $0.069_{-0.062}^{+0.137}$ & $0.061_{-0.054}^{+0.107}$ & $\kepvicirLCbsq$ & $\kepvieccLCbsq$ \\
$(\zeta/\rstar)$/days$^{-1}$ & - & $14.80_{-0.19}^{+0.19}$ & $14.80_{-0.13}^{+0.14}$ & $\kepvicirLCzeta$ & $\kepvieccLCzeta$ \\
$(F_{P,\mathrm{d}}/F_*)$/ppm & - & $5_{-36}^{+36}$ & $-4_{-24}^{+24}$ & $15 \pm 18$ & $21 \pm 35$ \\
$e\sin \omega$ & 0$^{*}$ & 0$^{*}$ & $0.042_{-0.050}^{+0.055}$ & 0$^{*}$ & $\kepvieccRVh$ \\
$e\cos \omega$ & 0$^{*}$ & 0$^{*}$ & $-0.019_{-0.019}^{+0.020}$ & 0$^{*}$ & $\kepvieccRVk$ \\
$K$/ms$^{-1}$ & $80.9 \pm 2.6$ & $78.4_{-6.0}^{+6.1}$ & $81.8_{-4.6}^{+4.7}$ & $\kepvicirRVK$ & $\kepvieccRVK$ \\
$\gamma$/ms$^{-1}$ & $-18.3 \pm 3.5$ & $-18.2_{-4.5}^{+4.5}$ & $-17.7_{-3.0}^{+3.0}$ & $\kepvicirRVgammarel$ & $\kepvieccRVgammarel$ \\
$\dot{\gamma}$/ms$^{-1}$day$^{-1}$ & 0$^{*}$ & 0$^{*}$ & 0$^{*}$ & 0$^{*}$ & 0$^{*}$ \\
$t_{\mathrm{troj}}$/days & 0$^{*}$ & $-0.041_{-0.043}^{+0.042}$ & 0$^{*}$ & 0$^{*}$ & 0$^{*}$ \\
$B$ & $1.033 \pm 0.004$ & $1.033 \pm 0.004$ $\dagger$ & $1.033 \pm 0.004$ $\dagger$ & $1.033 \pm 0.004$ $\dagger$ & $1.033 \pm 0.004$ $\dagger$ \\
\hline
\emph{SME derived params.} & & \\
\hline
$T_{\mathrm{eff}}$/K & $5647 \pm 44$ & \kepvicirSMEteff & \kepvicirSMEteff & \kepvicirSMEteff & \kepvicirSMEteff \\
log$(g/\mathrm{cgs})$ & $4.59 \pm 0.10$ & $4.59 \pm 0.10$ & $4.59 \pm 0.10$ & $4.59 \pm 0.10$ & $4.59 \pm 0.10$ \\
$\feh$ (dex) & \kepvicirSMEzfeh & \kepvicirSMEzfeh & \kepvicirSMEzfeh & \kepvicirSMEzfeh & \kepvicirSMEzfeh \\
$\vsini$ (\kms) & \kepvicirSMEvsin & \kepvicirSMEvsin & \kepvicirSMEvsin & \kepvicirSMEvsin & \kepvicirSMEvsin \\
$u_1$ & - & $0.55_{-0.11}^{+0.13}$ & $0.544_{-0.072}^{+0.084}$ & \kepvicirLBikep$^{*}$ & \kepvicirLBikep$^{*}$ \\
$u_2$ & - & $0.01_{-0.27}^{+0.26}$ & $0.02_{-0.18}^{+0.17}$ & \kepvicirLBiikep$^{*}$ & \kepvicirLBiikep$^{*}$ \\
\hline
\emph{Model indep.~params.} & & \\
\hline
$\rpl/\rstar$ & $0.09829_{-0.00050}^{+0.00014}$ & $0.0955_{-0.0015}^{+0.0024}$ & $0.0954_{-0.0011}^{+0.0018}$ & $\kepivcirLCrprstar$ & $\kepiveccLCrprstar$ \\
$\arstar$ & $7.05_{-0.06}^{+0.11}$ & $7.32_{-0.48}^{+0.22}$ & $7.00_{-0.46}^{+0.44}$ & $\kepvicirPPar$ & $\kepvieccPPar$ \\
$b$ & $0.398_{-0.039}^{+0.020}$ & $0.26_{-0.26}^{+0.19}$ & $0.25_{-0.25}^{+0.16}$ & $\kepvicirLCimp$ & $\kepvieccLCimp$ \\
$i$/$^{\circ}$ & $86.8 \pm 0.3$ & $87.9_{-1.7}^{+1.4}$ & $87.9_{-1.6}^{+1.4}$ & $\kepvicirPPi$ & $\kepvieccPPi$ \\
$e$ & 0$^{*}$ & 0$^{*}$ & $0.056_{-0.028}^{+0.044}$ & 0$^{*}$ & $\kepvieccRVeccen$ \\
$\omega$/$^{\circ}$ & - & - & $115_{-24}^{+86}$ & - & $\kepvieccRVomega$ \\
RV jitter (\ms) & - & 0.6 & 0.0 & \kepvicirRVjitter & \kepvieccRVjitter \\
$\rho_\star$/gcm$^{-3}$ & $0.634$ & $0.709_{-0.131}^{+0.067}$ & $0.619_{-0.114}^{+0.125}$ & \kepvicirISOrho & \kepvieccISOrho \\
log$(g_P/\mathrm{cgs})$ & $2.974_{-0.022}^{+0.016}$ & $3.010_{-0.081}^{+0.052}$ & $2.99_{-0.066}^{+ 0.054}$ & $\kepvicirPPlogg$ & $\kepvieccPPlogg$ \\
$S_{1,4}$/s & - & $12982_{-140}^{+160}$ & $15371_{-2832}^{+165}$ & $13003 \pm 60$ & $10022 \pm 829$ \\
$t_{sec}$/(BJD-2,454,000) & - & $959.33827_{-0.00034}^{+0.00034}$ & $957.68_{-0.04}^{+1.60}$ & $978.7470 \pm 0.0001$ & $978.772 \pm 0.048$ \\
\hline
\emph{Derived stellar params.} & & \\
\hline
$\mstar/\msun$ & $1.209_{-0.038}^{+0.044}$ & $1.127_{-0.065}^{+0.045}$ & $1.137_{-0.076}^{+0.057}$ & \kepvicirISOmlong & \kepvieccISOmlong \\ 
$\rstar/\rsun$ & $1.391_{-0.034}^{+0.017}$ & $1.306_{-0.046}^{+0.102}$ & $1.370_{-0.091}^{+0.109}$ & \kepvicirISOrlong & \kepvieccISOrlong \\
log$(g/\mathrm{cgs})$ & $4.236 \pm 0.011$ & $4.254_{-0.056}^{+0.026}$ & $4.218_{-0.057}^{+0.050}$ & \kepvicirISOlogg & \kepvieccISOlogg \\ 
$\lstar/\lsun$ & $1.99_{-0.21}^{+0.24}$ & $1.57_{-0.17}^{+0.26}$ & $1.71_{-0.25}^{+0.32}$ & \kepvicirISOlum & \kepvieccISOlum \\
$M_{V}$/mag & - & $4.35_{-0.17}^{+0.14}$ & $4.26_{-0.19}^{+0.18}$ & \kepvicirISOmv & \kepvieccISOmv \\
Age/Gyr & $3.8 \pm 1.0$ & $5.65_{-0.86}^{+2.68}$ & $5.65_{-0.90}^{+2.84}$ & \kepvicirISOage & \kepvieccISOage \\
Distance/pc & - & $604_{-37}^{+50}$ & $630_{-51}^{+59}$ & \kepvicirXdist & \kepvieccXdist \\
\hline
\emph{Derived planetary params.} & & \\
\hline
$\mpl/\mjup$ & $0.669_{-0.030}^{+0.025}$ & $0.617_{-0.051}^{+0.052}$ & $0.645_{-0.044}^{+0.047}$ & $\kepvicirPPmlong$ & $\kepvieccPPmlong$ \\ 
$\rpl/\rjup$ & $1.323_{-0.029}^{+0.026}$ & $1.208_{-0.049}^{+0.129}$ & $1.271_{-0.091}^{+0.116}$ & $\kepvicirPPrlong$ & $\kepvieccPPrlong$ \\ 
$\rho_P$/gcm$^{-3}$ & $0.352_{-0.022}^{+0.018}$ & $0.426_{-0.105}^{+0.069}$ & $0.389_{-0.081}^{+0.084}$ & $\kepvicirPPrho$ & $\kepvieccPPrho$ \\ 
$a$/AU & $0.04567_{-0.00046}^{+0.00055}$ & $0.04452_{-0.00088}^{+0.00058}$ & $0.04466_{-0.00101}^{+0.00074}$ & $\kepvicirPParel$ & $\kepvieccPParel$ \\ 
$T_{\mathrm{eq}}/K$ & $1500 \pm 200$ & $1480_{-33}^{+51}$ & $1511_{-51}^{+57}$ & $\kepvicirPPteff$ & $\kepvieccPPteff$ \\ [1ex]
\hline\hline  
\end{tabular}
\label{tab:kep6tab} 
\end{table*}

\subsection{Transit timing analysis}
\subsubsection{Analysis of variance}

We find TTV residuals of 17.4 seconds and TDV residuals of 25.4
seconds.  The TTV indicates timings consistent with a linear ephemeris,
producing a $\chi^2$ of 7.8 for 11 degrees of freedom. The TDV is also
consistent with a non-variable system exhibiting a $\chi^2$ of 11.4 for
12 degrees of freedom.

For the TTV, we note that for 11 degrees of freedom, excess scatter
producing a $\chi^2$ of 28.5 would be detected to 3-$\sigma$
confidence. This therefore excludes a short-period signal of
r.m.s.~amplitude $\geq 26.2$\,seconds. Similarly, for the TDV, we exclude
scatter producing a $\chi^2$ of 30.1, or a short-period
r.m.s.~amplitude of $\geq 31.5$\,seconds, to 3-$\sigma$ confidence. This
constitutes a 0.53\% variation in the transit duration.

These limits place constraints on the presence of perturbing planets, 
moons and Trojans. For the case a 2:1 mean-motion resonance perturbing planet,
the libration period would be $\sim 76.5$\,cycles and thus we do not
possess sufficient baseline to look for such perturbers yet. However, a
4:3 resonance would produce a libration period of $17.7$\,cycles and thus
we would expect the effects to be visible over the 14 observed cycles
so far. This therefore excludes a 4:3 resonant perturber of mass $\geq 0.38
\mearth$ to 3-$\sigma$ confidence.

The current TTV data rules out an exomoon at the maximum orbital radius
(for retrograde revolution) of $\geq 4.8 \mearth$.  The current TDV data
rules an exomoon of $\geq 8.2 \mearth$ at a minimum orbital
distance.

Using the expressions of \citet{for07} and assuming $\mpl \gg 
M_{\mathrm{Trojan}}$, the expected libration period of \kepvib\
induced by Trojans at L4/L5 is 16.7 cycles and therefore we can 
search for such TTVs. We find such Trojans of cumulative mass 
$\geq 0.67 M_{\oplus}$ at angular displacement $10^{\circ}$ are excluded
by our timings. Inspection of the folded photometry at $\pm P/6$ 
from the transit centre excludes Trojans of 
effective radius $\geq 0.87 R_{\oplus}$ to 3-$\sigma$ confidence.

\subsubsection{Periodograms}

The TTV periodogram reveals a significant peak of period $(5.34 \pm
0.26)$\,cycles with amplitude $(19.7 \pm 5.0)$\,seconds and FAP
2.6-$\sigma$, which is one of our highest significances found, yet
still below our formal detection threshold of 3-$\sigma$. Nevertheless, the
signal warrants a more detailed investigation as it does not appear to
land on any of the probable aliasing frequencies.

Given that Trojans should induce libration of period 16.7 cycles, this
hypothesis can be instantly discounted. For two planets in a
$j$:$j+1$ mean motion-resonance, we may set the TTV signal's period to
be equal to the expression for the expected libration period (from
\citet{agol:2005}) and solve for $j$. This yields $j =
7.37_{-0.26}^{+0.28}$, or a period ratio of $1.136 \pm 0.005$, which
does not appear to be a resonant configuration. \citet{bar06} used the
Hill criterion to provide an expression for estimating period ratios
which are not guaranteed to be dynamically stable (equation 2). Using
this expression, we estimate period ratios of $>0.75$ and $<1.34$ are
unlikely to be stable for masses above $1 \mearth$. This makes the
hypothesis of a mean-motion resonance unlikely.

For an out-of-resonant signal, the period of the TTV is more variable,
depending on the difference between the perturber's period and the
nearest resonant position, quantified by the parameter $\epsilon$ in
\citet{agol:2005}. For a completely non-resonant system, $\epsilon = 1$
and the period ratio is 6.34. As we move closer to resonance, this
period ratio decreases. For the maximally non-resonant configuration,
the perturber would require a mass of $M_{\mathrm{pert}} = (0.75 \pm
0.19) \mjup$ to generate the putative TTV signal. Given that this is
approximately the same mass as the transiting planet and the period
ratio is not large, it is clear that such a signal would be evident in
the RV data. This makes the hypothesis of a non-resonant perturber
unlikely but not completely excluded due to the possibility of
out-of-the-plane perturbers (see \citet{nesvorny:2010}; \citet{veras:2010}).

If such a signal was due to an aliasing of an exomoon's orbital period,
the maximum amplitude would occur for a moon at the maximum orbital
separation given by \citet{dom06}, for a retrograde configuration. A
moon of mass $(1.8 \pm 0.4) \mearth$ could generate such a signal
and closer moons require larger masses. Such a moon should produce a
photometric dip of between 40ppm and 170ppm, depending on its
composition. Following the method outlined in \citet{kip09c}, we may
use the phasing predicted by the TTV fit to folded the individual
transit residuals for cases where maximum planet-moon separation is
predicted; epochs 1, 7 \& 11 and the time-reversed epochs 4 \& 9.
Inspection of this folded residual rules out exomoons of $R > 1.6
\rearth$ to 2-$\sigma$ confidence.

An additional method of confirming a moon would be through a TDV
analysis. The TDV periodogram does have one significant peak at $(3.6
\pm 0.2)$\,days though but with a weaker significance of $2.2$-$\sigma$
and amplitude $(24.3 \pm 7.7)$\,seconds. Using the ratio of the TTV and
TDV amplitude, \citet{kip09a} showed a solution for the exomoon period
is possible. Following this method and assuming both the TTV peak and
the TDV peak are genuinely due to a moon, we estimate an exomoon period
of $(4.2 \pm 1.7)$\,hours which would place the moon at a semi-major axis
of $(0.8 \pm 0.2)$\,planetary radii, i.e.~within the planet's volume.
Although there is a large uncertainty on this, we conclude that the
hypothesis of an exomoon inducing both the TTV and TDV putative signals
is highly improbable.

The TDV peak is quite likely to be spurious due to the proximity of the
periodogram peak to twice the Nyquist frequency. The exomoon hypothesis
can be further interrogated by evaluating the range of dynamically
stable moons around Kepler-6b. Assuming Jovian-like tidal dissipation
values of $k_{2p} = 0.5$ and $Q_P = 10^5$, the expression of
\citet{bar02} indicate that a single moon could only be stable for the
age of the system if it has a mass below $0.25 \mearth$. In
contrast, the TTV could only be produced by moons $> 1.8 \mearth$
suggesting we would require $Q_P \geq 10^6$ to make the two compatible.
In conclusion, there is no confirming evidence for the moon hypothesis
and at thus we cannot assign this explanation to the TTV peak.

The TTV signal's period of 17.3 days is similar to that what we might
expect for the star's rotational period and thus could be due to the
motion of star spots across the stellar surface, as observed by
\citet{alo08} for CoRoT-2b. Based upon the projected $\vsini$ rotation
of the star and the inferred stellar radius, we estimate the rotation
period to be $23.5_{-5.9}^{+11.7}$\,days which is compatible with the
period of the TTV signal. A bisector analysis is not possible due to
the very low cadence but will be possible once short-cadence data
becomes available. We propose that this is the most likely explanation
for the signal, based upon the current evidence.

The phasing signal has a period of 4 cycles, which is close to both
periods found in the TTV and TDV periodograms. This therefore detracts
credence to the hypothesis that these signals are genuine. Further
observations, particularly in short-cadence mode, where phasing will
not be a significant issue, will clarify the reality of these signals.

\begin{figure*}
\begin{center}
\includegraphics[width=16.8 cm]{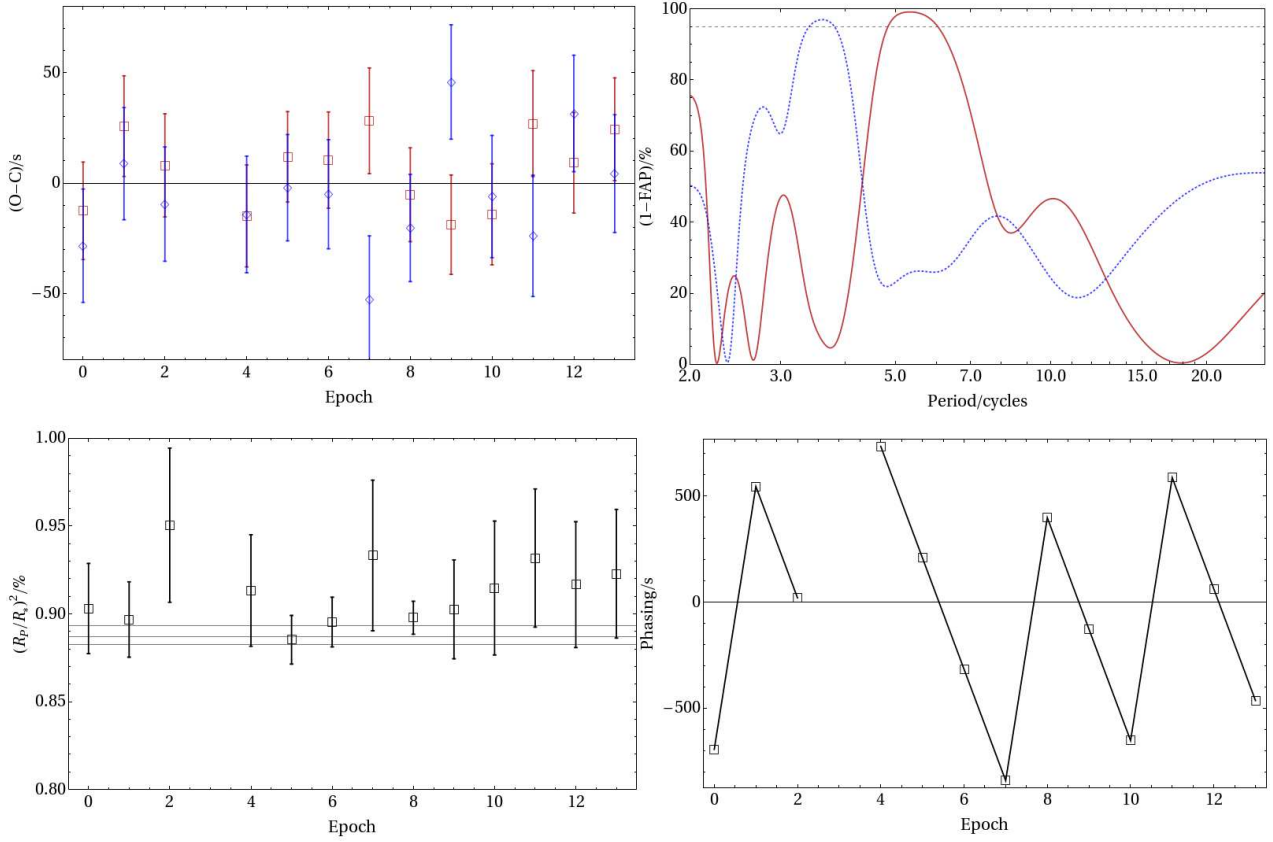}
\caption{\emph{
	{\bf Upper Left:} TTV (squares) and TDV (diamonds) for \kepvib\
	(see \S2.3.1 for details).
	{\bf Upper Right:} TTV periodogram (solid) and TDV periodogram
	(dashed) for \kepvib, calculated using a sequential F-test (see
	\S2.3.2 for description).
	{\bf Lower Left:} Transit depths from individual fits of \kepvib\
	(see \S2.3.1 for details).
	{\bf Lower Right:} ``Phasing'' of Kepler long cadence photometry
	for \kepvib\ (see \S2.3.4 for description).
}} \label{fig:kep6ttv}
\end{center}
\end{figure*}

\subsection{Depth and OOT Variations}

The transit depth is consistent with the globally fitted value yielding
a $\chi^2$ of 10.1 for 12 degrees of freedom.  Similarly, the OOT
levels are flat yielding a $\chi^2$ of 5.1 for 12 degrees of freedom.

\begin{table*}
\caption{\emph{Mid-transit times, transit durations, transit depths and out-of-transit (normalized) fluxes for Kepler-6b.}} 
\centering 
\begin{tabular}{c c c c c} 
\hline\hline 
Epoch & $t_c$/(BJD-2,454,000) & $T$/s & $(\rpl/\rstar)^2$/\% & $F_{oot}$ \\ [0.5ex] 
\hline
0 & $954.48624_{-0.00025}^{+0.00026}$ & $11758.5_{-52.1}^{+50.4}$ & $0.903_{-0.016}^{+0.035}$ & $0.999964_{-0.000015}^{+0.000015}$ \\
1 & $957.72141_{-0.00027}^{+0.00026}$ & $11832.7_{-50.6}^{+50.8}$ & $0.897_{-0.014}^{+0.029}$ & $0.999971_{-0.000015}^{+0.000015}$ \\
2 & $960.95593_{-0.00027}^{+0.00027}$ & $11796.1_{-51.1}^{+52.3}$ & $0.951_{-0.047}^{+0.041}$ & $0.999969_{-0.000015}^{+0.000015}$ \\
3 & - & - & - & - \\
4 & $967.42511_{-0.00026}^{+0.00027}$ & $11786.7_{-53.3}^{+52.2}$ & $0.913_{-0.022}^{+0.041}$ & $0.999950_{-0.000015}^{+0.000015}$ \\
5 & $970.66014_{-0.00024}^{+0.00024}$ & $11811.0_{-48.0}^{+48.2}$ & $0.885_{-0.010}^{+0.018}$ & $0.999987_{-0.000015}^{+0.000015}$ \\
6 & $973.89484_{-0.00025}^{+0.00025}$ & $11805.0_{-49.2}^{+49.6}$ & $0.895_{-0.010}^{+0.018}$ & $0.9999685_{-0.000015}^{+0.000015}$ \\
7 & $977.12977_{-0.00028}^{+0.00028}$ & $11709.6_{-59.0}^{+56.4}$ & $0.933_{-0.040}^{+0.046}$ & $0.999962_{-0.000015}^{+0.000015}$ \\
8 & $980.36411_{-0.00025}^{+0.00025}$ & $11774.5_{-48.0}^{+49.0}$ & $0.898_{-0.008}^{+0.011}$ & $0.999980_{-0.000015}^{+0.000015}$ \\
9 & $983.59867_{-0.00026}^{+0.00026}$ & $11906.7_{-51.4}^{+52.2}$ & $0.903_{-0.022}^{+0.034}$ & $0.999963_{-0.000015}^{+0.000015}$ \\
10 & $986.83345_{-0.00026}^{+0.00027}$ & $11802.9_{-56.0}^{+54.6}$ & $0.915_{-0.036}^{+0.041}$ & $0.999968_{-0.000015}^{+0.000015}$ \\
11 & $990.06865_{-0.00028}^{+0.00028}$ & $11767.4_{-55.2}^{+54.6}$ & $0.932_{-0.040}^{+0.039}$ & $0.999967_{-0.000015}^{+0.000015}$ \\
12 & $993.30317_{-0.00026}^{+0.00026}$ & $11878.1_{-52.0}^{+53.4}$ & $0.917_{-0.033}^{+0.039}$ & $0.999977_{-0.000015}^{+0.000015}$ \\
13 & $996.53807_{-0.00027}^{+0.00027}$ & $11823.7_{-52.8}^{+53.8}$ & $0.923_{-0.037}^{+0.036}$ & $0.999954_{-0.000015}^{+0.000015}$ \\ [1ex]
\hline
\end{tabular}
\label{tab:kep6ttv} 
\end{table*}

%% file: kep7.tex

\subsection{Global fits}
\subsubsection{Comparison to the discovery paper}

\begin{figure}[!ht]
\plotone{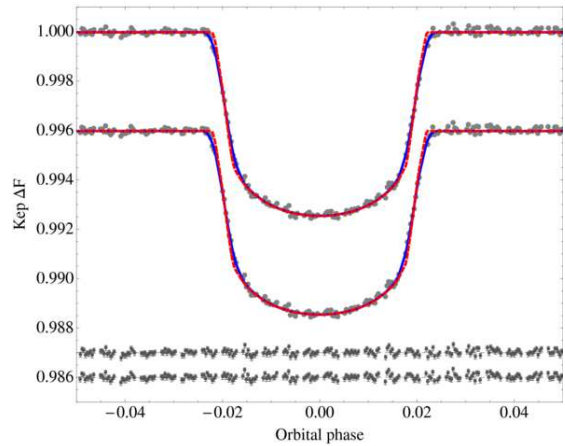}
\caption{
    {\bf Top:} Phase-folded primary transit \lc\ of \kepvii\ using
    method B. The upper curve shows the circular fit, the bottom curve
    the eccentric fit.  Solid (blue) lines indicate the best fit resampled
    model (with bin-number 4). The dashed (red) lines show the
    corresponding unbinned model, which one would get if the transit
    was observed with infinitely fine time resolution.
	Residuals from the best fit resampled model for the
	circular and eccentric solutions are shown below in
        respective order.
\label{fig:kep7prim}}
\end{figure}

\begin{figure}[ht]
\plotone{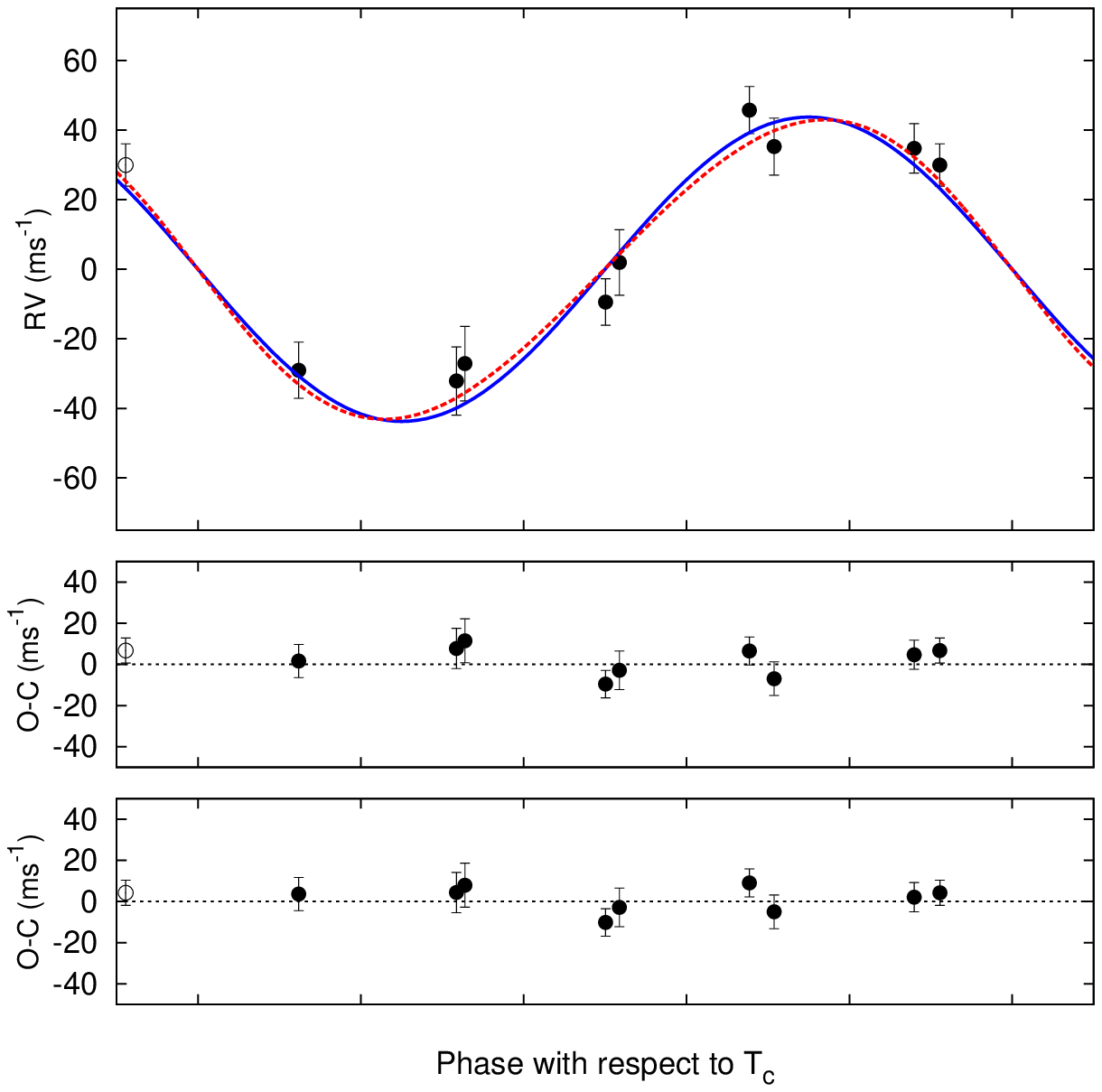}
\caption{
    {\bf Top:} RV measurements from Keck for \kepvii{}, along with an
    orbital fit, shown as a function of orbital phase, using our
    best-fit period. Solid (blue) line shows the circular orbital fit
    with binned RV model (3 bins, separated by 600\,seconds). The
    dashed (red) line shows the eccentric orbital fit (with the same
    bin parameters). Zero phase is defined by the transit midpoint. The
    center-of-mass velocity has been subtracted.  Note that the error
    bars include the stellar jitter (taken for the circular solution),
    added in quadrature to the formal errors given by the spectrum
    reduction pipeline. Fits from method B.
    {\bf Middle:} phased residuals after subtracting the orbital fit
    for the circular solution. The 
    r.m.s.~variation of the residuals is \kepviicirRVfitrms\,\ms, and the
    stellar jitter that was added in quadrature to the formal RV errors
    is $\kepviicirRVjitter$\,\ms.
	{\bf Bottom:} phased residuals after subtracting the orbital fit
    for the eccentric solution.  Here the r.m.s.~variation of the
    residuals is \kepviieccRVfitrms\,\ms, and the stellar jitter is
    $\kepviieccRVjitter$\,\ms.
\label{fig:kep7rv}}
\end{figure}

\kepviib\ was discovered by \citet{lat10}. The planet is noteworthy for
its very low density and thus significant inflation. Our global fits find highly
consistent results with the original discovery paper. We note that the
\citet{lat10} values are the most consistent with our own values out of
any of discovery papers discussed in this work. The fitted models on
the phase-folded \lcs\ are shown on \reffigl{kep7prim} for method B. Correlated
noise was checked for in the residuals using the Durbin-Watson statistic, which 
finds $d=1.866$, at the border of the 1\% critical level of $1.868$, suggesting 
correlated noise is marginal. The orbital fits to the RV points are shown in 
\reffig{kep7rv}, depicting both the circular and eccentric fits using method B. 
Unusually, both methods A and B found stellar jitter levels consistent with 
zero, which suggests the publicly available RV measurements have perhaps 
overestimated errors. The final planetary parameters are summarized at the 
bottom of Table~\ref{tab:kep7tab}.

\subsubsection{Eccentricity}
The global circular fit of method A yields a $\chi^2 = 6.87$ and the
eccentric fit obtains $\chi^2 = 4.89$ for the RVs. An F-test finds that 
the eccentric fit has a significance of 79.4\%. Therefore, we conclude 
that the eccentric fit does not meet our detection criteria and we 
constrain $e<0.31$ to 95\% confidence.

One obvious question at this point is whether the range of allowed
eccentricities could explain the bloated radius of \kepviib\ through
tidal heating. The $Q_P$ value of the planet is an element of large
uncertainity here but we can constrain $Q$'s minimum limit using the same
method we adopted for \kepivb. We find that 
$Q_P \geq 9.4_{-4.6}^{+14.0} \times 10^6$. One may argue that this is only
valid under the assumption that the eccentric model is selected over the
circular one, but we would counter that in the circular model case the
tidal heating is zero anyway and thus the hypothesis of tidal heating
being the origin of the bloated radius is invalid.

Using the posterior distribution of the minimum allowed $Q_P$ value, we
can translate this to a maximum allowed tidal energy dissipation using
the expressions of \citet{pea78}:

\begin{equation}
\dot{E}_{\mathrm{tidal}} \geq \frac{21}{2} \frac{k_{2}}{Q_{P,\mathrm{min}}} \frac{G M_*^2 R_P^5 n e^2}{a^6}
\end{equation}

We divide this value by the amount of power hitting the planetary surface,
found by $L_* R_P^2/(4 a^2)$ to give us the posterior of the fraction
of maximum allowed tidal heating relative to the incoming radiation and
find $\dot{E}_{\mathrm{tidal}}/\dot{E}_{\mathrm{stellar}} = 1.5_{-1.0}^{+2.5} \times 10^{-12}$.
We therefore conclude that tidal heating plays a negligible role in the
heating of \kepviib\ and the planet's inflated radius.

\subsubsection{Secondary eclipse}

\begin{figure}[!ht]
\plotone{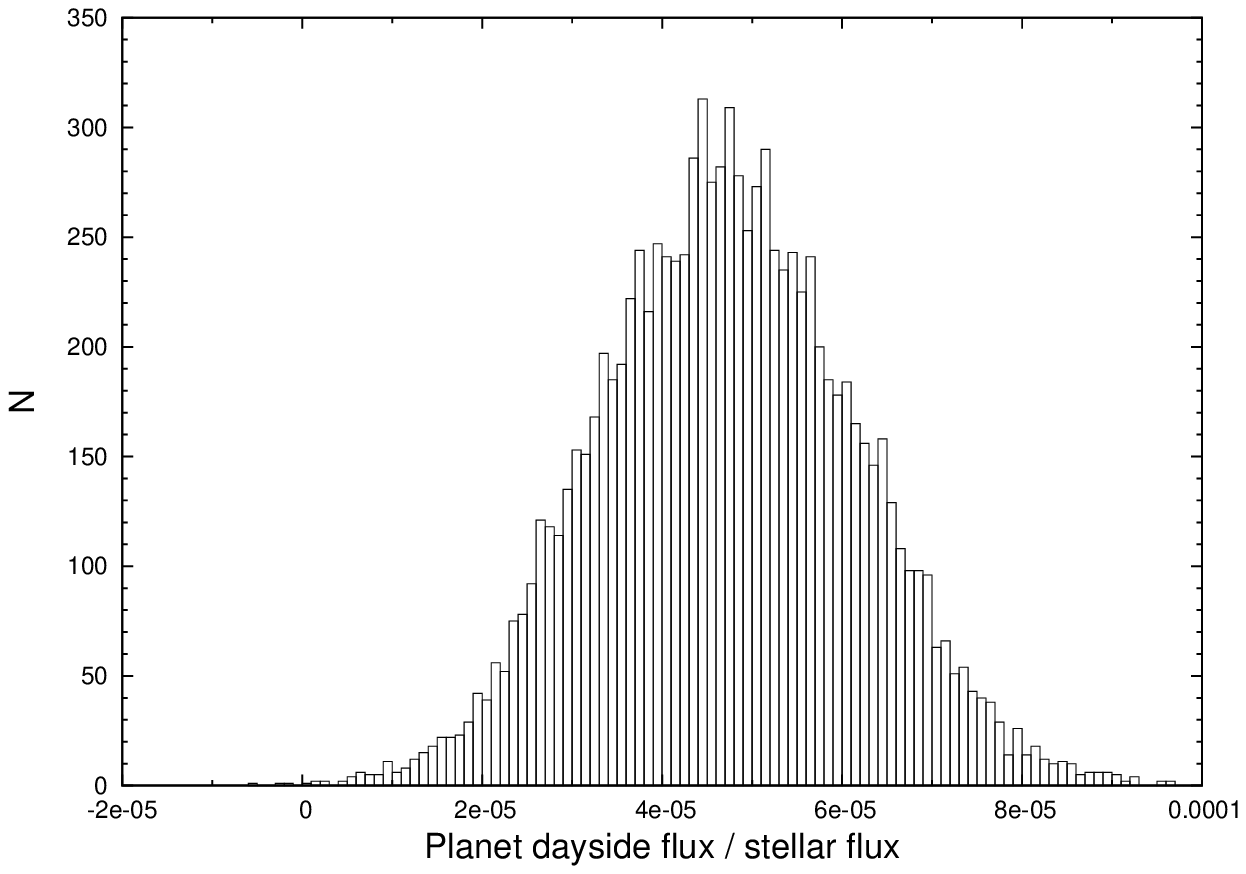}
\caption{
    Distribution of \fpdfs\ for for \kepvii\ from the global modeling
	of method B.  \fpdfs\ is primarily constrained by the depth of the
	secondary eclipse. Only results for the circular orbital solution
	are shown.
\label{fig:kep7sampid}}
\end{figure}

As before, method B is the better tool to constrain secondaries due to its
rescaling method. Method B detects a secondary eclipse in the circular
fits of depth $(47 \pm 14)$\,ppm and significance 3.5-$\sigma$ (99.96\%).
We note that \citet{lat10} reported a 2.4-$\sigma$ weakly detected
eclipse but did not report a depth estimate. The MCMC distribution of
the \fpdfs\ values for method B is shown in \reffigl{kep7sampid}.

Inspection of the method A and \citet{lat10} values relative to the method
B result reveals a clear difference and we must
consider why we obtain such different numbers. All of the fits
are circular and thus the orbital solutions are nearly identical with
the eclipse occuring at the same phase in each case. The only
difference between them is the definition of the out-of-eclipse
baseline. Method A assumes the same baseline level for both primary and
secondary transit whereas method B uses a simplified phase curve model
as described in \S2.2. If the secondary eclipse local baseline is
different from the primary transit baseline, method A would move
towards an average solution and actually attenuate the eclipse depth. The
same attenuation also occurs in the \citet{lat10} which utilizes the same
technique as method A (D. Latham, personal communication).

To investigate this futher, we may compare the three baseline levels found
by method B: $F_{oot} = (0.999976 \pm 0.000007)$, $F_{oom} = (0.999972 
\pm 0.000003)$ and $F_{oos} = (0.999993 \pm 0.000007)$. The first two 
values are compatible with a constant but the baseline surrounding the 
secondary eclipse is clearly larger. If the secondary eclipse baseline 
is larger, method A would favor a lower average baseline due to $F_{oot}$ 
and $F_{oom}$ and thus find a lower secondary eclipse. This seems to be the 
explanation with method A finding $F_{oot} = (0.9999764 \pm 0.0000024)$. 
We believe this explains the discrepancy between the two methods.

Using the method B results only, the difference in baselines indicates
a day-night contrast. Using our MCMC trials, we estimate a nightside flux
$(F_{\mathrm{n}}/F_\star) = 29_{-14}^{+15}$\,ppm which is of 2.4-$\sigma$
significance. We can also calculate the day-night difference explicitly
and we find $[(F_{P,\mathrm{d}}/F_\star) - (F_{P,\mathrm{n}}/F_\star)] =
(17 \pm 9)$\,ppm. Since nightside flux cannot be due to reflected light,
this supports the hypothesis of thermal emission as a source for both
the secondary eclipse and the phase curve. 

There is a very strong caveat to bear in mind for the day-night
contrast. The data we have analyzed has been processed by the Kepler
pipeline which involves numerous data manipulation processes. Time
trends of very low amplitude and of time periods of days may be
affected by this processing, with the most probable behaviour being
that of a long-cut filter. As a result, a strong-day night contrast
could be attenuated by the act of passing through this filter. It is
therefore still possible that the day-night contrast is 100\%
(i.e.~equal to the secondary eclipse depth) and thus the light we are
seeing is reflected light, rather than thermal emission, but the phase
curve is artificially distorted by the pipeline processing.

If the secondary eclipse was due to thermal emission, we may estimate
the brightness temperature by integrating the Planck function of the
planet over the Kepler response function, divided by that of the star.
We repeat this for planetary brightness temperatures from 2000K to
3000K in 1K steps and calculate $(F_{P,\mathrm{d}}/F_\star)$ in each
case. We find that a best-fit brightness temperature of
$2570_{-85}^{+108}$\,K, which is much larger than the equilibrium
temperature of $(1554 \pm 32)$\,K. Even the nightside flux of
$(F_{P,\mathrm{n}}/F_\star) = 29_{-14}^{+15}$\,ppm would require a
temperature for the nightside of $2441_{-179}^{+127}$\,K. These very
large temperatures do not seem consistent with the energy budget of the
planet, particularly as tidal heating has already been shown to offer a
negligible contribution to the energy budget of this planet.
From stellar irradiation only, the redistribution factor would have to
be 7.48 to re-create the dayside eclipse, which is greater than the
maximum physically permitted value of 8/3. The only remaining
possibility is internal heating, for which we find a heat source
responsible for 2440\,K could explain both the dayside and nightside
fluxes. This would require an internal heat source of 
$\mathrm{log}(L/\lsun) = -3.19$, which is equivalent to the luminosity of an
M6-M7 star. For example, LHS 292 (M6.5) and VB8 (M7) have luminosities of 
$\mathrm{log}(L/\lsun) = -3.19$ and $\mathrm{log}(L/\lsun) = -3.24$ respectively.

Whilst other sources of heating are possible apart from tidal, for example
radioactivity, the magnitude required detracts credence from the 
hypothesis of thermal emission as the principal origin for the eclipse 
depth.

If due to reflection, method B finds a geometric albedo of $A_g = (0.38
\pm 0.12)$, which is somewhat larger than the value found for \kepvb\
of $A_g = (0.15 \pm 0.10)$, but not greatly different from the albedos of
Jupiter and Saturn (0.52 and 0.47 respectively) and consistent with the
study of \citet{cow10}. We believe this latter hypothesis,
as with \kepvb, is a more plausible scenario for the secondary
eclipse measurement. As a consequence, we should not expect any
nightside flux, which bears some questions on the apparent non-zero
nightside flux detected in this study. There are two possible
explanations which are consistent with reflection: i) the nightside
flux is not statistically significant at 2.4-$\sigma$ ii) the phase
curve has been attenuated due to the Kepler pipeline producing effects
which mimic a long-cut filter iii) $F_{oos}$ is increased away from zero
nightside flux due to the phase curve of the planet i.e.~reflected
light from the viewable crescent. In reality, a combination of these
seems probable. Once more transits become available for analysis, a
more sophisticated phase curve model can be fitted to the data set to
resolve this question.

\begin{figure}[!ht]
\plotone{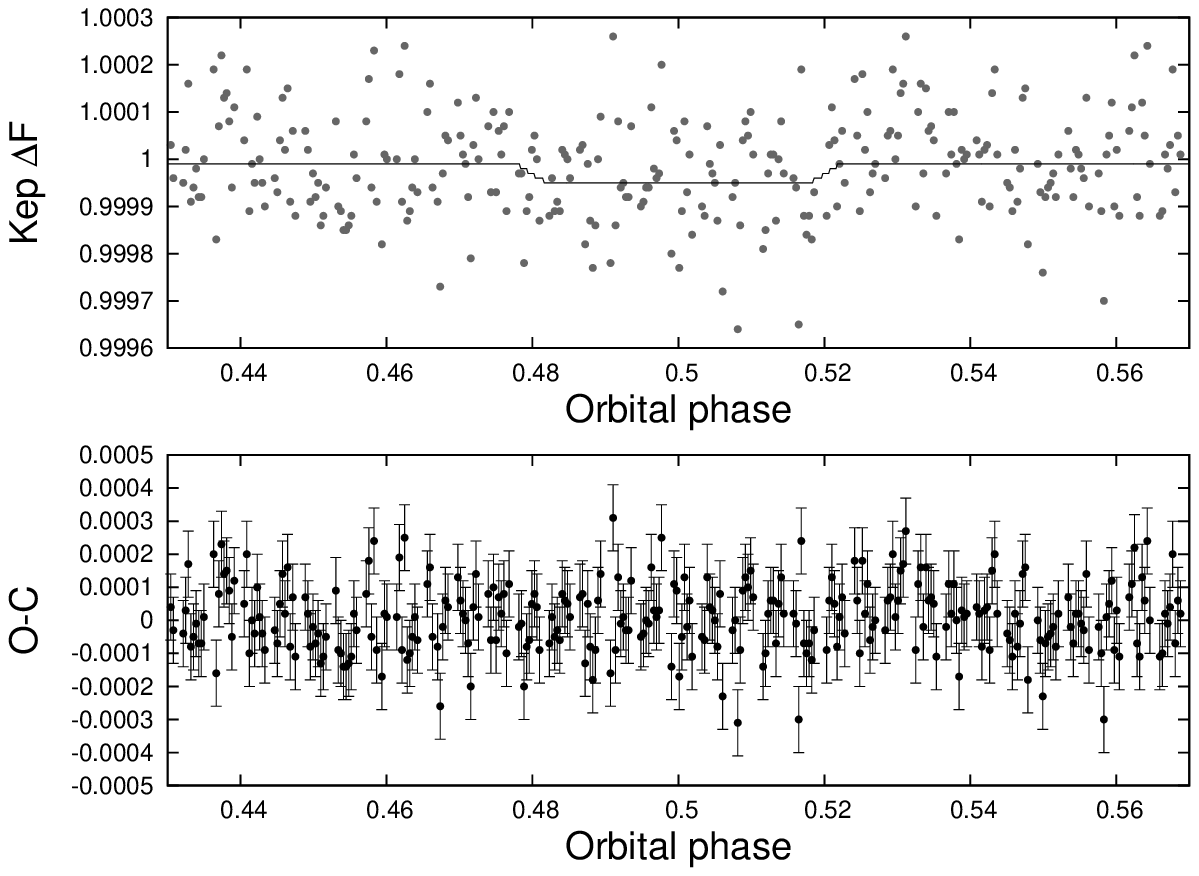}
\caption{
    Phase-folded secondary transit \lc\ of \kepvii\ using method B
    (top), and residuals from the best fit (bottom). Only the fit for
    the circular orbital solution is shown.
\label{fig:kep7sec}}
\end{figure}

\subsubsection{Properties of the parent star \kepvii}
\label{sec:stelparam}

\begin{figure}[!ht]
\plotone{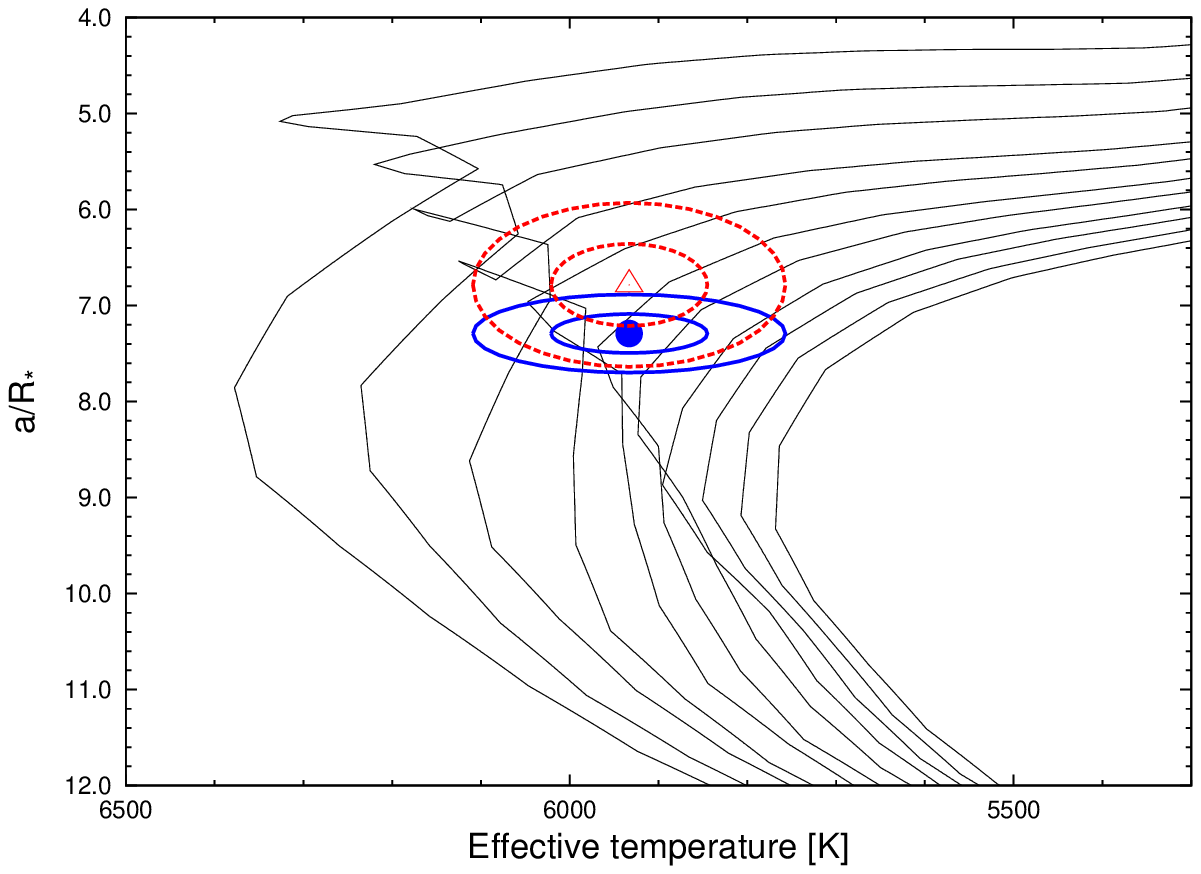}
\caption{
    Stellar isochrones from \citet{yi:2001} for metallicity
    \feh=\kepviicirSMEiizfehshort\ and ages 
	1.0, 1.4, 1.8, 2.2, 2.6, 3.0, 3.4 and 3.8\,Gyrs.
	The final choice of $\teffstar$ and \arstar\ for the circular
    solution are marked by a filled circle, and is encircled by the
    1$\sigma$ and 2$\sigma$ confidence ellipsoids (solid, blue lines). 
    Corresponding values and confidence ellipsoids for the eccentric
    solution are shown with a triangle and dashed (red) lines. Fits
    from method B.
\label{fig:kep7iso}}
\end{figure}

The Yonsei-Yale model isochrones from \citet{yi:2001} for metallicity
\feh=\kepviicirSMEzfehshort\ are plotted in \reffigl{kep7iso}, with the
final choice of effective temperature $\teffstar$ and \arstar\ marked,
and encircled by the 1$\sigma$ and 2$\sigma$ confidence ellipsoids,
both for the circular and the eccentric orbital solution. Here, the MCMC
distribution of \arstar\ comes from the global modeling of the data.

\begin{table*}
\caption{
	\emph{Global fits for Kepler-7b.  Quoted values are medians of
	MCMC trials with errors given by 1-$\sigma$ quantiles. Two independent
	methods are used to fit the data (A\&B) with both circular and eccentric 
	modes (c\&e).* = fixed parameter; $\dagger$ = parameter was floated but 
	not fitted.}
}
\centering 
\begin{tabular}{c c c c c c} 
\hline\hline 
Parameter & Discovery & Method A.c & Method A.e & Model B.c & Model B.e \\ [0.5ex] 
\hline
\emph{Fitted params.} & & \\
\hline 
$P/$days & $4.885525 \pm 0.000040$ & $4.88552_{-0.00010}^{+0.00010}$ & $4.885522_{-0.000061}^{+ 0.000062}$ & $\kepviicirLCP$ & $\kepviieccLCP$ \\
$E$ (BJD-2,454,000) & $967.27571 \pm 0.00014$ & $957.50468_{-0.00049}^{+0.00049}$ & $957.50481_{-0.00035}^{+0.00034}$ & $957.50461 \pm 0.00020$ & $957.50458 \pm 0.00021$ \\
$T_{1,4}$/s & - & $18694_{-216}^{+224}$ & $18704_{-150}^{+149}$ & $18541 \pm 95$ & $18516 \pm 95$ \\
$T$/s & - & $16951_{-237}^{+242}$ & $16916_{-156}^{+203}$ & $16935 \pm 22$ & $16917 \pm 31$ \\ 
$(T_{1,2} \simeq T_{3,4})$/s & - & $1722_{-297}^{+376}$ & $1812_{-299}^{+245}$ & $1624 \pm 104$ & $1616 \pm 104$ \\
$(\rpl/\rstar)^2$/\% & - & $0.661_{-0.034}^{+0.038}$ & $0.671_{-0.035}^{+0.024}$ & $0.6529 \pm 0.0081$ & $0.6529 \pm 0.0081$ \\
$b^2$ & - & $0.19_{-0.15}^{+0.13}$ & $0.230_{-0.137}^{+0.083}$ & $\kepviicirLCbsq$ & $\kepviieccLCbsq$ \\
$(\zeta/\rstar)$/days$^{-1}$ & - & $10.24_{-0.15}^{+0.16}$ & $10.28_{-0.14}^{+0.12}$ & $\kepviicirLCzeta$ & $\kepviieccLCzeta$ \\
$(F_{P,\mathrm{d}}/F_\star)$/ppm & - & $24_{-30}^{+30}$ & $-21_{-26}^{+32}$ & $47 \pm 14$ & $41 \pm 37$ \\
$e\sin \omega$ & 0$^{*}$ & 0$^{*}$ & $0.00_{-0.13}^{+0.13}$ & 0$^{*}$ & $\kepviieccRVh$ \\
$e\cos \omega$ & 0$^{*}$ & 0$^{*}$ & $0.047_{-0.043}^{+0.021}$ & 0$^{*}$ & $\kepviieccRVk$ \\
$K$/ms$^{-1}$ & $42.9 \pm 3.5$ & $42.8_{-10.6}^{+10.6}$ & $41.6_{-6.2}^{+6.3}$ & $\kepviicirRVK$ & $\kepviieccRVK$ \\
$\gamma$/ms$^{-1}$ & 0 & $-0.1_{-7.3}^{+7.3}$ & $0.7_{-4.3}^{+4.5}$ & $\kepviicirRVgammarel$ & $\kepviieccRVgammarel$ \\
$\dot{\gamma}$/ms$^{-1}$day$^{-1}$ & 0$^{*}$ & 0$^{*}$ & 0$^{*}$ & 0$^{*}$ & 0$^{*}$ \\
$t_{\mathrm{troj}}$/days & 0$^{*}$ & 0$^{*}$ & 0$^{*}$ & 0$^{*}$ & 0$^{*}$ \\
$B$ & $1.025 \pm 0.004$ & $1.025 \pm 0.004$ $\dagger$ & $1.025 \pm 0.004$ $\dagger$ & $1.025 \pm 0.004$ $\dagger$ & $1.025 \pm 0.004$ $\dagger$ \\
\hline
\emph{SME derived params.} & & \\
\hline
$T_{\mathrm{eff}}$/K & $5933 \pm 44$ & \kepviicirSMEteff & \kepviicirSMEteff & \kepviicirSMEteff & \kepviicirSMEteff \\
log$(g/\mathrm{cgs})$ & $3.98 \pm 0.10$ & $3.98 \pm 0.10$ & $3.98 \pm 0.10$ & $3.98 \pm 0.10$ & $3.98 \pm 0.10$ \\
$\feh$ (dex) & \kepviicirSMEzfeh & \kepviicirSMEzfeh & \kepviicirSMEzfeh & \kepviicirSMEzfeh & \kepviicirSMEzfeh \\
$\vsini$ (\kms) & \kepviicirSMEvsin & \kepviicirSMEvsin & \kepviicirSMEvsin & \kepviicirSMEvsin & \kepviicirSMEvsin \\
$u_1$ & - & $0.34_{-0.13}^{+0.16}$ & $0.35_{-0.10}^{+0.12}$ & \kepviicirLBikep$^{*}$ & \kepviieccLBikep$^{*}$ \\
$u_2$ & - & $0.33_{-0.34}^{+0.26}$ & $0.27_{-0.23}^{+0.24}$ & \kepviicirLBiikep$^{*}$ & \kepviieccLBiikep$^{*}$ \\
\hline
\emph{Model indep.~params.} & & \\
\hline
$\rpl/\rstar$ & $0.08241_{-0.00043}^{+0.00030}$ & $0.0813_{-0.0021}^{+0.0023}$ & $0.0819_{-0.0021}^{+0.0015}$ & $\kepviicirLCrprstar$ & $\kepviieccLCrprstar$ \\
$\arstar$ & $7.22_{-0.13}^{+0.16}$ & $7.14_{-0.53}^{+0.56}$ & $7.03_{-0.95}^{+1.10}$ & $\kepviicirPPar$ & $\kepviieccPPar$ \\
$b$ & $0.445_{-0.044}^{+0.032}$ & $0.44_{-0.23}^{+0.13}$ & $0.480_{-0.175}^{+0.080}$ & $\kepviicirLCimp$ & $\kepviieccLCimp$ \\
$i$/$^{\circ}$ & $86.5 \pm 0.4$ & $86.5_{-1.4}^{+2.0}$ & $86.3_{-1.9}^{+1.7}$ & $\kepviicirPPi$ & $\kepviieccPPi$ \\
$e$ & 0$^{*}$ & 0$^{*}$ & $0.102_{-0.047}^{+0.104}$ & 0$^{*}$ & $\kepviieccRVeccen$ \\
$\omega$/$^{\circ}$ & - & - & $129_{-80}^{+182}$ & - & $\kepviieccRVomega$ \\
RV jitter (\ms) & - & 0.0 & 0.0 & \kepviicirRVjitter & \kepviieccRVjitter \\
$\rho_\star$/gcm$^{-3}$ & $0.304$ & $0.288_{-0.060}^{+0.073}$ & $0.276_{-0.097}^{+0.151}$ & \kepviicirISOrho & \kepviieccISOrho \\
log$(g_P/\mathrm{cgs})$ & $2.691_{-0.045}^{+0.038}$ & $2.69_{-0.15}^{+0.13}$ & $2.66_{-0.16}^{+0.14}$ & $\kepviicirPPlogg$ & $\kepviieccPPlogg$ \\
$S_{1,4}$/s & - & $18694_{-216}^{+224}$ & $15406_{-1023}^{+5065}$ & $18541 \pm 95$ & $15535 \pm 2350$ \\
$t_{sec}$/(BJD-2,454,000) & - & $964.83265_{-0.00049}^{+0.00049}$ & $964.662_{-0.158}^{+0.074}$ & $975.8300 \pm 0.0002$ & $994.137 \pm 0.058$ \\
\hline
\emph{Derived stellar params.} & & \\
\hline
$\mstar/\msun$ & $1.347_{-0.054}^{+0.072}$ & $1.257_{-0.073}^{+0.087}$ & $1.268_{-0.090}^{+0.115}$ & \kepviicirISOmlong & \kepviieccISOmlong \\ 
$\rstar/\rsun$ & $1.843_{-0.066}^{+0.048}$ & $1.83_{-0.14}^{+0.17}$ & $1.86_{-0.28}^{+0.34}$ & \kepviicirISOrlong & \kepviieccISOrlong \\
log$(g/\mathrm{cgs})$ & $4.030_{-0.019}^{+0.018}$ & $4.012_{-0.063}^{+0.061}$ & $4.00_{-0.12}^{+0.12}$ & \kepviicirISOlogg & \kepviieccISOlogg \\ 
$\lstar/\lsun$ & $4.15_{-0.54}^{+0.63}$ & $3.72_{-0.60}^{+0.78}$ & $3.8_{-1.1}^{+1.6}$ & \kepviicirISOlum & \kepviieccISOlum \\
$M_{V}$/mag & - & $3.38_{-0.21}^{+0.20}$ & $3.34_{-0.39}^{+0.37}$ & \kepviicirISOmv & \kepviieccISOmv \\
Age/Gyr & $3.5 \pm 1.0$ & $4.82_{-1.20}^{+0.82}$ & $4.45_{-0.96}^{+1.11}$ & \kepviicirISOage & \kepviieccISOage \\
Distance/pc & - & $772_{-67}^{+79}$ & $784_{-122}^{+153}$ & \kepviicirXdist & \kepviieccXdist \\
\hline
\emph{Derived planetary params.} & & \\
\hline
$\mpl/\mjup$ & $0.433_{-0.041}^{+0.040}$ & $0.42_{-0.11}^{+0.11}$ & $0.405_{-0.064}^{+0.067}$ & $\kepviicirPPmlong$ & $\kepviieccPPmlong$ \\ 
$\rpl/\rjup$ & $1.478_{-0.051}^{+0.050}$ & $1.45_{-0.15}^{+0.18}$ & $1.48_{-0.24}^{+0.29}$ & $\kepviicirPPrlong$ & $\kepviieccPPrlong$ \\ 
$\rho_P$/gcm$^{-3}$ & $0.166_{-0.020}^{+0.019}$ & $0.167_{-0.056}^{+0.074}$ & $0.155_{-0.064}^{+0.099}$ & $\kepviicirPPrho$ & $\kepviieccPPrho$ \\ 
$a$/AU & $0.06224_{-0.00084}^{+0.00109}$ & $0.0608_{-0.0012}^{+0.0014}$ & $0.0609_{-0.0015}^{+0.0018}$ & $\kepviicirPParel$ & $\kepviieccPParel$ \\ 
$T_{\mathrm{eq}}/K$ & $1540 \pm 200$ & $1569_{-58}^{+65}$ & $1582_{-110}^{+127}$ & $\kepviicirPPteff$ & $\kepviieccPPteff$ \\ [1ex]
\hline\hline 
\end{tabular}
\label{tab:kep7tab} 
\end{table*}

\subsection{Transit timing analysis}
\subsubsection{Analysis of variance}

We find TTV residuals of 32.1 seconds and TDV residuals of 22.2
seconds. The TTV indicates timings consistent with a linear ephemeris,
producing a $\chi^2$ of 9.1 for 7 degrees of freedom (75.1\%
significance of excess).  The TDV is also consistent with a
non-variable system exhibiting a $\chi^2$ of 4.1 for 8 degrees of
freedom (84.9\% significance of being too low).

For the TTV, we note that for 7 degrees of freedom, excess scatter
producing a $\chi^2$ of 21.8 would be detected to 3-$\sigma$
confidence. This therefore excludes a short-period signal of
r.m.s.~amplitude $\geq 38.0$\,seconds. Similarly, for the TDV, we exclude
scatter producing a $\chi^2$ of 23.6, or a short-period
r.m.s.~amplitude of $\geq 42.6$\,seconds, to 3-$\sigma$ confidence. This
consitutes a 0.50\% variation in the transit duration.

These limits place constraints on the presence of perturbing planets, 
moons and Trojans. For the case a 2:1 mean-motion resonance perturbing planet,
the libration period would be $\sim 111$\,cycles and thus we do not
possess sufficient baseline to look for such perturbers yet. After a
further 102 cycles have been obtained at the same photometric quality,
the presence of a 2:1 resonant perturber of mass $0.06 \mearth$
may be constrained.

The current TTV data rules out an exomoon at the maximum orbital radius
(for retrograde revolution) of $\geq 2.5 \mearth$.  The current TDV data
rules an exomoon of $\geq 5.9 \mearth$ at a minimum orbital
distance.

Using the expressions of \citet{for07} and assuming $\mpl \gg 
M_{\mathrm{Trojan}}$, the expected libration period of \kepviib\
induced by Trojans at L4/L5 is 21.4 cycles and therefore we do
not yet possess sufficient baseline to search for these TTVs.
Inspection of the folded photometry at $\pm P/6$ 
from the transit centre excludes Trojans of 
effective radius $\geq 1.11 R_{\oplus}$ to 3-$\sigma$ confidence.

What is peculiar about this system is that the TTV produces a slight
excess variance whilst the TDV does the opposite. The error on the TDV
data points, and thus the scatter, should be exactly the same as the
TTV points, or in fact slightly larger due to limb darkening effects.
Thus, the observation of the TTV scatter being much larger than the TDV
scatter is curious. We note for all other planets we studied we found
the TDV scatter is greater than or equal to the TTV scatter. Also, for
several planets we find the observed scatter is much lower than the
uncertainties predict, suggesting we have overestimated our errors. If
this is true for the \kepviib\ system, the TDV scatter indicates that
the measurement uncertainties should be scaled by a factor of 0.71.
This factor should be common to the TTV, which would change the
observed $\chi^2$ from 9.1 to 17.7, for 7 degrees of freedom. The
scaled $\chi^2$ would have a significance of 98.7\%, or 2.5-$\sigma$,
which is still below our formal detection threshold. Nevertheless, we
identify \kepviib\ as an important target for continued transit timing
analysis.

\subsubsection{Periodograms}

The TTV F-test periodogram presents one significant peak very close to
Nyquist frequency, which we may discard. The TDV periodogram presents
no significant peaks.

\begin{figure*}
\begin{center}
\includegraphics[width=16.8 cm]{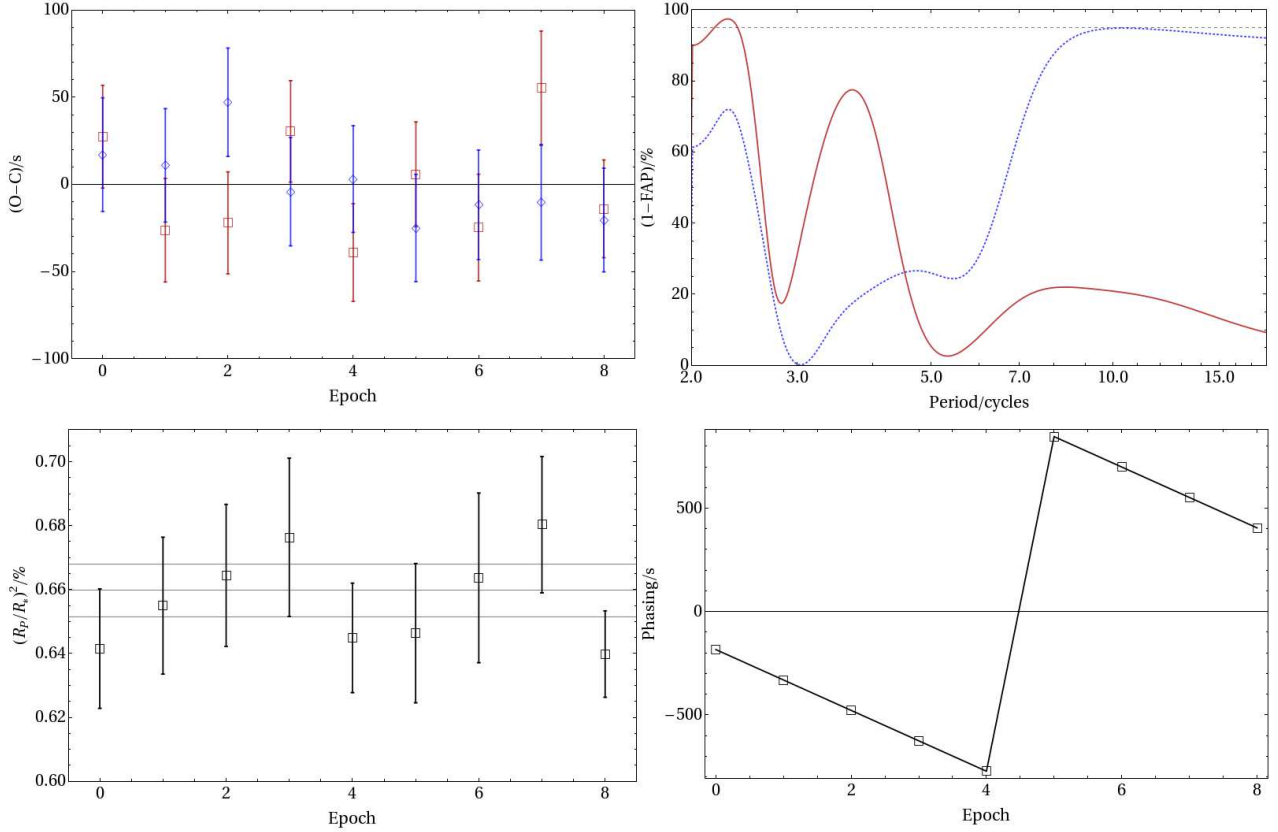}
\caption{\emph{
	{\bf Upper Left:} TTV (squares) and TDV (diamonds) for \kepviib\
	(see \S2.3.1 for details).
	{\bf Upper Right:} TTV periodogram (solid) and TDV periodogram
	(dashed) for \kepviib, calculated using a sequential F-test (see
	\S2.3.2 for description).
	{\bf Lower Left:} Transit depths from individual fits of \kepviib\
	(see \S2.3.1 for details).
	{\bf Lower Right:} ``Phasing'' of Kepler long cadence photometry
	for \kepviib\ (see \S2.3.4 for description).
}} \label{fig:kep7ttv}
\end{center}
\end{figure*}

\subsection{Depth and OOT Variations}

The transit depth is consistent with the globally fitted value yielding
a $\chi^2$ of 5.8 for 8 degrees of freedom.  Similarly, the OOT levels
are flat yielding a $\chi^2$ of 3.4 for 8 degrees of freedom.

\begin{table*}
\caption{
	\emph{Mid-transit times, transit durations, transit depths and
	out-of-transit (normalized) fluxes for Kepler-7b.}
} 
\centering 
\begin{tabular}{c c c c c} 
\hline\hline 
Epoch & $t_c$/(BJD-2,454,000) & $T$/s & $(\rpl/\rstar)^2/\%$ & $F_{oot}$ \\ [0.5ex] 
\hline
0 & $957.50508_{-0.00034}^{+0.00034}$ & $17006.2_{-65.6}^{+64.8}$ & $0.641_{-0.014}^{+0.023}$ & $0.999970_{-0.000010}^{+0.000010}$ \\
1 & $962.39000_{-0.00034}^{+0.00035}$ & $16993.8_{-65.1}^{+65.0}$ & $0.655_{-0.020}^{+0.023}$ & $0.999967_{-0.000011}^{+0.000011}$ \\
2 & $967.27558_{-0.00033}^{+0.00034}$ & $17066.2_{-62.3}^{+62.0}$ & $0.664_{-0.023}^{+0.022}$ & $0.999975_{-0.000010}^{+0.000010}$  \\
3 & $972.16172_{-0.00033}^{+0.00034}$ & $16963.6_{-61.9}^{+62.3}$ & $0.676_{-0.026}^{+0.024}$ & $0.999974_{-0.000010}^{+0.000010}$ \\
4 & $977.04645_{-0.00032}^{+0.00032}$ & $16978.2_{-60.7}^{+61.7}$ & $0.645_{-0.013}^{+0.021}$ & $0.999973_{-0.000010}^{+0.000010}$ \\
5 & $981.93250_{-0.00034}^{+0.00035}$ & $16922.0_{-61.2}^{+61.9}$ & $0.646_{-0.020}^{+0.024}$ & $0.999976_{-0.000011}^{+0.000011}$ \\
6 & $986.81768_{-0.00035}^{+0.00035}$ & $16948.6_{-62.2}^{+63.3}$ & $0.664_{-0.028}^{+0.025}$ & $0.999980_{-0.000010}^{+0.000010}$ \\
7 & $991.70414_{-0.00038}^{+0.00038}$ & $16951.3_{-65.7}^{+66.6}$ & $0.680_{-0.023}^{+0.020}$ & $0.999976_{-0.000010}^{+0.000010}$ \\
8 & $996.58887_{-0.00032}^{+0.00033}$ & $16931.2_{-59.7}^{+59.3}$ & $0.640_{-0.009}^{+0.018}$ & $0.999957_{-0.000010}^{+0.000010}$ \\ [1ex]
\hline
\end{tabular}
\label{tab:kep7ttv} 
\end{table*}

%% file: kep8.tex

\subsection{Global Fits}
\subsubsection{Comparison to the discovery paper}

\kepviiib\ was discovered by \citet{jen10a}. \kepviiib\ posed several
unique problems not encountered with the other Kepler planets.  The
data format of the public data was quite different from the other sets. 
Typical values of the photometric data was $\sim0$ whereas the other
data sets where $\sim1$.  This suggests that either the photometry was
raw differential magnitudes or normalized fluxes with unity subtracted. 
We tried both assumptions and that the normalized fluxes minus unity
assumption led to system parameters more consistent with the original
discovery paper.  This assumption has been since confirmed (personal
communication with J. Jenkins).

Another peculiarity of the data is that the r.m.s.~scatter is
not constant but actually increases significantly with time, as
discussed in \S3.2. 
The source of this varying scatter is an instrumental effect,
caused by oscillations in the photometry as a result of a heater on one
of the two reaction wheel assemblies that modulates the focus by 
$\sim1$\,$\mu$m (see \citet{jen10c} for more details). Since the
discovery of this effect in the Q1 photometry, the Kepler Team
have subsequently trimmed the bounding temperatures for this heater 
to reduce the amplitude and heater cycle (J. Jenkins, personal
communication). Therefore, future data sets
will not suffer from this effect as accutely as the photometry
analyzed here. Concordantly, the results obtained in our analysis
may not be as reliable as for the other Kepler planets.

In our analysis, we disregard radial velocity points occuring 
during the transit to avoid the Rossiter-McLaughlin effect.
This is justified since our prinicpal goal here is to produce
refined system parameters rather measure the spin-orbit
alignment.

\begin{figure}[ht]
\plotone{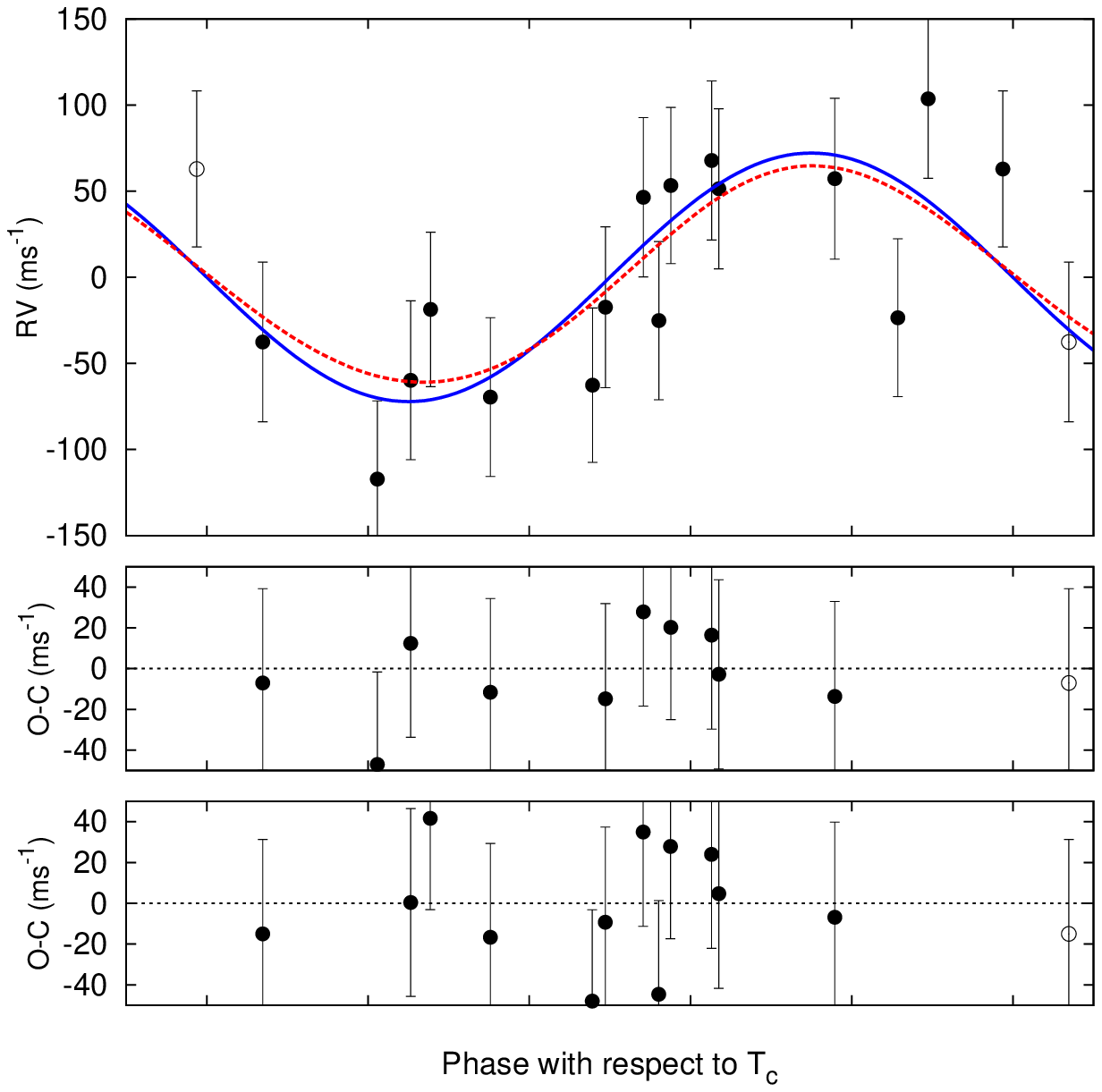}
\caption{
    {\bf Top:} RV measurements from Keck for \kepviii{}, along with an
    orbital fit, shown as a function of orbital phase, using our
    best-fit period. Solid (blue) line shows the
    circular orbital fit with binned RV model (3 bins, separated by
    600\,seconds). The dashed (red) line shows the eccentric orbital
    fit (with the same bin parameters). Zero phase is defined by the
    transit midpoint. The center-of-mass velocity has been subtracted. 
    Note that the error bars include the stellar jitter (taken for the
    circular solution), added in quadrature to the formal errors given
    by the spectrum reduction pipeline.
    {\bf Middle:} phased residuals after subtracting the orbital fit
    for the circular solution. The r.m.s.~variation of the residuals is
    \kepviiicirRVfitrms\,\ms, and the stellar jitter that was added in
    quadrature to the formal RV errors is $\kepviiicirRVjitter$\,\ms.
	{\bf Bottom:} phased residuals after subtracting the orbital fit
    for the eccentric solution.  Here the r.m.s.~variation of the
    residuals is \kepviiieccRVfitrms\,\ms, and the stellar jitter is
    $\kepviiieccRVjitter$\,\ms.
\label{fig:kep8rv}}
\end{figure}

We find the orbit is consistent with a circular orbit and derive
a significantly lower planetary mass than that reported by \citet{jen10a}.
The fitted models on the phase-folded \lcs\ are
shown on \reffigl{kep8prim}. The orbital fits to the RV points are
shown in \reffig{kep8rv}, depicting both the circular and eccentric
fits. The final planetary parameters are summarized at the bottom of
Table~\ref{tab:kep8tab}.

The residuals of the lightcurve fits from method A (fitted limb
darkening) were inspected using the Durbin-Watson statistic, which
finds $d = 1.344$, indicating time correlated noise is certainly
present. The origin of this is very likely the heater oscillations
described earlier. An F-test periodogram finds the largest peak
occuring at 3.27\,hours with amplitude 50.6\,ppm, significance
4.25-$\sigma$. The rich forest of perdiodic lines make a correction
unfeasible, as seen in Figure~\ref{fig:kep8prim}.

\begin{figure}[!ht]
\plotone{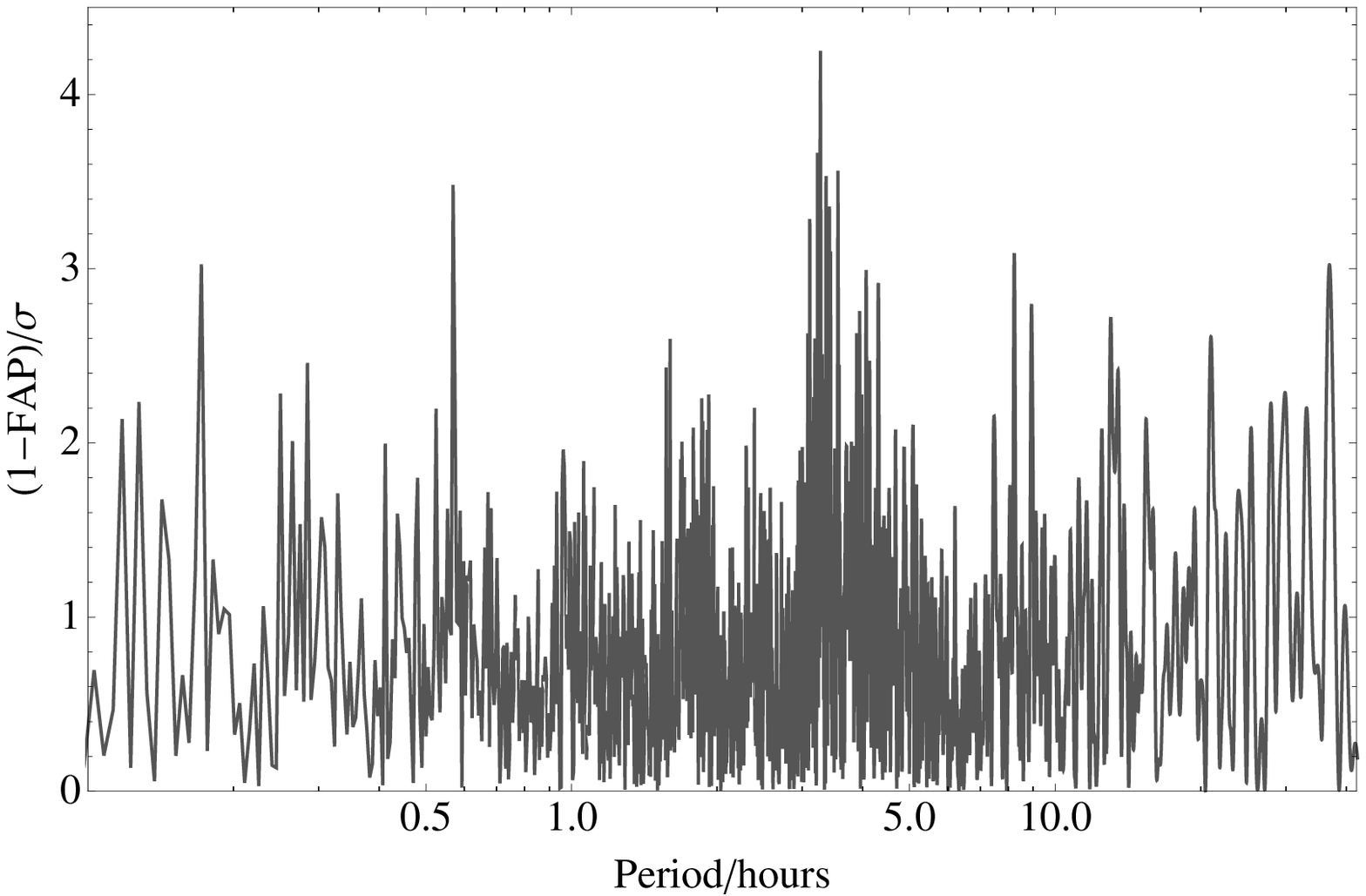}
\caption{
    F-test periodogram of residuals from primary transit \lc\ of 
    \kepviii. The y-axis gives the false-alarm-probability
    subtracted from unity, in sigmas confidence. The highest peak
    occurs for a period of 3.27\,hours with amplitude 50.6\,ppm.
\label{fig:kep8prim}}
\end{figure}

\subsubsection{Eccentricity}

The global circular fit yields a $\chi^2 = 20.96$ and the eccentric
fit obtains $\chi^2 = 20.55$. Based on an F-test, the inclusion of
two new degrees of freedom to describe the eccentricity is justified at
a significance of 13.1\% and thus is not accepted.

\subsubsection{Secondary eclipse}

Using the circular fit, there is no evidence for a secondary eclipse; we
can exclude secondary eclipses of depth $101.1$\,ppm to 3-$\sigma$
confidence, or equivalently geometric albedos $\geq 0.63$. We can also
exclude a brightness temperature $\geq 2932$\,K to the same confidence
level, which places no constraints on the redistribution of heat around
the planet.

\begin{figure}[!ht]
\plotone{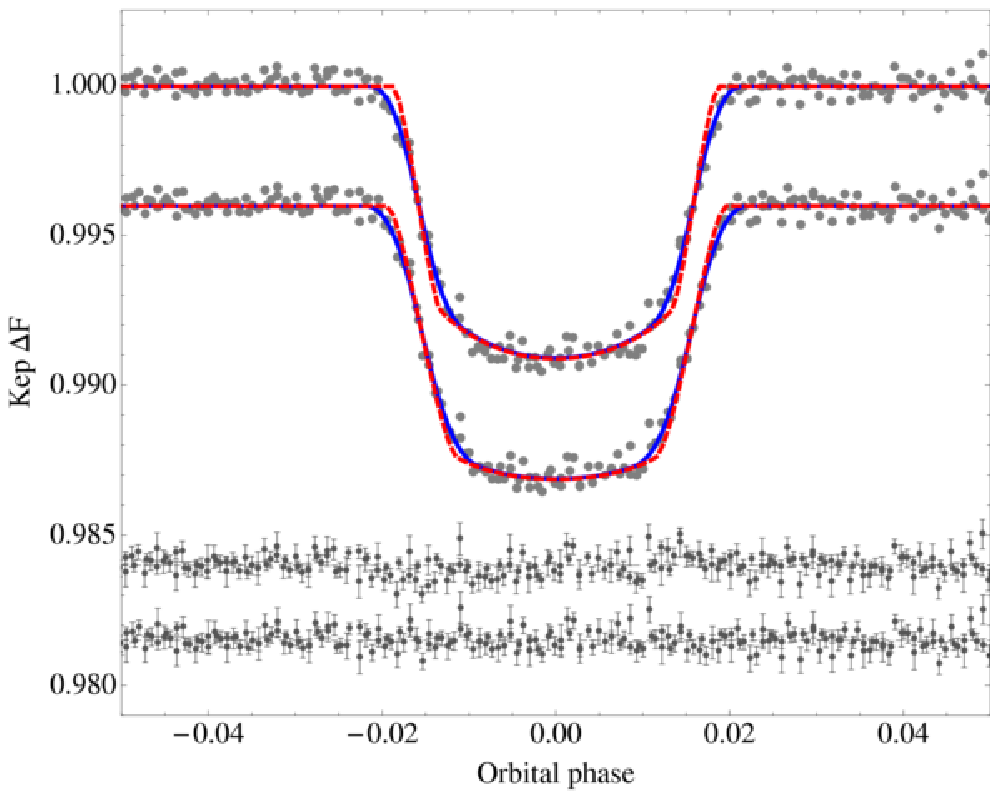}
\caption{
    {\bf Top:} Phase-folded primary transit \lc\ of \kepviii. The upper
    curve shows the circular fit, the bottom curve the eccentric fit. 
    Solid (blue) lines indicate the best fit resampled
    model (with bin-number 4). The dashed (red) lines show the
    corresponding unbinned model, which one would get if the transit
    was observed with infinitely fine time resolution.
	Residuals from the best fit resampled model for the
	circular and eccentric solutions are shown below in
        respective order.
\label{fig:kep8prim}}
\end{figure}

\subsubsection{Properties of the parent star \kepviii}
\label{sec:stelparam}

\begin{figure}[!ht]
\plotone{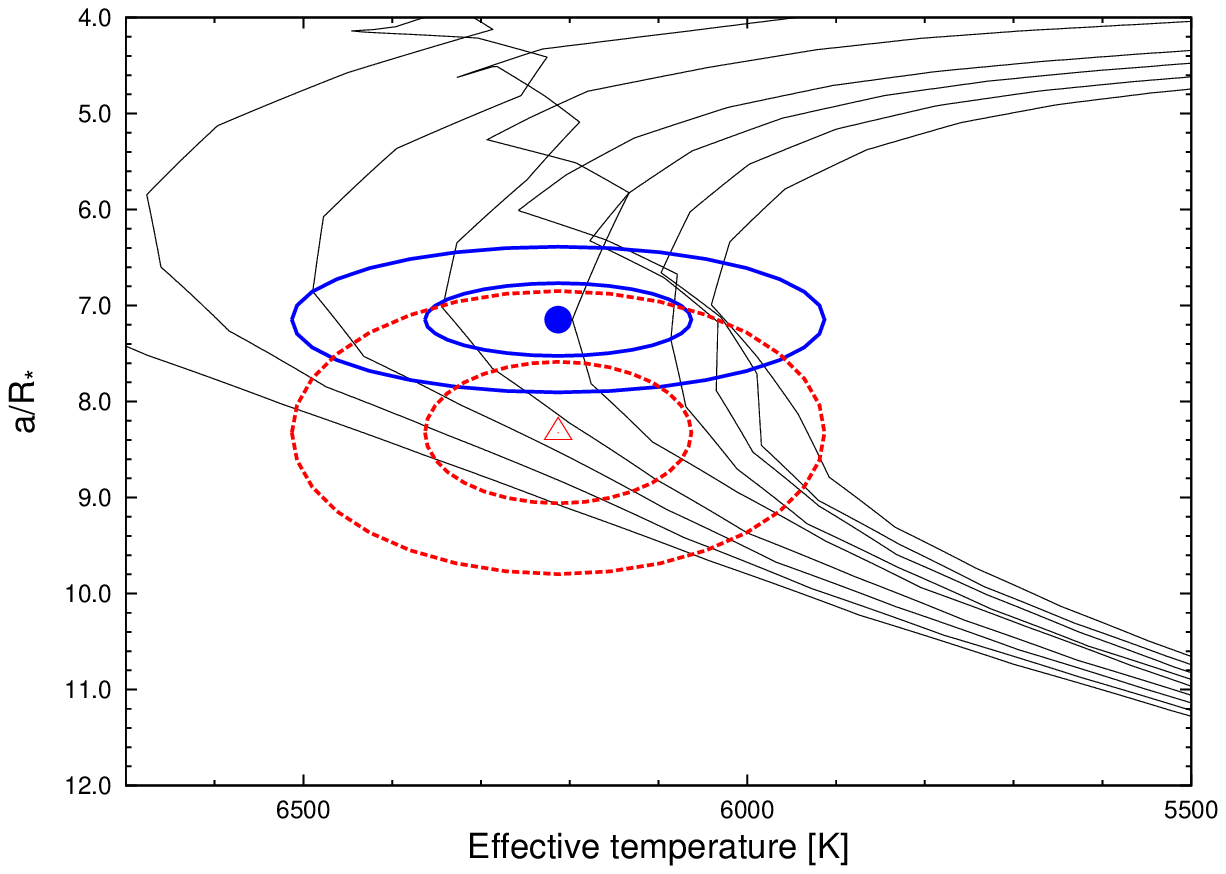}
\caption{
    Stellar isochrones from \citet{yi:2001} for metallicity
    \feh=\kepviiicirSMEiizfehshort\ and ages 
	1.0, 1.4, 1.8, 2.2, 2.6, 3.0, 3.4 and 3.8\,Gyrs.
	The final choice of $\teffstar$ and \arstar\ for the circular
    solution are marked by a filled circle, and is encircled by the
    1$\sigma$ and 2$\sigma$ confidence ellipsoids (solid, blue lines). 
    Corresponding values and confidence ellipsoids for the eccentric
    solution are shown with a triangle and dashed (red) lines.
\label{fig:kep8iso}}
\end{figure}

The Yonsei-Yale model isochrones from \citet{yi:2001} for metallicity
\feh=\kepviiicirSMEzfehshort\ are plotted in \reffigl{kep8iso}, with the
final choice of effective temperature $\teffstar$ and \arstar\ marked,
and encircled by the 1$\sigma$ and 2$\sigma$ confidence ellipsoids,
both for the circular and the eccentric orbital solution. Here the MCMC
distribution of \arstar\ comes from the global modeling of the data.

\begin{table*}

\caption{
	\emph{Fits for Kepler-8b.  Quoted values are medians of
	MCMC trials with errors given by 1-$\sigma$ quantiles. Two independent
	methods are used to fit the data (A\&B) with both circular and eccentric 
	modes (c\&e).* = fixed parameter; $\dagger$ = parameter was floated but 
	not fitted.}
} 
\centering 
\begin{tabular}{c c c c c c} 
\hline\hline 
Parameter & Discovery & Method A.c & Method A.e & Model B.c & Model B.e \\ [0.5ex] 
\hline
\emph{Fitted params.} & & \\
\hline 
$P/$days & $3.52254_{-0.00005}^{+0.00003}$ & $3.52226_{-0.00013}^{+0.00013}$ & $3.522260_{-0.000065}^{+0.000066}$ & $\kepviiicirLCP$ & $\kepviiieccLCP$ \\
$E$ (BJD-2,454,000) & $954.1182_{-0.0004}^{+0.0003}$ & $954.11849_{-0.00062}^{+0.00062}$ & $954.1184_{-0.0018}^{+0.0021}$ & $954.11912 \pm 0.00040$ & $954.11926 \pm 0.00065$ \\
$T_{1,4}$/s & - & $11554_{-360}^{+354}$ & $11780_{-143}^{+139}$ & $11664 \pm 173$ & $11673 \pm 173$ \\
$T$/s & - & $9837_{-373}^{+358}$ & $9693_{-233}^{+293}$ & $9994 \pm 46$ & $10000 \pm 41$ \\ 
$(T_{1,2} \simeq T_{3,4})$/s & - & $1802_{-720}^{+737}$ & $2417_{-323}^{+274}$ & $1745 \pm 207$ & $1745 \pm 207$ \\
$(\rpl/\rstar)^2$/\% & - & $0.887_{-0.106}^{+0.091}$ & $0.942_{-0.041}^{+0.092}$ & $0.870 \pm 0.022$ & $0.870 \pm 0.022$ \\
$b^2$ & - & $0.48_{-0.31}^{+0.14}$ & $0.600_{-0.057}^{+0.035}$ & $\kepviiicirLCbsq$ & $\kepviiieccLCbsq$ \\
$(\zeta/\rstar)$/days$^{-1}$ & - & $17.68_{-0.66}^{+0.75}$ & $18.70_{-0.51}^{+0.68}$ & $\kepviiicirLCzeta$ & $\kepviiieccLCzeta$ \\
$(F_{P,\mathrm{d}}/F_\star)$/ppm & - & $-12_{-105}^{+103}$ & - & $-27 \pm 51$ & $-20 \pm 62$   \\
$e\sin \omega$ & 0$^{*}$ & 0$^{*}$ & $0.32_{-0.10}^{+0.14}$ & 0$^{*}$ & $\kepviiieccRVh$ \\
$e\cos \omega$ & 0$^{*}$ & 0$^{*}$ & $-0.01_{-0.14}^{+0.15}$ & 0$^{*}$ & $\kepviiieccRVk$ \\
$K$/ms$^{-1}$ & $68.4 \pm 12.0$ & $67.9_{-36.8}^{+36.7}$ & $71.1_{-21.1}^{+22.1}$ & $\kepviiicirRVK$ & $\kepviiieccRVK$ \\
$\gamma$/ms$^{-1}$ & $-52.72 \pm 0.10$ & $-11.6_{-25.6}^{+25.6}$ & $-10.8_{-13.9}^{+13.7}$ & $\kepviiicirRVgammarel$ & $\kepviiieccRVgammarel$ \\
$\dot{\gamma}$/ms$^{-1}$day$^{-1}$ & 0$^{*}$ & 0$^{*}$ & 0$^{*}$ & 0$^{*}$ & 0$^{*}$ \\
$t_{\mathrm{troj}}$/days & 0$^{*}$ & 0$^{*}$ & 0$^{*}$ & 0$^{*}$ & 0$^{*}$ \\
$B$ & 1$^{*}$ & $1.000 \pm 0.004$ $\dagger$ & $1.000 \pm 0.004$ $\dagger$ & $1.000 \pm 0.004$ $\dagger$ & $1.000 \pm 0.004$ $\dagger$ \\
\hline
\emph{SME derived params.} & & \\
\hline
$T_{\mathrm{eff}}$/K & \kepviiicirSMEteff & \kepviiicirSMEteff & \kepviiicirSMEteff & \kepviiicirSMEteff & \kepviiicirSMEteff \\
log$(g/\mathrm{cgs})$ & $4.28 \pm 0.10$ & $4.28 \pm 0.10$ & $4.28 \pm 0.10$ & $4.28 \pm 0.10$ & $4.28 \pm 0.10$ \\
$\feh$ (dex) & \kepviiicirSMEzfeh & \kepviiicirSMEzfeh & \kepviiicirSMEzfeh & \kepviiicirSMEzfeh & \kepviiicirSMEzfeh \\
$\vsini$ (\kms) & \kepviiicirSMEvsin & \kepviiicirSMEvsin & \kepviiicirSMEvsin & \kepviiicirSMEvsin & \kepviiicirSMEvsin \\
$u_1$ & - & $0.41_{-0.25}^{+0.55}$ & $0.56_{-0.39}^{+0.78}$ & \kepviiicirLBikep$^{*}$ & \kepviiieccLBikep$^{*}$ \\
$u_2$ & - & $0.11_{-0.83}^{+0.44}$ & $-0.26_{-0.89}^{+0.51}$ & \kepviiicirLBiikep$^{*}$ & \kepviiieccLBiikep$^{*}$ \\
\hline
\emph{Model indep.~params.} & & \\
\hline
$\rpl/\rstar$ & $0.09809_{-0.00046}^{+0.00040}$ & $0.0942_{-0.0058}^{+0.0047}$ & $0.0971_{-0.0021}^{+0.0046}$ & $\kepviiicirLCrprstar$ & $\kepviiieccLCrprstar$ \\
$\arstar$ & $6.97_{-0.24}^{+0.20}$ & $7.16_{-0.82}^{+1.57}$ & $4.71_{-0.61}^{+0.51}$ & $\kepviiicirPPar$ & $\kepviiieccPPar$ \\
$b$ & $0.724 \pm 0.020$ & $0.691_{-0.281}^{+0.095}$ & $0.775_{-0.038}^{+0.022}$ & $\kepviiicirLCimp$ & $\kepviiieccLCimp$ \\
$i$/$^{\circ}$ & $84.07 \pm 0.33$ & $84.5_{-1.6}^{+2.8}$ & $75.8_{-6.8}^{+3.2}$ & $\kepviiicirPPi$ & $\kepviiieccPPi$ \\
$e$ & 0$^{*}$ & 0$^{*}$ & $0.35_{-0.11}^{+0.15}$ & 0$^{*}$ & $\kepviiieccRVeccen$ \\
$\omega$/$^{\circ}$ & - & - & $91_{-24}^{+24}$ & - & $\kepviiieccRVomega$ \\
RV jitter (\ms) & - & 19.7 & 20.7 & \kepviiicirRVjitter & \kepviiieccRVjitter \\
$\rho_*$/gcm$^{-3}$ & $0.522$ & $0.56_{-0.17}^{+0.45}$ & $0.160_{-0.054}^{+0.057}$ & \kepviiicirISOrho & \kepviiieccISOrho \\
log$(g_P/\mathrm{cgs})$ & $2.871 \pm 0.119$ & $2.91_{-0.36}^{+0.27}$ & $2.50_{-0.18}^{+0.15}$ & $\kepviiicirPPlogg$ & $\kepviiieccPPlogg$ \\
$S_{1,4}$/s & - & $11554_{-360}^{+354}$ & $0_{-0}^{+0}$ & $11664 \pm 173$ & $13046 \pm 2592$ \\
$t_{sec}$/(BJD-2,454,000) & - & $959.40166_{-0.00062}^{+0.00062}$ & $957.624_{-0.32}^{+1.64}$ & $955.88033 \pm 0.00040$ & $955.95 \pm 0.21$ \\
\hline
\emph{Derived stellar params.} & & \\
\hline
$\mstar/\msun$ & $1.213_{-0.063}^{+0.067}$ & $1.214_{-0.087}^{+0.092}$ & $1.46_{-0.13}^{+0.12}$ & \kepviiicirISOmlong & \kepviiieccISOmlong \\ 
$\rstar/\rsun$ & $1.486_{-0.062}^{+0.053}$ & $1.46_{-0.26}^{+0.21}$ & $2.33_{-0.26}^{+0.40}$ & \kepviiicirISOrlong & \kepviiieccISOrlong \\
log$(g/\mathrm{cgs})$ & $4.174 \pm 0.026$ & $4.192_{-0.094}^{+0.149}$ & $3.860_{-0.110}^{+0.084}$ & \kepviiicirISOlogg & \kepviiieccISOlogg \\ 
$\lstar/\lsun$ & $4.03_{-0.54}^{+0.52}$ & $2.81_{-0.94}^{+0.98}$ & $7.3_{-1.7}^{+2.9}$ & \kepviiicirISOlum & \kepviiieccISOlum \\
$M_{V}$/mag & $3.28 \pm 0.15$ & $3.66_{-0.34}^{+0.45}$ & $2.61_{-0.37}^{+0.31}$ & \kepviiicirISOmv & \kepviiieccISOmv \\
Age/Gyr & $3.84 \pm 1.5$ & $3.23_{-0.88}^{+1.37}$ & $2.78_{-0.60}^{+0.99}$ & \kepviiicirISOage & \kepviiieccISOage \\
Distance/pc & $1330 \pm 180$ & $935_{-174}^{+158}$ & $1516_{-199}^{+278}$ & \kepviiicirXdist & \kepviiieccXdist \\
\hline
\emph{Derived planetary params.} & & \\
\hline
$\mpl/\mjup$ & $0.603_{-0.19}^{+0.13}$ & $0.58_{-0.31}^{+0.32}$ & $0.66_{-0.20}^{+0.21}$ & $\kepviiicirPPmlong$ & $\kepviiieccPPmlong$ \\ 
$\rpl/\rjup$ & $1.419_{-0.058}^{+0.056}$ & $1.35_{-0.31}^{+0.26}$ & $2.23_{-0.27}^{+0.39}$ & $\kepviiicirPPrlong$ & $\kepviiieccPPrlong$ \\ 
$\rho_P$/gcm$^{-3}$ & $0.261 \pm 0.071$ & $0.28_{-0.17}^{+0.36}$ & $0.0693_{-0.028}^{+0.040}$ & $\kepviiicirPPrho$ & $\kepviiieccPPrho$\\ 
$a$/AU & $0.0483_{-0.0012}^{+0.0006}$ & $0.0483_{-0.0012}^{+0.0012}$ & $0.0514_{-0.0016}^{+0.0014}$ & $\kepviiicirPParel$ & $\kepviiieccPParel$ \\ 
$T_{\mathrm{eq}}/K$ & $1764 \pm 200$ & $1645_{-146}^{+108}$ & $2059_{-126}^{+190}$ & $\kepviiicirPPteff$ & $\kepviiieccPPteff$ \\ [1ex]
\hline\hline 
\end{tabular}
\label{tab:kep8tab} 
\end{table*}

\subsection{Transit timing analysis}
\subsubsection{Analysis of variance}

We find TTV residuals of 68.1 seconds and TDV residuals of 92.2
seconds.  The TTV indicates timings consistent with a linear ephemeris,
producing a $\chi^2$ of 10.7 for 11 degrees of freedom.  The TDV is
also consistent with a non-variable system exhibiting a $\chi^2$ of
18.8 for 12 degrees of freedom (84.9\% significance of excess).

For the TTV, we note that for 11 degrees of freedom, excess scatter
producing a $\chi^2$ of 28.5 would be detected to 3-$\sigma$
confidence. This therefore excludes a short-period signal of
r.m.s.~amplitude $\geq 58.8$\,seconds. Similarly, for the TDV, we exclude
scatter producing a $\chi^2$ of 30.1, or a short-period
r.m.s.~amplitude of $\geq 73.0$\,seconds, to 3-$\sigma$ confidence. This
consitutes a 1.46\% variation in the transit duration.

These limits place constraints on the presence of perturbing planets,
moons and Trojans. For a 2:1 mean-motion resonance, the libration period would
be $\sim83$\,cycles and thus we do not possess sufficient baseline yet
to look for such perturbers.  However, a 4:3 resonance system would
yield a libration period of $\sim19$\,cycles and therefore we would
expect to see such a signal with the 13 cycles observed.  This means a
4:3 perturber of $\geq 0.50 \mearth$ is ruled out to $3$-$\sigma$
confidence by the current TTV data for this object.  After 83 cycles
have been obtained at the same photometric quality, the presence of a
2:1 resonant perturber of mass $\geq 0.17 \mearth$ may be
constrained.

The current TTV data rules out an exomoon at the maximum orbital radius
(for retrograde revolution) of $\geq 2.1 \mearth$.  The current TDV data
rules an exomoon of $\geq 21.5 \mearth$ at a minimum orbital
distance.

Using the expressions of \citet{for07} and assuming $\mpl \gg 
M_{\mathrm{Trojan}}$, the expected libration period of \kepviiib\
induced by Trojans at L4/L5 is 25.5 cycles and therefore we do
not yet possess sufficient baseline to search for these TTVs.
Inspection of the folded photometry at $\pm P/6$ 
from the transit centre excludes Trojans of 
effective radius $\geq 1.58 R_{\oplus}$ to 3-$\sigma$ confidence.

\subsubsection{Periodograms}

The TTV periodogram reveals a peak at twice the Nyquist period, which
we disregard. No other significant peaks are present. The TDV also has
no significant peaks. The phasing signal appears to have a period of 2
cycles.

\begin{figure*}
\begin{center}
\includegraphics[width=16.8 cm]{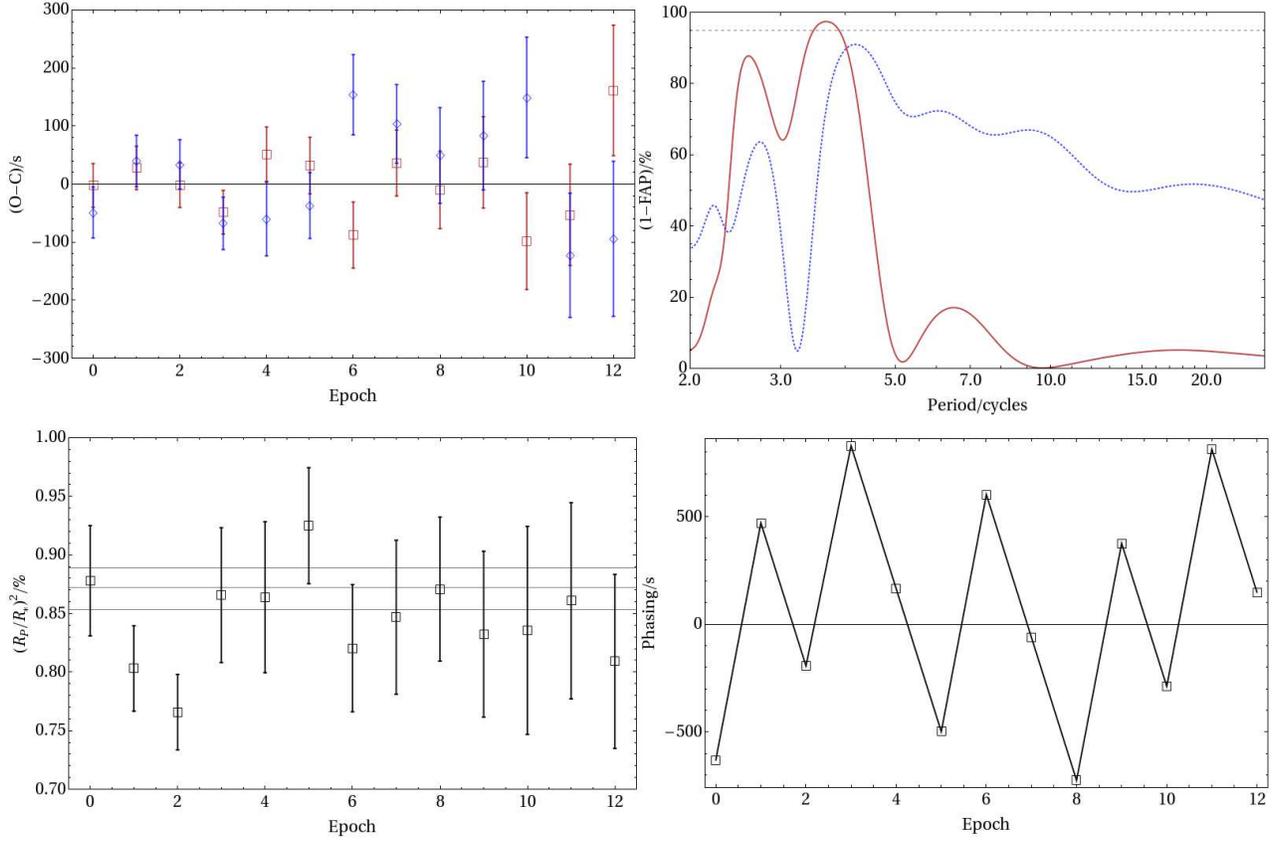}
\caption{\emph{
	{\bf Upper Left:} TTV (squares) and TDV (diamonds) for \kepviiib\
	(see \S2.3.1 for details).
	{\bf Upper Right:} TTV periodogram (solid) and TDV periodogram
	(dashed) for \kepviiib, calculated using a sequential F-test (see
	\S2.3.2 for description).
	{\bf Lower Left:} Transit depths from individual fits of \kepviiib\
	(see \S2.3.1 for details).
	{\bf Lower Right:} ``Phasing'' of Kepler long cadence photometry
	for \kepviiib\ (see \S2.3.4 for description).
}}
\label{fig:kep8ttv}
\end{center}
\end{figure*}

\subsection{Depth and OOT Variations}

The transit depth produces a slightly excess scatter in reference to
the global fit.  We find a $\chi^2$ of 18.0 for 12 degrees of freedom,
which is 1.6-$\sigma$ significant.  The OOT is consistent with the
global value yielding $\chi^2 = 13.3$ for 12 degrees of freedom.  As
for the TDV, the varying depths are neither statistically significant
nor particularly robust in light of the irregularities encountered with
this data set.

\begin{table*}
\caption{
	\emph{Mid-transit times, transit durations, transit depths and
	out-of-transit (normalized) fluxes for Kepler-8b.}
}
\centering 
\begin{tabular}{c c c c c} 
\hline\hline 
Epoch & $t_c$/(BJD-2,454,000) & $T$/s & $(\rpl/\rstar)^2/\%$ & $F_{oot}$ \\ [0.5ex] 
\hline
0 & $953.61846_{-0.00043}^{+0.00043}$ & $9887.5_{-86.6}^{+88.5}$ & $0.878_{-0.053}^{+0.041}$ & $0.999971_{-0.000022}^{+0.000022}$ \\
1 & $957.14108_{-0.00043}^{+0.00043}$ & $10064.9_{-88.3}^{+88.4}$ & $0.803_{-0.031}^{+0.042}$ & $0.999999_{-0.000022}^{+0.000022}$ \\
2 & $960.66300_{-0.00044}^{+0.00045}$ & $10052.9_{-84.9}^{+84.7}$ & $0.766_{-0.024}^{+0.040}$ & $0.999984_{-0.000023}^{+0.000023}$ \\
3 & $964.18472_{-0.00044}^{+0.00043}$ & $9850.0_{-88.2}^{+91.8}$ & $0.866_{-0.060}^{+0.055}$ & $0.999974_{-0.000026}^{+0.000026}$ \\
4 & $967.70814_{-0.00054}^{+0.00054}$ & $9865.0_{-130.1}^{+123.6}$ & $0.864_{-0.072}^{+0.056}$ & $0.999946_{-0.000026}^{+0.000027}$ \\
5 & $971.23018_{-0.00056}^{+0.00056}$ & $9911.2_{-112.5}^{+113.7}$ & $0.925_{-0.054}^{+0.045}$ & $0.999951_{-0.000029}^{+0.000029}$ \\
6 & $974.75106_{-0.00066}^{+0.00066}$ & $10292.7_{-136.6}^{+139.2}$ & $0.820_{-0.055}^{+0.054}$ & $0.999993_{-0.000032}^{+0.000032}$ \\
7 & $978.27476_{-0.00065}^{+0.00067}$ & $10192.1_{-138.6}^{+132.1}$ & $0.847_{-0.055}^{+0.076}$ & $1.000071_{-0.000036}^{+0.000036}$ \\
8 & $981.79649_{-0.00079}^{+0.00076}$ & $10083.6_{-160.2}^{+168.9}$ & $0.871_{-0.062}^{+0.061}$ & $0.999998_{-0.000040}^{+0.000039}$ \\
9 & $985.31931_{-0.00092}^{+0.00090}$ & $10151.9_{-_190.0}^{+183.6}$ & $0.832_{-0.069}^{+0.073}$ & $1.000013_{-0.000045}^{+0.000044}$ \\
10 & $988.84000_{-0.00095}^{+0.00098}$ & $10283.4_{-218.5}^{+196.2}$ & $0.836_{-0.073}^{+0.105}$ & $1.000024_{-0.000051}^{+0.000051}$ \\
11 & $992.36279_{-0.00099}^{+0.00102}$ & $9740.2_{-206.6}^{+220.9}$ & $0.861_{-0.071}^{+0.097}$ & $0.999981_{-0.000056}^{+0.000056}$ \\
12 & $995.88753_{-0.00128}^{+0.00132}$ & $9796.6_{-278.6}^{+255.2}$ & $0.809_{-0.062}^{+0.087}$ & $0.999909_{-0.000063}^{+0.000063}$ \\ [1ex]
\hline
\end{tabular}
\label{tab:tab8ttv} 
\end{table*}
